\documentclass[fleqn,usenatbib]{mnras}

\usepackage{newtxtext,newtxmath}

\usepackage[T1]{fontenc}
\usepackage{lineno}

\DeclareRobustCommand{\VAN}[3]{#2}
\let\VANthebibliography\thebibliography
\def\thebibliography{\DeclareRobustCommand{\VAN}[3]{##3}\VANthebibliography}

\usepackage{graphicx}	
\usepackage{amsmath}	
\usepackage{hyperref}
\usepackage{multirow}
\usepackage{wasysym}
\usepackage{soul}

\newcommand{\ulp}{ASKAP J1448$-$6856}
\newcommand{\ulplong}{ASKAP J144834$-$685644}

\title[\ulp]{ASKAP J144834$-$685644: a newly discovered long period radio transient detected from radio to X-rays}

\author[A. Anumarlapudi et al.]{
Akash Anumarlapudi,$^{1}$\thanks{E-mail: aakash@uwm.edu}
David L. Kaplan,$^{1}$
Nanda Rea,$^{2,3}$
Nicolas Erasmus,$^{4,5}$
Daniel Kelson,$^6$
\newauthor
Stella~Koch~Ocker,$^{7,6}$
Emil Lenc,$^8$
Dougal Dobie,$^{9,10}$
Natasha~Hurley-Walker,$^{11}$
Gregory Sivakoff,$^{12}$
\newauthor
David A.~H.~Buckley,$^{4,13,14}$
Tara Murphy,$^{9,10}$
Joshua Pritchard,$^{15}$
Laura Driessen,$^9$
Kovi Rose,$^{9,15}$
\newauthor
Andrew Zic$^{15,10}$
\\
$^{1}$Department of Physics, University of Wisconsin-Milwaukee, P.O. Box 413, Milwaukee, WI 53201, USA\\
$^{2}$Institute of Space Sciences (ICE), CSIC, Campus UAB, Carrer de Can Magrans s/n, E-08193, Barcelona, Spain\\
$^{3}$Institut d’Estudis Espacials de Catalunya (IEEC), 
Castelldefels (Barcelona), Spain\\
$^4$South African Astronomical Observatory, PO Box 9, Observatory, 7935, Cape Town, South Africa\\
$^5$Department of Physics, Stellenbosch University, Stellenbosch, 7602, South Africa\\
$^6$Observatories of the Carnegie Institution for Science, Pasadena, CA 91101, USA\\
$^7$Cahill Center for Astronomy and Astrophysics, California Institute of Technology, Pasadena, CA 91125, USA\\
$^8$CSIRO Space and Astronomy, PO Box 76, Epping, NSW, 1710, Australia\\
$^9$ Sydney Institute for Astronomy, School of Physics, University of Sydney, NSW, 2006, Australia\\
$^{10}$ARC Centre of Excellence for Gravitational Wave Discovery (OzGrav), Hawthorn, Victoria, Australia\\
$^{11}$ International Centre for Radio Astronomy Research, Curtin University, Kent Street, Bentley WA, 6102, Australia\\
$^{12}$Department of Physics, University of Alberta, CCIS 4-181, Edmonton AB T6G 2E1, Canada\\
$^{13}$Department of Astronomy, University of Cape Town, Private Bag X3, Rondebosch 7701, South Africa\\
$^{14}$Department of Physics, University of the Free State, PO Box 339, Bloemfontein 9300, South Africa\\
$^{15}$Australia Telescope National Facility, CSIRO Space and Astronomy, PO Box 76, Epping, NSW 1710, Australia\\
}

\date{Accepted XXX. Received YYY; in original form ZZZ}

\begin{document}
\label{firstpage}
\pagerange{\pageref{firstpage}--\pageref{lastpage}}
\maketitle

\begin{abstract}
Long-period radio transients (LPTs) are an emerging group of radio transients that show periodic polarised radio bursts with periods varying from a few minutes to a few hours. Fewer than a dozen LPTs have been detected so far, and their origin (source and emission mechanism) remains unclear. Here, we report the discovery of a 1.5\,hr LPT, \ulplong, adding to the current sample of sources. \ulplong\  is one of the very few LPTs that has been detected from X-rays to radio. It shows a steep radio spectrum and polarised radio bursts, which resemble the radio emission in known LPTs. In addition, it also shows highly structured and periodic narrow-band radio emission. Multi-wavelength properties suggest that the spectral energy distribution (SED) peaks at near ultraviolet wavelengths, indicating the presence of a hot magnetic source. Combining multi-wavelength information, we infer that \ulplong\ may be a near edge-on magnetic white dwarf binary (MWD), although we can not fully rule out \ulplong\  being an isolated white dwarf pulsar\ or even a transitional millisecond pulsar (despite the lack of radio pulsations). 
If \ulplong\ is a MWD binary, the observed broadband spectral energy distribution can be explained by emission from an accretion disk. This hints that some fraction of optically bright LPTs may be accreting binaries with the radio period being the orbital period.  It might further suggest a connection between optically bright synchronized LPTs, such as polars, and non-accreting asynchronous WD pulsars, such as AR Sco and J1912$-$4410. 
\end{abstract}




\section{Introduction} \label{sec:intro}
The timescales over which radio sources show variability vary from nanoseconds (the `nano-shots' from the Crab pulsar; \citealt{Hankins2003}) to years in active galactic nuclei (AGNs) and radio afterglows of extragalactic explosions such as supernovae, gamma-ray bursts, and tidal disruption events. However, not all of these timescales have been uniformly sampled by large-scale radio surveys, owing to the narrow fields of view of the radio interferometers and a low cadence ($\sim$years). With the advent of wide-field interferometers like the Australian SKA Pathfinder \citep[ASKAP;][]{askap}, Murchison Wide-field Array \citep[MWA;][]{Tingay2013,mwa}, and more broadly the LOw Frequency ARray \citep[LOFAR;][]{lofar}, and MeerKAT \citep{meerkat}, the cadence of the large-scale surveys has tremendously improved, with current surveys probing the minute--hour timescale regime, something that has been poorly explored historically. 

Surveys from these wide-field interferometers have started to discover radio sources that are periodic on timescales of few minutes to hours --- dubbed  ``long period transients" \citep[LPTs, potentially also including Galactic Center radio transients or GCRTs;][]{hyman_powerful_2005,hurley-walker_radio_2022,caleb_discovery_2022,hurley-walker_long-period_2023,caleb_emission-state-switching_2024,dong_discovery_2024,hurley-walker_29-hour_2024,de_ruiter_white_2024,Dobie2024,lee2025,Wang2025,Li2024,Dong2025,Bloot2025}.
Many of these sources are reported to have steep spectra (power-law index $\alpha<-1$, where $S_\nu\propto \nu^\alpha$), indicating that the emission is not thermal in nature. All of these sources show high levels of polarisation,  linear and/or circular, although the relative contribution of these to the total polarisation is source-specific. These levels of polarisation imply very strong and highly structured magnetic fields. The inferred brightness temperature from the dispersion-measure (DM) based distances exceeds the limit for incoherent radiation ($10^{12}$\,K; \citealt{Kellermann1969}), indicating the emission mechanism is coherent. For the sources in which the period derivative was constrained \citep{hurley-walker_radio_2022,hurley-walker_long-period_2023,caleb_emission-state-switching_2024,Wang2025}, the inferred radio luminosity exceeded the spin-down luminosity, demanding an alternative source of energy if the observed period was due to the rotation of the source. In addition, these sources occupy a region in the period-period derivative ($P$-$\dot{P}$; see \citealt{handbook}) phase space that is close to or beyond the pulsar `death-valley' \citep{rea_long-period_2024}.

Until recently, none of the LPTs (or GCRTs) were detected at wavelengths other than radio, some despite very deep observations \citep{kaplan_search_2008,rea_constraining_2022}, although in many cases, these observations were not simultaneous in time with the radio bursts. However, three of the six LPTs discovered in the past year were detected at either optical or X-ray wavelengths. ASKAP J1832$-$0911 was detected at X-rays \citep{Wang2025}, and ILT J1101+5521 and GLEAM-X J0704$-$37 were detected at optical wavelengths \citep{de_ruiter_white_2024,hurley-walker_29-hour_2024}. 
ILT J1101+5521 and GLEAM-X J0704$-$37 were established as white dwarf (WD)+M-dwarf binaries with the periodicity of the radio pulses coinciding with the orbital period \citep{de_ruiter_white_2024,Rodriguez2025}. The optical emission is shown to be dominated by the thermal emission from the M-dwarf companion and, to a lesser extent, from the WD \citep{Rodriguez2025}. In ASKAP J1832$-$0911 (with no optical counterpart), the X-ray emission is pulsed at the radio period, and the X-ray and radio pulse profiles are shown to align well \citep{Wang2025}. Moreover, the X-ray and radio fluxes vary considerably over timescales of weeks.

These multi-wavelength detections can be interpreted under two different paradigms: emission from an isolated magnetic compact object (magnetized white dwarf or magnetar; \citealt{caleb_discovery_2022,caleb_emission-state-switching_2024,dong_discovery_2024,Dobie2024,Wang2025,lee2025}) or emission from WD+M-dwarf binary \citep[with optical and radio emission powered by different mechanisms;][]{de_ruiter_white_2024,hurley-walker_29-hour_2024}. 
The similarity of the radio and the optical period in the WD systems led the authors to propose that the systems might be detached binaries with strong (10--300\,MG) magnetic fields, known as \textit{polars}, although direct evidence for the strength of the magnetic field is missing. 

However, in either of the scenarios, the emission mechanism that powers the radio pulses remains unclear. If the underlying source is an isolated compact object, explaining the radio emission through canonical pulsar emission models \citep{ruderman76} can be challenging due to the long periods and the radio luminosities greater than spin-down luminosities \citep{hurley-walker_long-period_2023,caleb_emission-state-switching_2024}. An isolated white dwarf is sometimes invoked to increase the available spin-down energy budget but faces challenges given some of the observed pulse widths \citep[$\sim$10\,ms in GPM J1839$-$10;][]{Men2025}. A more favoured proposal involves a highly slowed-down magnetar, possibly due to fallback accretion after the supernova explosion \citep{ronchi2022,Fan2024}. The high degree of linear polarisation in LPTs; the mode switching in ASKAP J1935+2154 \citep{caleb_emission-state-switching_2024}, 
the close phase alignment of X-ray and radio pulses, the flat spectrum  
in ASKAP J1832$-$0911 \citep{Li2024,Wang2025}, and the observed Faraday conversion in GPM J1839$-$10 \citep{Men2025} provide some resemblance to the properties of known magnetars, but there is no direct link.

In the case of WD binaries, the radio pulses are likely due to the interaction of the WD's magnetic field with the companion, powering coherent radio pulses, for example, via electron cyclotron maser emission \citep[ECME;][]{Treumann2006}. These systems are hypothesized to be associated with so-called ``WD pulsars," (these are detached WD binaries in which the WD rotates asynchronously and emits beamed radiation like a pulsar, like AR~Sco and J1912$-$4410; \citealt{marsh_radio-pulsing_2016,pelisoli_53-min-period_2023}), although the WD temperatures in LPTs and WD pulsars seem to be distinctly different \citep{Castro2025}. The interaction of the WD's magnetic field with the companion is thought to synchronize the WD spin period with the orbital period, and as such, LPTs are thought to be evolved stages of WD pulsars \citep{Rodriguez2025b}. If there are episodes of mass transfer after the synchronization is complete, they can appear as magnetic cataclysmic variables (MCVs), and hence, LPTs are also thought to be associated with polars. However, given the very limited number of  LPT and WD pulsar discoveries, the evolutionary path of these systems and the origin of the radio emission are still uncertain. Discovering more LPTs (through radio) that show emission at other wavelengths and characterizing their multi-wavelength behaviour will help find the missing links between these systems, LPTs, WD pulsars, and MCVs.


In this article, we report the ASKAP discovery of a 1.5\,hr periodic polarised radio source that is detected from X-ray to radio wavelengths. Discovered in a search for circularly polarised sources, \ulplong\ (hereafter referred to as \ulp) shows highly variable linearly and circularly polarised emission. It also shows circular polarisation sign flips, both between the bursts and within a single burst, in addition to an inter-pulse. Here, we describe the multi-wavelength behaviour of \ulp\ to understand its nature and its emission, from X-rays to radio wavelengths. 
Based on the periodic nature of the radio bursts, we provisionally call \ulp\ a new LPT, although, as discussed at l ength, it shares some properties with other types of objects as well, so it may gain a more specific classification as more data are taken.
We describe our multi-wavelength observations and key findings in \S\ref{sec:data}. Detailed modelling of these data to estimate the source parameters is presented in \S\ref{sec:modeling}. In \S\ref{sec:discussion} we talk about the probable nature of \ulp\  and the emission mechanism that powers the multi-wavelength radiation, concluding in \S\ref{sec:conclusions}.  

\section{Observations and Data Analysis}\label{sec:data}
\ulp\ was observed with a slew of telescopes from radio to X-rays. A high-level observation log is provided in Table~\ref{tab:obs}. The subsequent subsections detail our observations and findings at each wavelength. 


\begin{table}
\centering
\caption{Observation log of \ulp.}
\label{tab:obs}
\begin{tabular}{ccc}
\hline
Date & Instrument & Band(s)\\
\hline
\multicolumn{3}{l}{Archival data (prior to first detection)\dotfill}\\
2014-04-13 -- 2021-08-16 & SkyMapper & $u, v, g, r, i, z$\\
2015-05-10 & VISTA & $J$\\
2015-05-10 & VISTA & $K_s$\\
2018-05-21 -- 2019-07-20 & DECam & $g, r, i, z, Y$ \\ \hline
\multicolumn{3}{l}{New data (after first detection)\dotfill}\\
2023-06-15 & ASKAP & 799--1090\,MHz \\
2024-05-26 & ASKAP & 799--1090\,MHz \\
2024-06-27 & ATCA & 1100--3100\,MHz \\
2024-07-09 -- 2024-07-11 & \textit{Swift} XRT &  0.2--10\,keV\\
 & \textit{Swift} UVOT & UVW2 (1928\,\AA) \\
2024-07-18 -- 2024-07-23 & Magellan & $J$\\
2024-07-30 -- 2024-07-31 & MeerKAT & 544--1712\,MHz \\
2024-07-30 & Lesedi & $g$\\
2024-07-31 -- 2024-08-01 & \textit{XMM-Newton} & 0.2--12\,keV \\ 
\hline
\end{tabular}
\end{table}

\subsection{Radio}
\ulp\ was discovered in a search for circularly polarised sources, with a flux density of $0.76\pm0.04$\,mJy in the data taken as a part of the Evolutionary Map of the Universe \citep[EMU;][]{emu,emu2} project at the ASKAP. Upon discovery, we performed follow-up observations of \ulp\ with the Australia Telescope Compact Array  \citep[ATCA;][]{cabb} and the MeerKAT radio telescope \citep{meerkat} spanning a 13-month baseline. Detailed information on the observational setup, instrumental calibration, and other radio data reduction techniques to generate images for each instrument are provided in Appendix~\ref{sec:radio_data_analysis} (\ref{sec:emudata} for ASKAP, \ref{sec:atcadata} for ATCA, and \ref{sec:mktdata} for MeerKAT). We start here with the time and frequency integrated images and the time and frequency-resolved intensities (\textit{the dynamic spectra}) for all four Stokes parameters at hand. 

\subsubsection{ASKAP EMU}\label{sec:emu}

\begin{figure*}
    \centering
    \includegraphics[width=\textwidth]{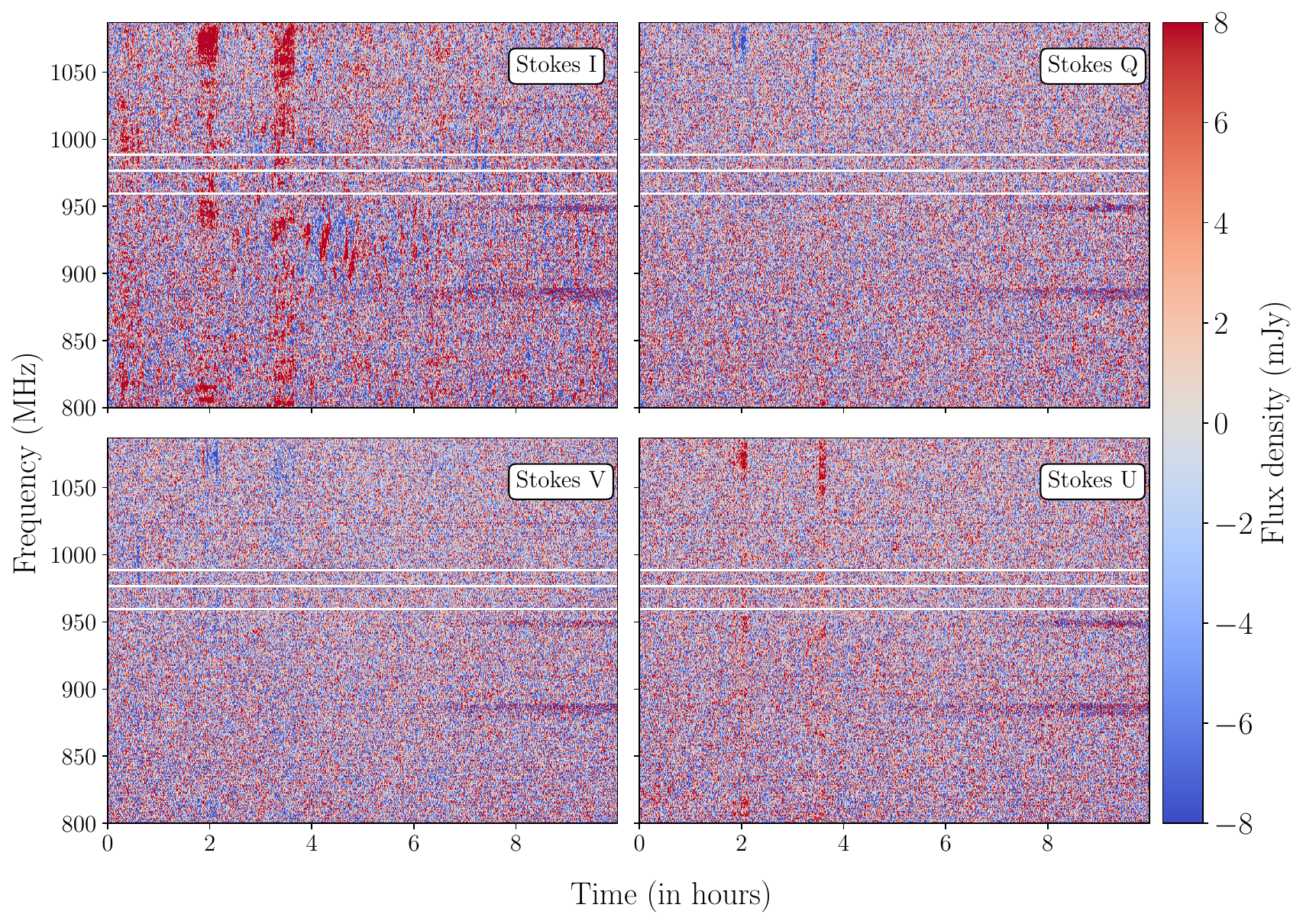}\\
    \caption{Dynamic spectra of all four Stokes parameters with a 10\,s time resolution and 1\,MHz frequency resolution from a 10\,hr EMU observation on June 15, 2023, and 799--1090\,MHz bandpass. We can see multiple radio bursts in all four polarisations. The color scale on the right represents the flux density. The horizontal gaps (white lines) in the spectra are the flagged frequency channels. The two persistent narrow-band features (around 900 and 950\,MHz) are likely due to un-flagged radio frequency interference (RFI).}
    \label{fig:emu_ds}
\end{figure*}
\begin{figure}
    \centering
    \includegraphics[width=\columnwidth]{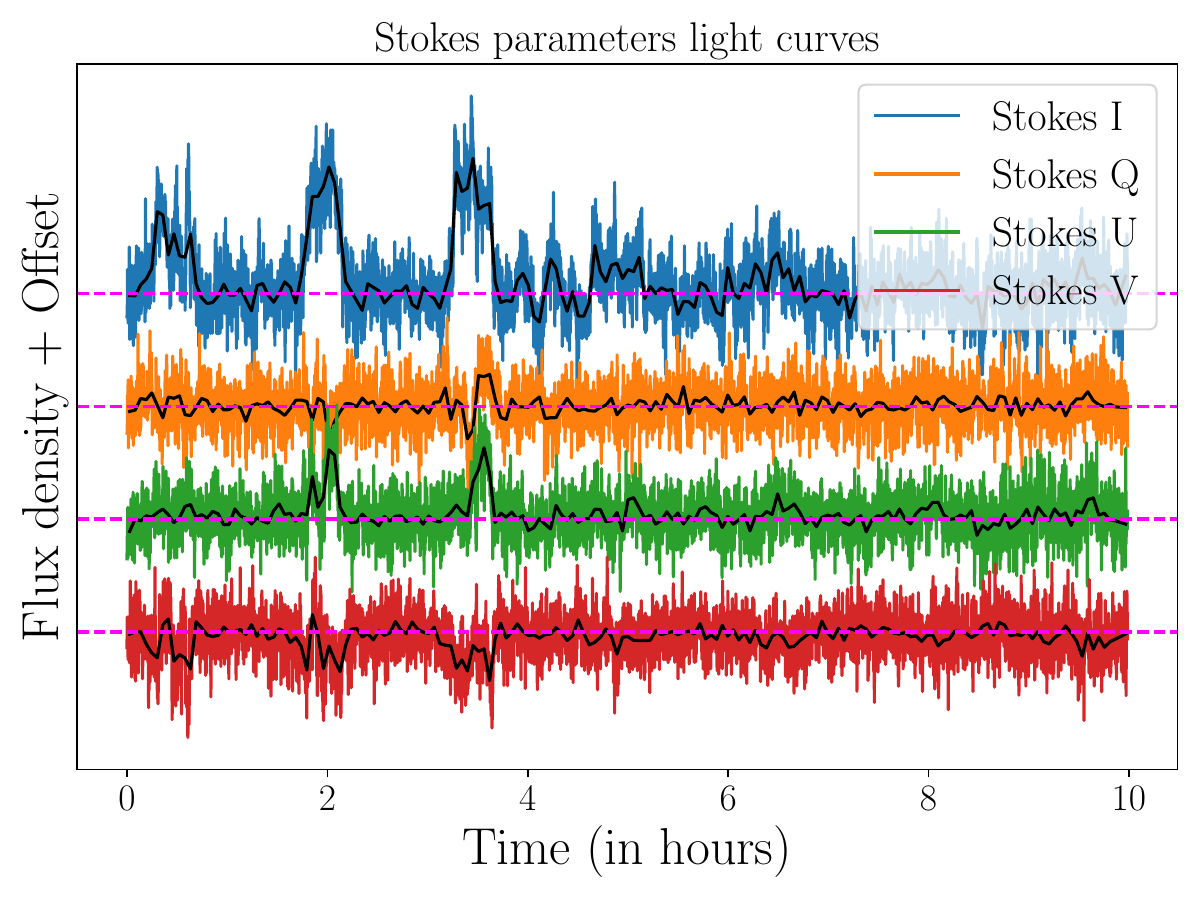}
    \caption{Light curves of the Stokes parameters from the first EMU observation obtained by averaging the dynamic spectra along the frequency axis. The native resolution of the data is 10\,s and the overlaid black lines show the light curves rebinned at 200\,s. The dashed magenta line shows zero intensity levels.}
    \label{fig:emu_lc}
\end{figure}
\begin{figure}
    \centering
    \includegraphics[width=\columnwidth]{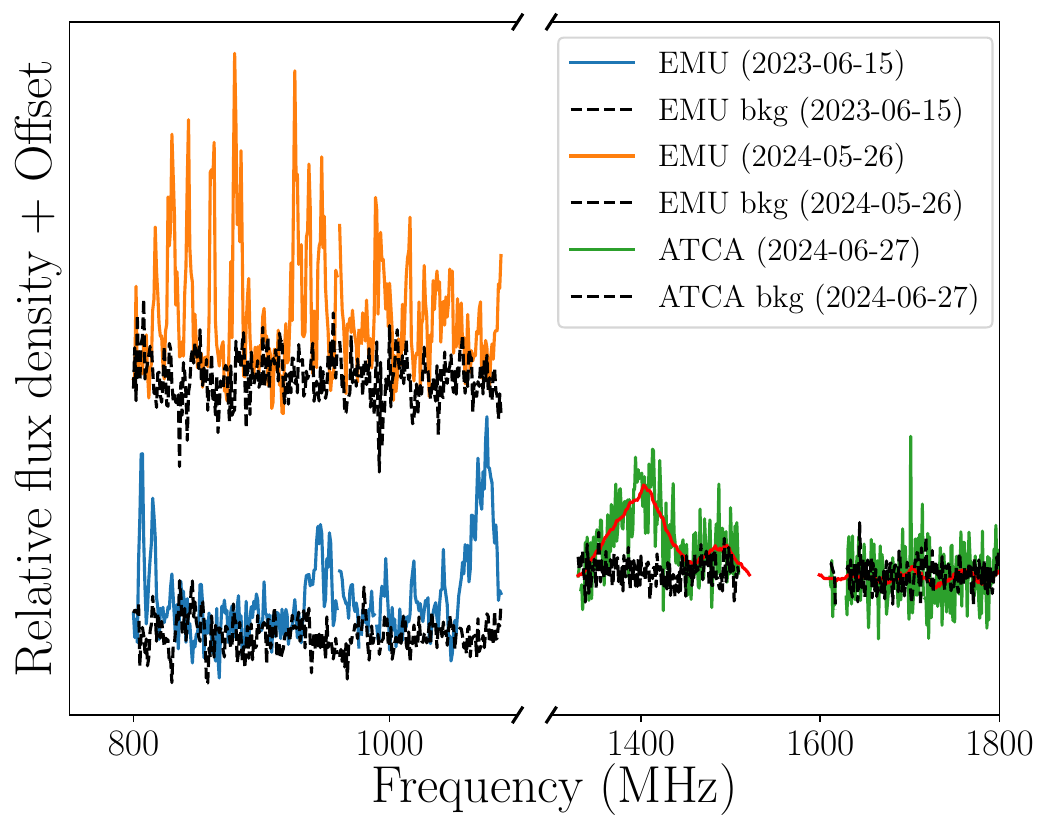}
    \caption{Frequency spectra of the bursts detected in EMU and ATCA observations. The blue curve shows the spectrum of the brightest burst in the first EMU observation, the orange curve shows the same for the second EMU observation. The green curve shows the spectrum of the only bright burst in the ATCA observation, showing a clear cut-off above $\sim$1500\,MHz. The dashed black lines are the spectra from the off-pulse regions. For the ATCA observation, the overlaid red line is the spectrum rebinned at 30\,MHz resolution.}
    \label{fig:spec}
\end{figure}

\ulp\ was discovered in the EMU data taken on June 15, 2023, between 799--1090\,MHz. We inspected the dynamic spectra of all four Stokes parameters (\textit{I, Q, U, V}), and found repeating polarised bursts (Figure~\ref{fig:emu_ds}). The light curves, shown in Figure~\ref{fig:emu_lc}, revealed a periodicity of $1.52\pm0.3$\,hr, and a burst width of $\sim$ 30\,min. In a 10-hour observation, we detected five bursts with varying intensities. Given the period, seven bursts were expected if the bursts were uniform in intensity, yielding a duty cycle of $\sim$70\%. The frequency spectrum of the bursts exhibited emission over multiple narrow bands instead of a continuum (see Figure~\ref{fig:emu_ds} and the blue curve in Figure~\ref{fig:spec}). The bursts were detected in Stokes Q, and U as well, but the low signal-to-noise (SNR) ratio per time and frequency bin and varying Stokes Q/U intensity across the burst (likely due to changing polarisation position angle), limited us from measuring  a non-zero Faraday rotation.

A second EMU observation taken 11\,months after the discovery epoch also resulted in the detection of a point source in the images, at a consistent flux level with the first EMU observation. However, the dynamic spectra showed that the pulses were narrower (15\% duty cycle) and also showed little to no linear polarisation. Instead, we observed higher levels of circular polarisation (50--100\%). In addition, we observed circular polarisation inversion, with the Stokes V light curve showing right-handed circular polarisation\footnote{In this article, we follow the definition from the International Astronomical Union convention for Stokes parameters \citep{iauconvetion,ieee}.} in one of the bursts. Given the narrower pulse widths, we obtained a better estimate of the period, $P=1.52\pm0.15$\,hr. The duty cycle was consistent with the discovery observation (five of the expected seven bursts were detected). The frequency spectrum of the bursts still showed rich narrow-band features but was a lot sharper than the first EMU observation (see the orange curve in Figure~\ref{fig:spec}). Detailed plots for the dynamic spectra, light curves, and frequency correlations are provided in Appendix~\ref{sec:emudata}.

A third EMU observation, taken on December 25, 2024, has also resulted in the detection of a point source in the total intensity image. However, the dynamic spectra (Figure~\ref{fig:emu_ds_3_app}) revealed that the bursts were weaker and hence the intensity variations resulted in the detection of only three bursts. They were still elliptically polarised and exhibited narrowband spectral features, consistent with the first two EMU observations.

\subsubsection{Australia Telescope Compact Array}\label{sec:atca}
Motivated by \ulp's behavior, we followed it up with the Australia Telescope Compact Array  \citep[ATCA;][]{cabb} at L/S band (1.1--3.1\,GHz) under the project code C3363 (PI: Tara Murphy) on July 15, 2024, two months after the second EMU observation. No point source was detected in any of the Stokes parameters in the full-time integrated images. But, the dynamic spectra revealed a single burst, with a fluence SNR of 7, that was almost 100\% circularly polarised. The burst was similar to the ones observed in the discovery EMU data in terms of its width, $\sim$ 30\,min. However, it showed a frequency cut-off around 1500\,MHz above which no detectable signal was found (see Figure~\ref{fig:spec}). The detection of a single burst coupled with the spectral cut-off was the likely reason for the non-detection in the integrated images. Detailed information on the data reduction, the dynamic spectra, and the light curves are provided in Appendix~\ref{sec:atcadata}.

\subsubsection{MeerKAT}\label{sec:mkt}

We observed \ulp\ with the MeerKAT radio telescope for 8\,hrs on July 30, and July 31, 2024 (16\,hr in total), under the proposal DDT-20240719-AA-01. We obtained the data in the split array mode, with half the array observing at UHF band (544--1088\,MHz) and the other half observing at L-band (856--1712\,MHz). Data in these frequency bands were analyzed independently (a more detailed description is provided in Appendix~\ref{sec:mktdata}). Using the Stokes I image in the UHF band, we constrained the position of \ulp\  (by fitting a 2D Gaussian) to (J2000) $\alpha=14^{\rm h}48^{\rm m}34\fs29\pm0\fs 02$ and $\delta=-68\degr 56\arcmin 44\farcs1\pm0\farcs 3$, consistent between the two MeerKAT observations.

We observed elliptically polarised bursts with varying degrees of linear and circular polarisations similar to the first EMU observation. We also observed flips in the circular polarisation within multiple bursts (in both observations). More importantly, we observed a circularly polarised interpulse in the first observation. Both MeerKAT observations showed a similar narrow-band frequency structure to ASKAP observations. Detailed data reduction, dynamic spectra, and the light curves are provided in Appendix~\ref{sec:mktdata}.

We also recorded simultaneous beamformed data at a higher time resolution (100\,$\mu$s) to estimate the dispersion measure of the sources. Unfortunately, the source was not bright enough to be detected in the beamformed data. 

\subsection{Multi-wavelength Observations}\label{sec:mw}
In what follows, we provide our multi-wavelength observations, going from X-rays to infrared (IR), rather than arranging them chronologically. 

\subsubsection{X-rays -- Swift}\label{sec:xrt}
We observed the field of \ulp\ between July 9, 2024, and July 11, 2024, for 8\,ks in nine different scans with \textit{Neil Gehrels Swift Observatory} X-ray telescope (XRT; \citealt{swift,xrt}) under Director's Discretionary Time (Target ID 16696). A 3-$\sigma$ point source was detected at at $\alpha=14^{\rm h}48^{\rm m}34\fs4\pm1\fs3$ and $\delta=-68\degr56\arcmin45\arcsec\pm5\arcsec$, 3\arcsec\ away from the Meerkat position, with a count rate of $(1.5\pm0.6) \times 10^{-3}$ counts s$^{-1}$ (or 12 counts in 8\,ks, over 0.2--10\,keV). A spectral estimation was not possible due to the low count rate. More details on the data reduction are provided in Appendix~\ref{sec:swiftdata}.

\subsubsection{X-rays -- XMM-Newton}\label{sec:xmm}
\begin{figure}
    \centering
    \includegraphics[width=\columnwidth]{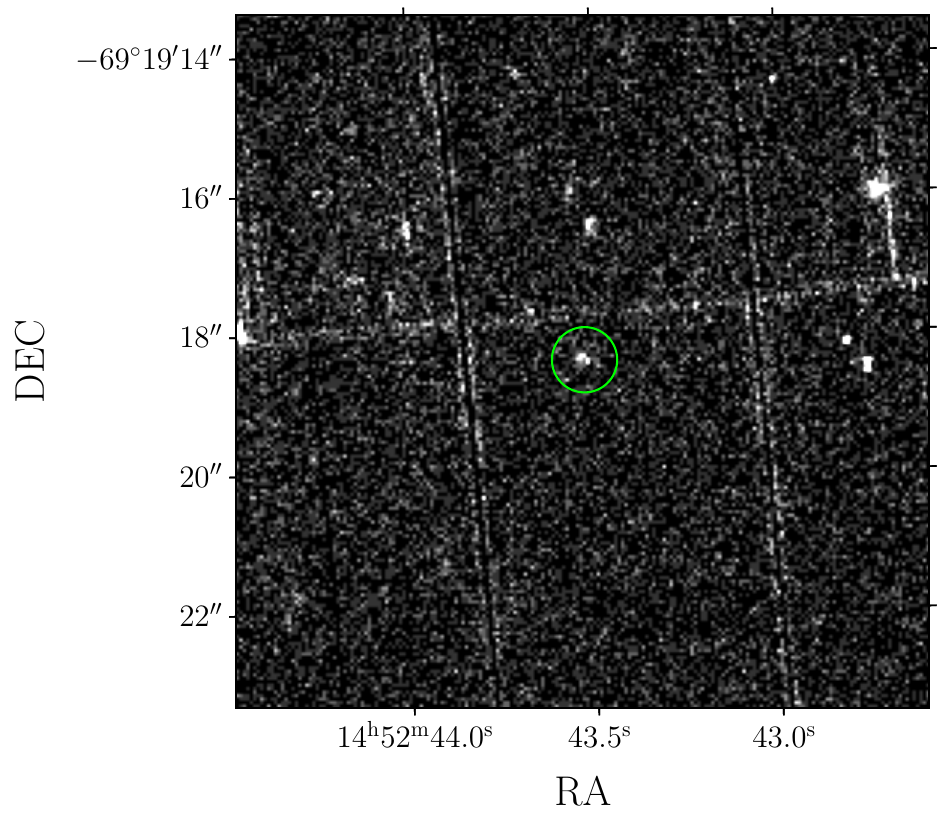}
    \caption{XMM-Newton EPIC-PN image cutout of the field of \ulp\  showing a point source detected 2\arcsec\, away from the Meerkat position. The green circle around the source is 30\arcsec\, in radius.}
    \label{fig:xmm}
\end{figure}

We observed the field of \ulp\  on July 31, 2024, for 30\,ks with the European Photon Imaging Camera (EPIC; \citealt{epic}) on board \textit{XMM-Newton} \citep{xmm} under Director's Discretionary Time (ObsID 0953011101). A point source was detected at $\alpha=14^{\rm h}48^{\rm m}34\fs 3\pm 0\fs3$ and $\delta=-68\degr 56\arcmin 43\arcsec\pm 1\arcsec$, 2\arcsec\, away from the Meerkat position (and consistent with the \textit{XRT} position; see Figure~\ref{fig:xmm}) with a total of 98$\pm$13\,counts, yielding a count rate of $(11\pm1) \times 10^{-3}$\,counts s$^{-1}$ (0.2--12\,keV). Simultaneous Optical/UV Monitor (OM) data from XMM were also taken, but the position of \ulp\, was affected by instrumental artefacts, which degraded the sensitivity and resulted in a non-detection. More details on the data reduction are provided in Appendix~\ref{sec:xmmdata}.

\subsubsection{Ultraviolet -- Swift}\label{sec:uvot}
Simultaneous ultraviolet optical telescope \citep[UVOT;][]{uvot} data were obtained for \ulp\ for 8\,ks in nine different scans alongside XRT data in the {UVW2} filter (central wavelength of 1928\,\AA). No point source was detected in the individual images (per scan), but a 4-$\sigma$ source was detected in the combined image with magnitude 22.6$\pm$0.3\,mag\footnote{We also looked at the Galaxy Evolution Explorer (GALEX; \citealp{galex}) data but the position of \ulp\ was not covered by the all-sky survey.} (AB; all the magnitudes are quoted in AB magnitude system unless explicitly stated otherwise). More details on the data reduction are provided in Appendix~\ref{sec:swiftuvotdata}.

\subsubsection{Optical -- Dark Energy Camera (DECam)}\label{sec:decam}
The field of \ulp\  was covered on multiple visits by DECam as a part of the DECam Plane Survey \citep[DECaPS;][]{decaps} between May 2018 and May 2019. We undertook standard data reduction techniques to estimate the source's magnitudes in $g, r, i, z, Y$ bands (see Appendix~\ref{sec:decamdata} for more details on our methods).

\begin{figure*}
    \centering
    \includegraphics[width=0.95\columnwidth]{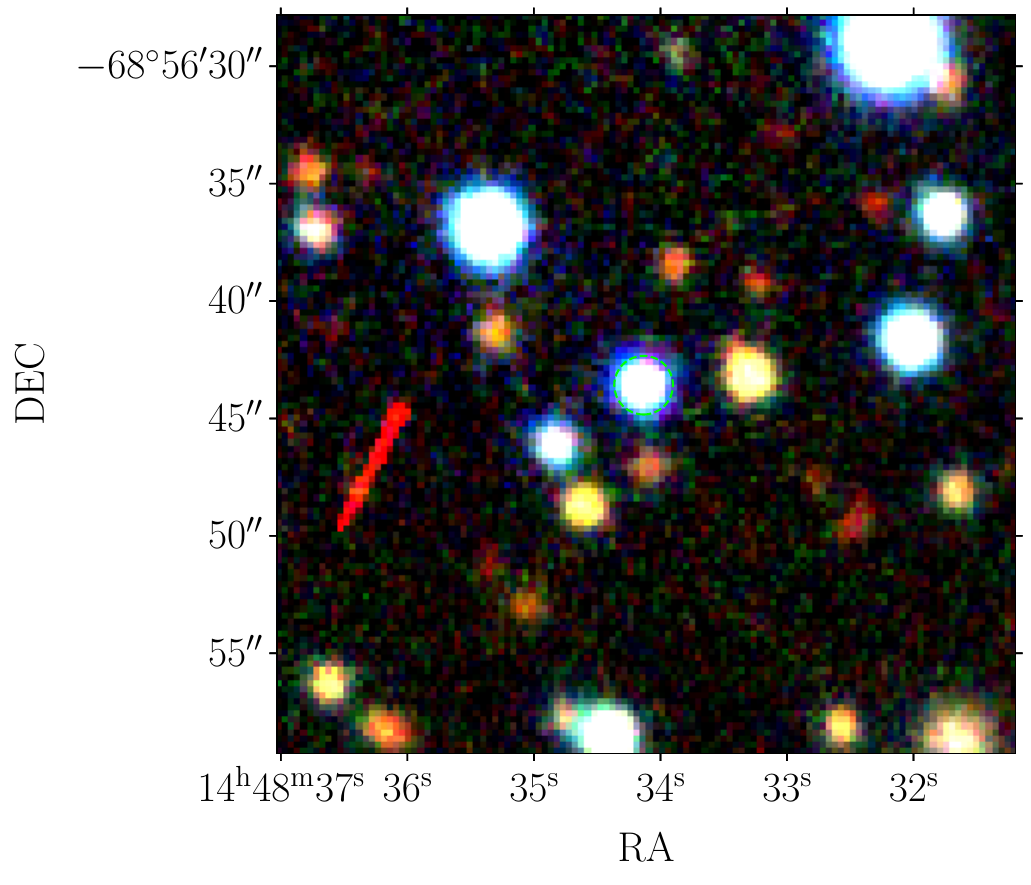}
    \includegraphics[width=0.95\columnwidth]{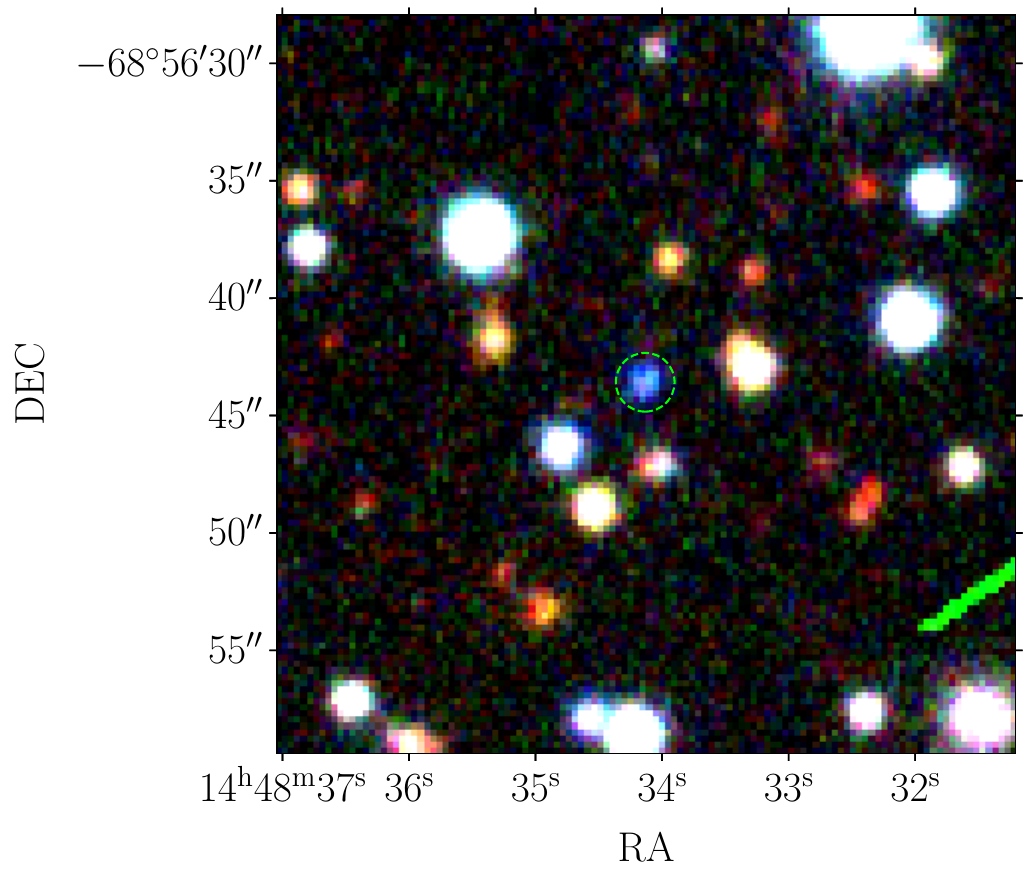}
    \caption{30\arcsec\, cut-outs of the \textit{g}, \textit{r}, \textit{i} composite images of the field of \ulp\ using the DECaPS data, during the flaring state (on the left) and the quiescent state (on the right). The lime dashed circle shows the error circle (2.5\arcsec) of \ulp. East is to the left and north is to the top of the image.}
    \label{fig:j1448_composite}
\end{figure*}

A variable, blue source at the position of \ulp. Figure~\ref{fig:j1448_composite} shows the \textit{gri} composite images of \ulp, both in its bright state (left) and in its faint state (right), where the bluish appearance of the source can be seen. We used the $z$ band magnitude in the bright state to estimate the probability of association by chance coincidence resulting in a false alarm rate of $<0.1$\%; together with the highly variable nature of the source, we conclude that the optical source is the counterpart to \ulp. Figure~\ref{fig:decaps} shows the optical light curve of \ulp\ showing that the source rose to peak ($i=18.3$, ${z}=18.0$) before declining to a quiescent level of ${g}=22.1$, ${r}=22$, and ${i}=22.4$, one year after the peak. Both in the bright and the faint state, although the source appeared blue compared to the rest of the field sources, strictly speaking, it is brighter in the redder bands (e.g, $g-r=0.2$ in the bright and $g-r=0.1$ in the faint state). 

Motivated by its optical variability, we searched the \textit{Gaia} \citep{gaiadr3} catalogue, but we did not find any optical counterpart. We checked the observation forecast tool\footnote{From \url{https://gaia.esac.esa.int/gost/}.} to retrieve the approximate pointing information of the satellite. The golden arrows in Figure~\ref{fig:decaps} show the times where \ulp\ was observed.

\subsubsection{Optical -- SkyMapper}
\begin{figure}
    \centering
    \includegraphics[width=\columnwidth]{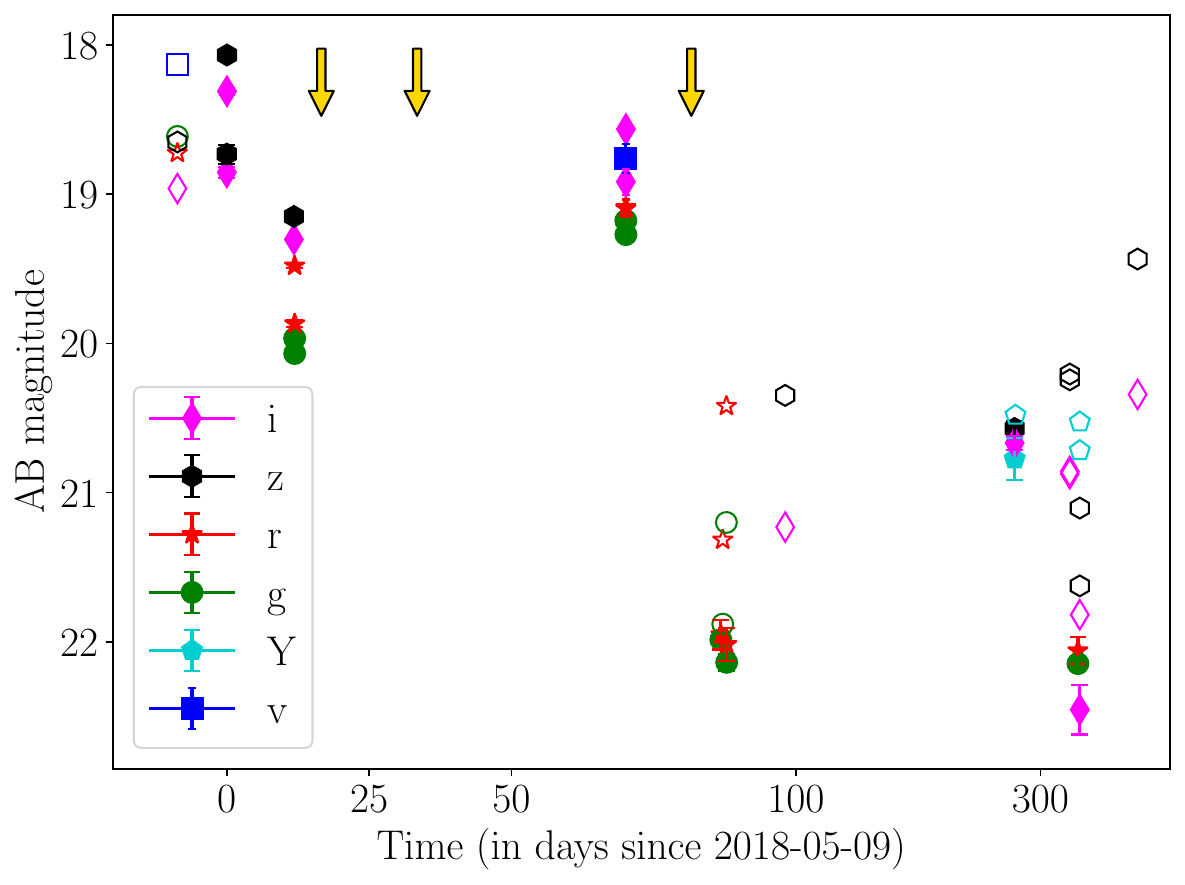}
    \caption{Optical light curve of \ulp\  using the data from DECaPS (in \textit{g, r, i, z, Y} bands) and SMSS (in \textit{v, g, r, i} bands) showing the evidence for optical variability. Data for different bands are shown as different symbols. Filled symbols represent detections and open symbols represent 5-$\sigma$ upper limits estimated from the local background. Error bars on the detections are too small to see. The golden arrows show the \textit{Gaia} observation times, where no source was detected.}
    \label{fig:decaps}
\end{figure}

The SkyMapper Southern Survey \citep[SMSS;][]{smss} is an optical multi-filter (\textit{u, v, g, r, i, z}) survey of the southern sky conducted between 2014 and 2021. SMSS visited the field of \ulp\ seven times   during this period. We retrieved the instrument-calibrated images in all six filters. A point source was detected in multiple filters in the images taken on 2018-07-18. We retrieved the photometric measurements\footnote{From \url{https://skymapper.anu.edu.au/}} to estimate the apparent magnitude of the source. Figure~\ref{fig:decaps} shows the SMSS observations in addition to the DECaPS data, indicating that the source brightened in between two DECaPS epochs. All the rest of the epochs (prior and post) resulted in non-detections (5-$\sigma$; see Figure~\ref{fig:decaps}). In particular, the non-detection on 2018-04-30 (SMSS has a typical depth of $g\approx21$, $r\approx21$, and $i\approx20$) implies that the optical outburst occurred between 2018-04-30 and 2018-05-19 (when the DECaPS observations started). The subsequent observations (on 2019-04-16 in $i$-band and on 2020-05-15 in $g, r$ bands), which resulted in non-detections, can be reconciled with the single epoch limiting magnitude of SMSS being higher than the source's quiescent magnitude (Figure~\ref{fig:decaps}).

\subsubsection{Optical -- Lesedi Telescope}\label{sec:mookodi}
We imaged \ulp\  with the Mookodi low-resolution spectrograph and imaging instrument \citep{mookodi} mounted on the 1-meter Lesedi telescope at the South African Astronomical Observatory on July 30, 2024, as part of the SAAO's Intelligent Observatory (``IO") rapid follow-up program \citep{saaoio,saaoobs}. This was undertaken to examine whether the observed variation in the DECaPS light curve has a contribution from the orbital variation. Contiguous 480\,s images throughout the orbit were obtained in the $g$ band, but no point source (5-$\sigma$) was detected in any single image. However, a faint source could be visibly seen in a few images, and the stacked image revealed a \textit{g}-band magnitude of 22.0$\pm$0.2. This is consistent with the brightness of the source in the faint state as observed by DECaPS. More details on the data reduction methods are provided in Appendix~\ref{sec:lesedidata}.

\subsubsection{Near infrared -- VISTA Hemisphere Survey (VHS)}
\label{sec:vhs}
We retrieved the archival near-infrared (NIR) observations from the VISTA Hemisphere Survey \citep[VHS;][]{vhs} survey from May 2015. In the 15\,sec $J, K_s$ images, no point source was detected down to a limiting magnitude of $J=20.2$ and $K_s=19.3$. 

\subsubsection{Near infrared -- Magellan}
\label{sec:magellan}
We also obtained three \textit{J}-band exposures, for 54\,min on July 18, 2024, 96\,min on July 22, 2024, and 52\,min on July 23, 2024, with the FourStar imager \citep{fourstar} on the 6.5-m Baade Telescope at Las Campanas Observatory. The observing conditions (seeing) varied from 0.7-2\arcsec\, between the observations with the worst seeing on Jul 22, which resulted in the non-detection of the source on July 22. A very faint source (4-$\sigma$) was visually identified in the remaining two epochs, with a better seeing on July 18. We measured the apparent magnitude to be 22.0$\pm$0.2, consistent between the two epochs (July 18, 23) and with the VHS $J$-band nondetection. More details on the data reduction methods are provided in Appendix~\ref{sec:magellandata}.  

\section{Modeling}\label{sec:modeling}
In this section, we describe our modeling of the multi-wavelength data, radio through X-ray, to derive various source parameters. We model emission at each wavelength independently, without reference to any particular model, and discuss various emission models that can simultaneously explain the multi-wavelength behavior in \S\ref{sec:discussion}.

\subsection{Radio Timing}
We fit the individual pulses using multiple Gaussian components (similar to the standard template generation method for pulsars) using a simple least-squares fit to extract the times of arrival (TOAs) of the pulses and generate a phase-connected solution. Individual bursts and their model fits are provided in Appendix~\ref{sec:bursts}. We estimate the individual TOAs as the weighted mean of the individual components. In general, many LPTs show pulse-to-pulse variability \citep{hurley-walker_radio_2022,hurley-walker_long-period_2023,lee2025,Wang2025}, similar to single pulse variations in pulsars \citep{handbook}. Pulse variations can result in the underestimation of the total error to the TOA, and are usually accounted for phenomenologically by using additional noise terms, one of which adds in quadrature to the fitting error on the TOA. This noise parameter is called ``EQUAD'' \citep{nanograv_noise_budget}. 

We extracted a total of 18 TOAs spanning a $\sim$ year. Given the small number of TOAs and their separation in time, we do not fit for the position. Additionally, we estimate the EQUAD parameter independently for both ASKAP and MeerKAT\footnote{We do this by ensuring that a model containing $F_0$, EQUAD as free parameters has a reduced $\chi^2$ of 1 for observatory-specific TOAs.} observations and then perform a global fit for the spin frequency $F_0$ using the pulsar timing software \texttt{pint} \citep{pint}. The resulting parameters are given in Table~\ref{tab:fit}. The EQUAD values are 570\,s and 273\,s for ASKAP and MeerKAT, respectively, which represent 30\% and 15\% of the pulse widths roughly. Even in the pulsars that have a very low timing noise, this effect can be 1--2\% \citep{nanograv_timing}. But given that we use individual pulses, as opposed to stacked pulse profiles (of thousands of pulses) commonly used for pulsars, this may explain the higher EQUAD values.


\begin{table}
\centering
\caption{Parameters for \ulp}
\label{tab:fit}
\begin{tabular}{lr}
\hline
Parameter & Value \\
\hline
Source name                  \dotfill & \ulp      \\ 
MJD range                    \dotfill & 60110---60522 \\ 
Data span (yr)               \dotfill & 1.53    \\ 
Number of TOAs               \dotfill & 20      \\ 
\hline 
RAJ, Right Ascension (J2000) ($\mathrm{{}^{h}}$)\dotfill &  $14^{\rm h}48^{\rm m}34\fs29$$^a$ \\ 
DECJ, Declination (J2000) ($\mathrm{{}^{\circ}}$)\dotfill &  $-68\degr 56\arcmin 44\farcs1$$^a$ \\ 
PEPOCH, Reference epoch ($\mathrm{d}$)\dotfill &  60456.444811 \\ 
F$_0$, Spin-frequency ($\mathrm{Hz}$)\dotfill &  $0.000177586(1)$ \\ 
$\rm |F_1|$ $^b$, Spin-frequency derivative ($\mathrm{Hz} \mathrm{s}^{-1}$)\dotfill & $<7\times 10^{-16}$\\
EQUAD (MKT; s)\dotfill &  273 \\ 
EQUAD (ASKAP; s)\dotfill &  570 \\ 
RMS TOA residuals (s) \dotfill & 533.963   \\ 
$\chi^2$                         \dotfill & 20.94    \\ 
Reduced $\chi^2$                 \dotfill & 1.31    \\ 
\hline
\end{tabular}
\text{$^a$Position is derived from the MeerKAT imaging data (UHF band).}
\text{$^b$This is the 5-$\sigma$ upper limit on F$_1$.}
\end{table}

\subsection{Optical burst}
It is evident from Figure~\ref{fig:decaps} that \ulp\  showed considerable optical variability. This could either be due to flaring episode(s) (during the start of the DECaPS observations), orbital variations, or both. However, our $g$-band observations with Lesedi revealed that at no instance within an orbit was the source brighter than $g=22.0$, which implies that the orbital variation, if any, is not brighter than this. So, we do not consider the bright state in the DECaPS light curve to be due to orbital variation but due to a flaring episode that faded towards the end of the DECaPS observations. The consistent $g$-band magnitudes between the DECaPS data (towards the end) and the Lesedi data also support this. Considering the DECaPS data alone, a simple power law fit to the data yields a temporal power law index of $\alpha=-1.02(2), -1.18(4), -0.42(1), \text{and} -0.45(2)$ in \textit{g, r, i,} and \textit{z} bands respectively and an eruption time of 2018-05-06T23\,hr$\pm 6$\,hr. 
On top of this secular evolution, the data shows rapid variations ($\sim$0.5\,mags in 40\,min in $i,z$ bands, $-$0.3\,mags in 20\,min in $g,r$ bands; see Figure \ref{fig:decaps}) on several minute timescales. However, the SMSS observations indicate that the optical variation is not a simple steady decline, which hints that there might be multiple flaring episodes or orbital variations in the flaring state. Given the sparse data set, we can not conclusively establish the history of this optical variation, but once the source settled to its quiescent state (end of DECaPS observations), our Leesedi observations constrained the peak brightness due to orbital variation  (if any) to $g\geq22.0$.


\subsection{Source spectral and polarimetric properties}
We present the spectral energy distribution (SED) of \ulp\  from radio to X-rays, along with our model fits to the SED. We model the SED at each wavelength independently, as the emission regions and mechanisms can vary by wavelength.

\subsubsection{Radio spectral and polarimetric properties}\label{sec:radio_burst_prop}
Although bursts were detected in the ASKAP (EMU) observations, they were not bright enough for a robust in-band spectral index. Hence we compare the flux densities at UHF and L-bands using the MeerKAT data to estimate the wide-band spectral properties. Modeling the spectrum to be a power law ($S_{\nu}\propto \nu^{\alpha}$), we get the best-fit estimate for $\alpha$ to $-2.44\pm0.06$ from the first observation and $-2.61\pm0.05$ from the second observation. 

\begin{figure}
    \includegraphics[width=\columnwidth]{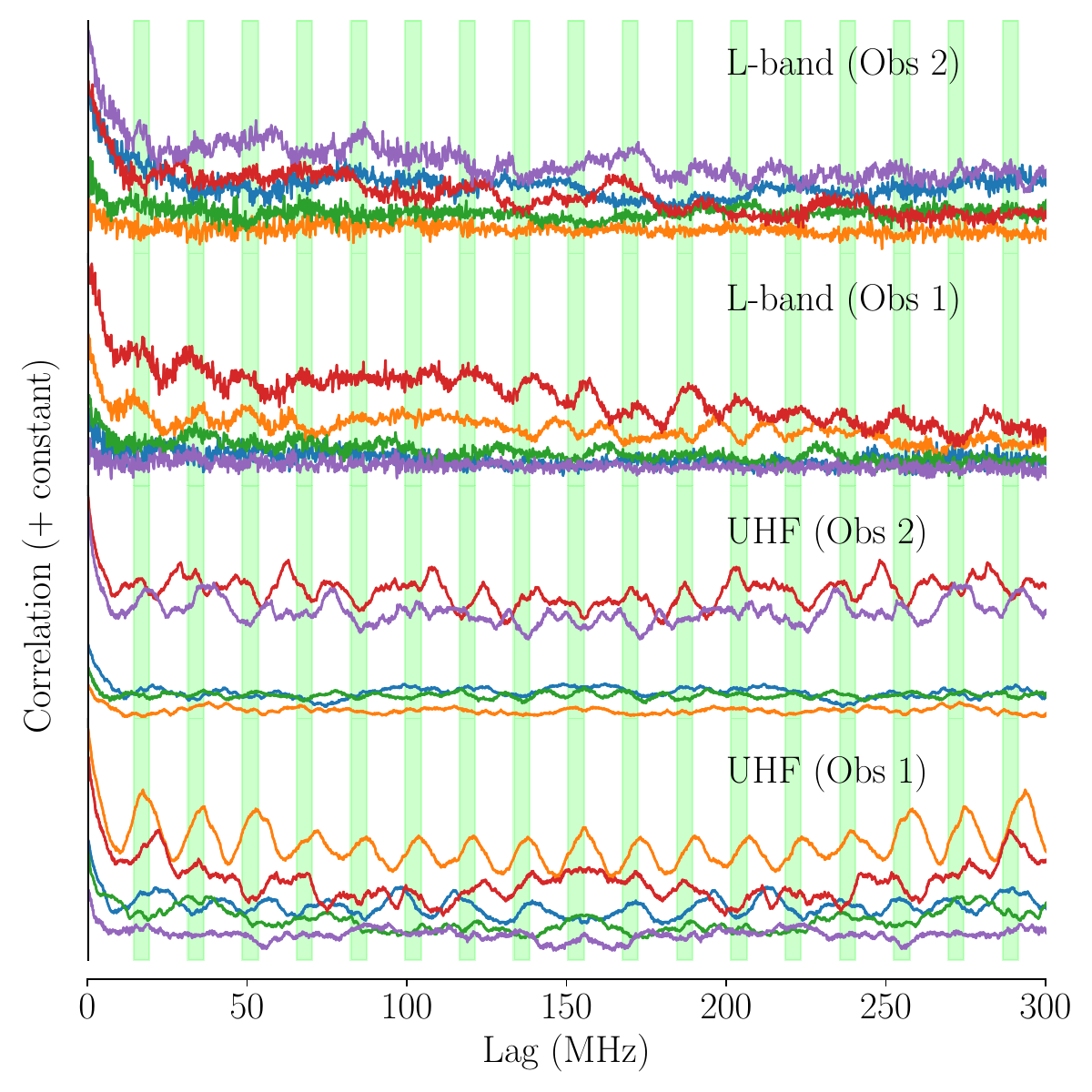}
    \caption{The correlation coefficient (shifted by an arbitrary constant for visualization) of the 1D frequency spectrum of all the bursts in the MeerKAT data. The color scheme represents the bursts in order: blue, orange, green red, and violet represent the five bursts seen in each of the MeerKAT observations. A periodic feature at the 17\,MHz (fundamental) and its harmonics (marked by the lime stripes) can be seen in the UHF and L band data for multiple bursts.}
    \label{fig:burst_period_mkt}
\end{figure}

In addition to this secular trend, there are also rich narrow-band features in the radio spectrum. To see if the narrow band emission is structured (periodic/quasi-periodic), we compute the autocorrelation function (ACF) for the 1D (time-scrunched) frequency spectrum (per burst)\footnote{The frequency spectrum is wrapped around after every shift to avoid introducing signal artifacts due to finite bandwidth.}. The resulting correlations are shown in Figure~\ref{fig:burst_period_mkt}, which show strong evidence for harmonic narrow-band emission. Using the observed harmonics, we derive a fundamental period of 17\,MHz and the observed harmonic number to be 40--60.

In terms of the polarimetric properties, the observed total polarisation fraction changes from $\approx$35\% to 100\% between the radio observations (ASKAP/ATCA/MeerKAT). The relative contributions of the linear and circular polarisations vary between the bursts as well (detailed pulse properties are provided in Table~\ref{tab:pol}). It is to be noted that 100\% circular polarisation is seen in some of the bursts (but not 100\% linearly polarised) while elliptically polarised bursts are more common in general.

\subsubsection{UV to NIR}\label{sec:mm_spec}
\begin{figure}
    \centering
    \includegraphics[width=\columnwidth]{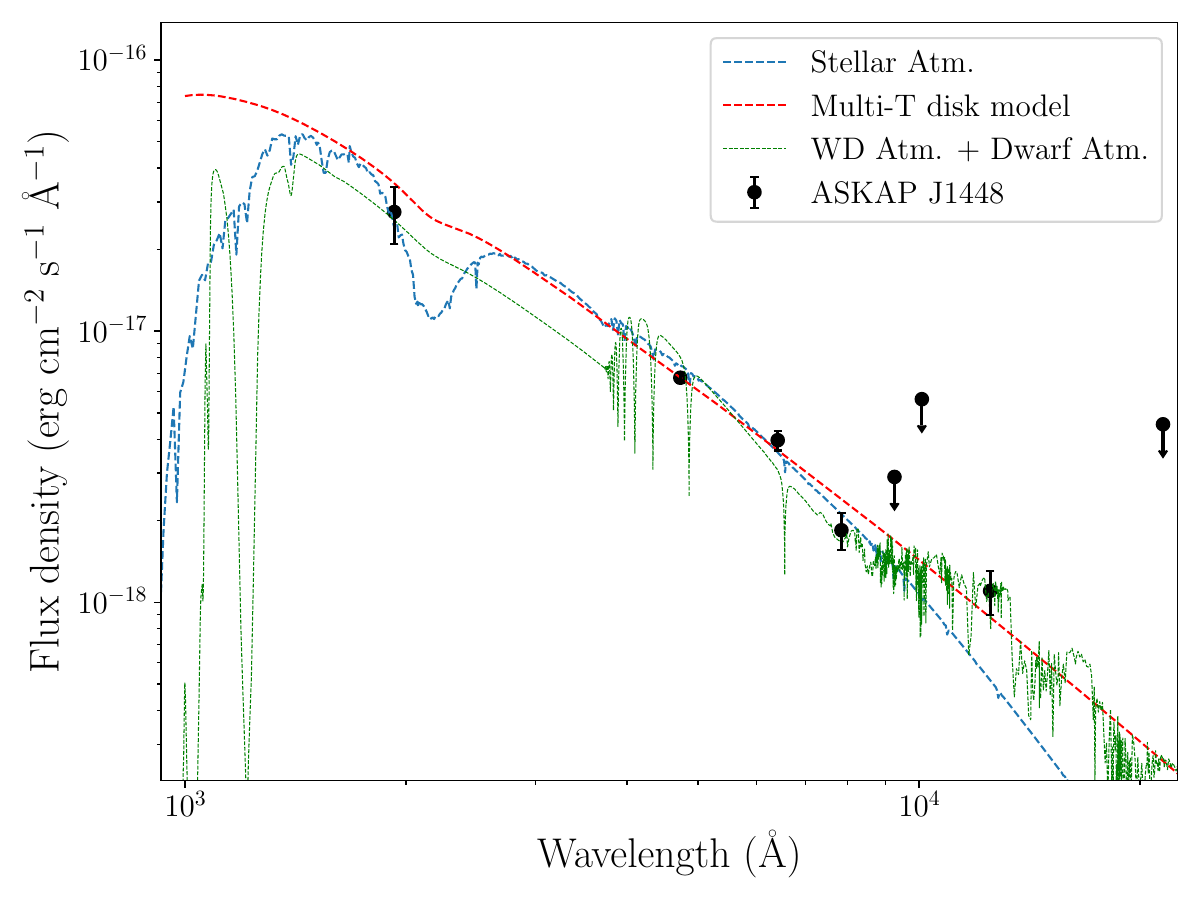}
    \caption{Broad-band optical/near-infrared SED of \ulp, based on the models from Appendix~\ref{sec:mm_spec_app}. The black data points show the observed flux density measurements (Table~\ref{tab:sed}), the blue line is the best fit stellar atmospheric model, the red line is the best-fit disk model and the green line is the best-fit WD+M-dwarf model.}
    \label{fig:sed_j1448}
\end{figure}

We then modeled the multi-wavelength SED of the source using our observations from UV to NIR. Table~\ref{tab:sed} shows the broadband flux density measurements from UV to NIR. We fit this  SED using three different models --- i) an isolated stellar atmosphere, ii) an interacting binary with a disk-like spectrum, and iii) a detached binary.  Figure~\ref{fig:sed_j1448} shows the resultant model fits. Below, we highlight the key results from these fits. Detailed information on these model fits and synthetic photometry is provided in Appendix~\ref{sec:mm_spec_app}.

When we fit the observed SED with a single atmospheric model, the resultant fit revealed the presence of a hot source ($T_{\rm eff}>35000$\,K). However, the fit is not well constrained as $T_{\rm eff}$ pushes against the upper edge of the available model atmospheres (40000\,K; see Appendix~\ref{sec:bb} for more details). In addition, trying to simultaneously model the UV and NIR data, these models predict a large $A_V$ of $\approx 1.5$, much greater than the maximum estimate from a 3D extinction map in that direction \citet{Zucker2025}, likely trying to flatten the model at UV to fit the data. Using the resultant amplitude, we estimate the distance to the source to be $d=110 (R_{\star}/R_{\astrosun})$\,kpc where $R_{\star}$ is the radius of the star.

In contrast, when we fit the SED with a multi-temperature disk-like spectrum, we infer that the lower limit on the temperature at the inner edge to be $T_{\rm in}>92\ 000$\,K. Given the sampling of our observed SED, we fall in a regime where the dependence on the temperature just scales the spectrum similar to the amplitude, and this covariance implies that we could just derive lower limits on $T_{\rm in}$ and the distance (see Appendix~\ref{sec:disk} for more details). We derive a lower limit on the distance to be $d=15 \ \sqrt{\cos i}\ (r_{\rm in}/R_{\oplus})$\,kpc where $r_{\rm in}$ is the radius of the inner edge of the disk.  

Finally, we attempt to fit the observed SED using a combination of WD and dwarf star atmospheric models. This is representative of a detached WD binary where the emission is powered by the thermal emission from the individual components. We found that (see Appendix~\ref{sec:wd_md} for more details) a 13\,000\,K, $1\,R_{\oplus}$ WD and a 2\,300\,K, $0.01\,R_{\astrosun}$ dwarf star binary placed at a kpc reasonably explained the observed SED. This distance estimate is consistent with the DM-based distance estimated from the lack of dispersion sweep in the radio pulses. Given the sparse data set, we reiterate that the goal here is to probe whether such a system is plausible given the data, and not to constrain the system parameters using the SED.


\begin{table}
\centering
\caption{Broadband photometry of \ulp\  using observations from UVOT, DECaPS, VHS, FourStar (\S\ref{sec:mw})}
\label{tab:sed}
\begin{tabular}{crrl}
\hline
Band & $\lambda$ & $F_{\nu}$$^a$ & $F_{\nu, \rm err}$\\
 & (nm) & ($\mu$Jy) & ($\mu$Jy)\\
\hline
UVW2 & 192.8 & 3.41 & 0.82 \\
$g$ & 473.0 & 5.03 & 0.2 \\
$r$ & 642.0 & 5.45 & 0.46 \\
$i$ & 784.0 & 3.79 & 0.59 \\
$z$ & 926.0 & $<$8.3 & ... \\
$Y$ & 1009.0 & $<$19 & ... \\
$J$ & 1250.0 & 5.75 & 1.06 \\
$K_s$ & 2150.0 & $<$70 & ... \\
\hline
\end{tabular}
\text{$^a$ Upper limits are at 5-$\sigma$ confidence.}
\end{table}

\subsubsection{X-rays}
\label{sec:xrayfit}
\begin{figure}
    \centering
    \includegraphics[width=\columnwidth]{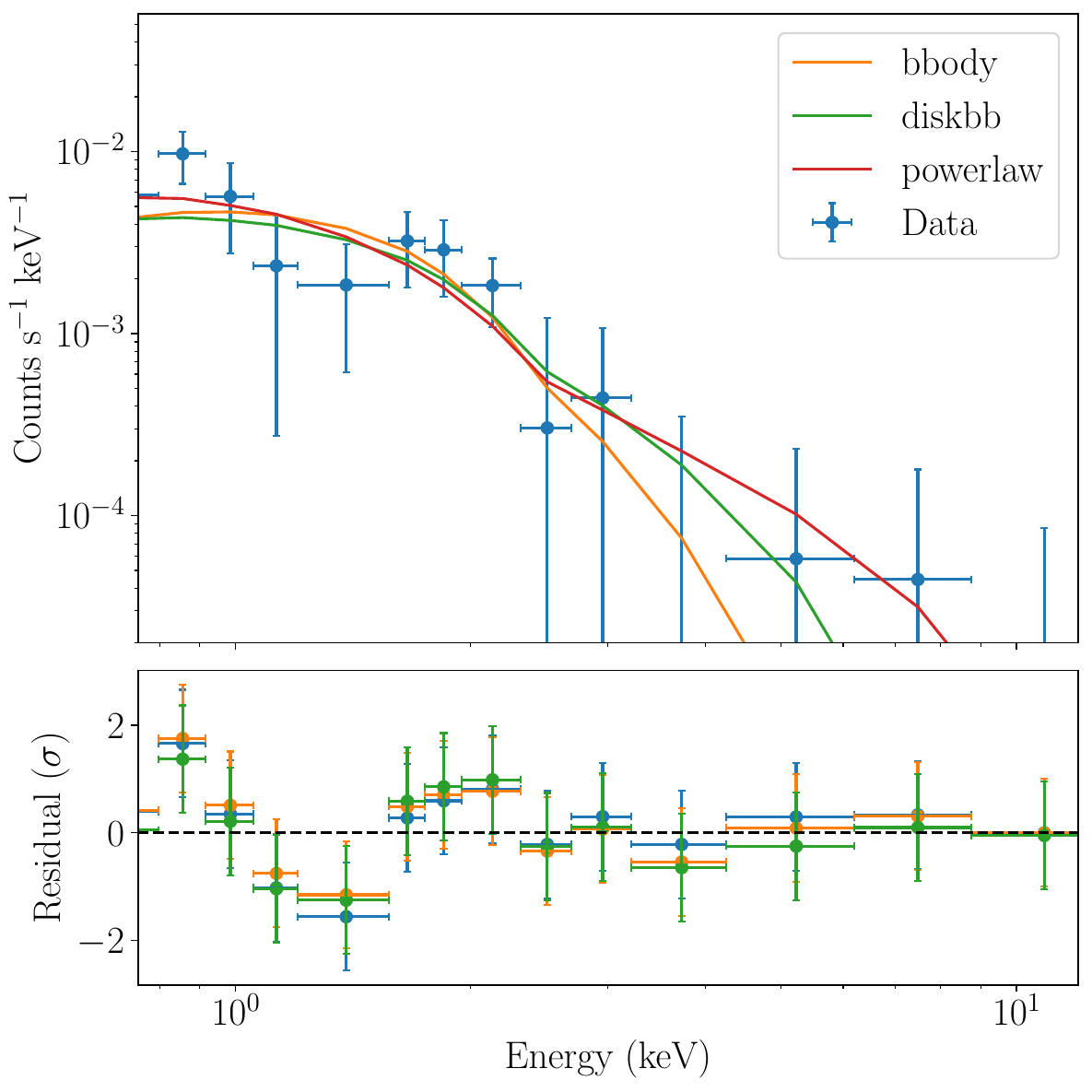}
    \caption{EPIC-PN spectrum of \ulp. The blue points in the top plot show the data, and the multicolored lines show different model fits to the data. The bottom panel shows the model residuals (in terms of the error). The dashed black line is the zero baseline. The color scheme in both panels is the same, with  \texttt{bbody}=blackbody (orange), \texttt{diskbb}=disk (multicolor blackbody, in green), and \texttt{powerlaw}=power law spectrum (red).}
    \label{fig:xmm-spec}
\end{figure}

Figure~\ref{fig:xmm-spec} (top panel) shows the source spectrum of \ulp\  as observed by EPIC-PN. We used \texttt{xspec} \citep{xspec} to model the spectrum. We tried three different models for the spectrum: i) blackbody (\texttt{bbody}), ii) Multicolor black body (relevant for accretion disks; \texttt{diskbb}), and iii) power law (\texttt{powerlaw}). The resultant fits are shown in Figure~\ref{fig:xmm-spec} and the model parameters, along with the observed fluxes and the fit statistic are given in Table~\ref{tab:xmm_fit}. All three models provide decent fits to the data.

\begin{table*}
\centering
\caption{Model fits to the EPIC-PN X-ray spectrum.}
\label{tab:xmm_fit}
\begin{tabular}{lc|rr|rrrrr}
\hline
Model & Parameter $^a$ & \multicolumn{2}{c}{Value} & \multicolumn{4}{c}{Flux ($\times 10^{-14}$\,erg\,cm$^{-2}$\,s$^{-1}$)} & $\chi^2$(DOF) \\
\hline
 &  & Free $N_H$ & Fixed $N_H$ $^b$ & \multicolumn{2}{c}{Free $N_H$} & \multicolumn{2}{c}{Fixed $N_H$ $^b$} &  \\
\cline{3-8}\\
&  &  &  & Absorbed & Unabsorbed & Absorbed & Unabsorbed & \\
\hline
 \multirow{3}{*}{\texttt{phabs*bb}} & $T_{\rm bb}$ & 0.4(1) \phantom{$\times 10^{-8}$} & 0.35(0.06) \phantom{$\times 10^{-8}$} & \multirow{3}{*}{1.6$^{+0.2}_{-0.5}$} & \multirow{3}{*}{1.6$^{+0.7}_{-0.5}$} & \multirow{3}{*}{1.3$^{+0.3}_{-0.2}$} & \multirow{3}{*}{1.8$^{+0.5}_{-0.5}$} & \multirow{3}{*}{11.52(18)}\\
  & \texttt{norm} & 19(4) $\times 10^{-8}$ & 22(3) $\times 10^{-8}$ & & & & & \\
  & $N_H$ & 2(70) $\times 10^{-3}$ & 13 $\times 10^{-2}$ & & & & & \\
\hline
\multirow{3}{*}{\texttt{phabs*diskbb}} & $T_{\rm in}$ & 0.7(3) \phantom{$\times 10^{-8}$} & 0.6(2) \phantom{$\times 10^{-8}$} & \multirow{3}{*}{1.8$^{+1.6}_{-1.8}$} & \multirow{3}{*}{2.2$^{+0.9}_{-0.7}$} & \multirow{3}{*}{1.6$^{+1.5}_{-1.6}$} & \multirow{3}{*}{2.3$^{+0.8}_{-0.6}$} & \multirow{3}{*}{10.55(18)}\\
  & \texttt{norm} & 44(74) $\times 10^{-4}$ & 102(104) $\times 10^{-4}$ & & & & & \\
  & $N_H$ & 7(8) $\times 10^{-2}$ & 13 $\times 10^{-2}$  & & & & & \\
\hline
\multirow{3}{*}{\texttt{phabs*powerlaw}} & $\Gamma$ & 2.5(8) \phantom{$\times 10^{-8}$} & 2.1(3) \phantom{$\times 10^{-8}$} & \multirow{3}{*}{2.6$^{+0.7}_{-0.6}$} & \multirow{3}{*}{4.6$^{+24}_{-1.9}$} &  \multirow{3}{*}{2.6$^{+1.0}_{-0.5}$} & \multirow{3}{*}{3.6$^{+1.0}_{-1.1}$} & \multirow{3}{*}{9.89(18)}\\
  & \texttt{norm} & 9(5) $\times 10^{-6}$ & 6(1) $\times 10^{-6}$ & & & & & \\
  & $N_H$ & 22(17) $\times 10^{-2}$ & 13 $\times 10^{-2}$ & & & & & \\
\hline
\end{tabular}
\text{$^a$ $N_H$ is defined in term of $10^{22}$\,cm$^{-2}$. The units of $T$ are keV, and of \texttt{norm} are counts\,s$^{-1}$\,cm$^{-2}$\,keV$^{-1}$.}
\text{$^b$ Using the phenomenological relation between $N_H$ and $A_V$ (e.g., \citealt{guver2009}), we derive $N_H$=1.3$\times 10^{21}$\,cm$^{-2}$ for $A_V=0.62$.}
\end{table*}

\section{Discussion}\label{sec:discussion}

\subsection{What is known about \ulp?}
\ulp\  is an LPT with a period of 1.56\,hr. It has a steep radio spectrum ($\alpha<-2$) up to a spectral cutoff at 1.5\,GHz, shows emission with a harmonic frequency structure, and polarised bursts. The emission is elliptically polarised with the polarisation fraction varying from 35\% to 100\% between observations. The duty cycle is around 70\% which resulted in the non-detection of a few bursts. This could either be due to intensity variations or complete cessation of bursts, like pulsar nulling \citep{backer1970}. Intensity variations are not uncommon in LPTs \citep{caleb_emission-state-switching_2024,Wang2025}. As an example, we consider the X-ray bright LPT, ASKAP J1832$-$0911 \citep{Wang2025}, which shows 2--3 orders of magnitude variation in the pulse intensities, albeit on multiple pulse timescales rather than pulse-to-pulse timescales. If \ulp\ has a similar single pulse fluence distribution, a factor of 5 is sufficient to miss the bursts to the instrumental noise, and hence the non-detection of a few pulses in \ulp\ is consistent with intensity variations in LPTs rather than complete cessation of pulses.

\begin{figure}
    \centering
    \includegraphics[width=0.95\columnwidth]{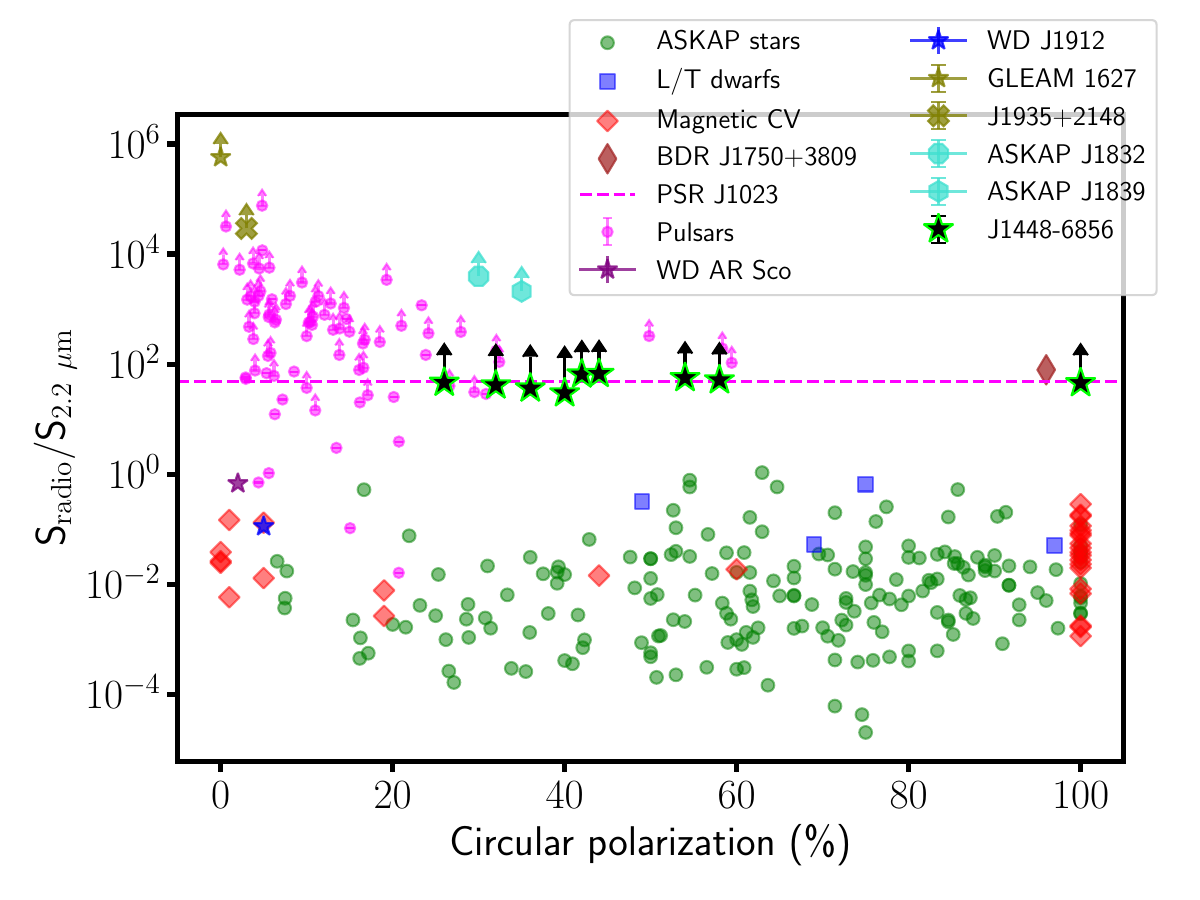}
    \includegraphics[width=0.95\columnwidth]{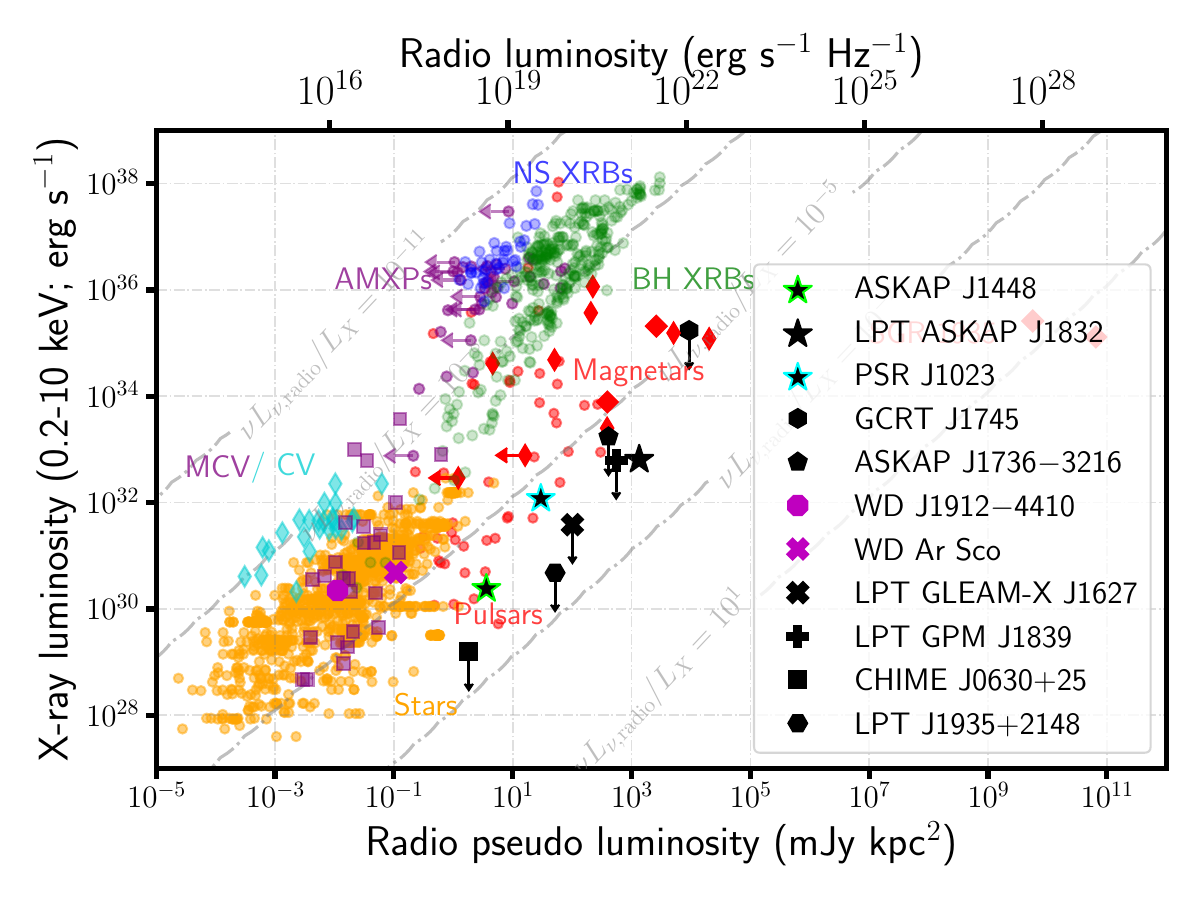}
    \caption{Properties of \ulp\  in the context of population properties of radio-bright sources. \textbf{\textit{Top}}: Phase space of circular polarisation versus radio-to-$K_s$-band flux ratio showing different source classes. The black stars show the IR upper limits obtained from the VHS data and the circular polarisation from our observations. Multiple measurements are shown as different points on the plot. Data for ASKAP stars is taken from \citet{SRSC}, from \citet{Barrett2017,Barrett2020} for MCVs, from \citet{Kao2016} for L/T-dwarfs, from \citet{Vedantham2020} for the radio-discovered T dwarf (BD J1750+3809), from \citet{marsh_radio-pulsing_2016,pelisoli_53-min-period_2023} for the two WD pulsars, and \citet{archibald2009} for the tMSP PSR J1023+0038. For the tMSP, polarization information in the radio pulsar state was unavailable, hence, the source is shown as the dashed line. \textbf{\textit{Bottom}}: Phase space of X-ray vs radio luminosity showing different classes of sources. Individual sources of interest, including different LPTs/GCRTs are highlighted to show how they differ from known source classes. Adapted from \citet{Wang2025}.
    }
    \label{fig:j1448_xray_ir_pop}
\end{figure}  


Unlike most other LPTs, \ulp\ is detected across wavelengths, from X-rays to radio, and shows variability at optical wavelengths. The multi-wavelength SED shows the presence of a hot source, peaking at NUV wavelengths. However, the distance to \ulp\ is not well constrained. For most LPTs, a well-constrained DM estimate was used to derive a DM-based distance using Galactic electron density maps \citep{ne2001,ymw16}. In the case of \ulp, a robust estimate of DM was not possible --- we did not detect \ulp\ in the high-time resolution MeerKAT data and no pulsar-like periodic candidates were found (see Appendix~\ref{sec:mktdata}). Hence, we exploit the simultaneous detection of \ulp\ at UHF and L-band to contain the dispersion delay to less than one sample, or $<8$\,s. This provides an upper limit of DM$<720$\,pc cm$^{-3}$. The maximum value of DM along the line of sight to \ulp, from the Galactic contribution, is 260\,pc cm$^{-3}$ \citep{ne2001}, which leaves the distance to \ulp\ unconstrained.

Examining the multi-wavelength properties of \ulp, we derive a radio to X-ray luminosity ratio of $\nu L_{\nu, \rm radio}/L_{\rm X}$ $\approx 10^{-3}$ and a lower limit on the radio to $K_s$ band flux density ratio of $>100$. Figure~\ref{fig:j1448_xray_ir_pop} shows where \ulp\  stands in comparison with known classes of sources in terms of its X-ray and IR properties. In terms of its X-ray and radio properties, \ulp\  is brighter at radio than  X-rays by at least 4--5 orders of magnitude compared to systems like CVs and X-ray binaries (XRBs), and at least 2 orders of magnitude compared to radio-bright stars. In terms of the IR and radio properties, it is at least 4 orders of magnitude brighter at radio than at $K_s$ band, compared to radio-bright CVs and stars. It is consistent in terms of its IR and X-ray properties with pulsars, but given the pulse widths (and period), a standard pulsar emission mechanism might be difficult to explain the emission in \ulp. We also infer a 5-$\sigma$ limit on the spin-down luminosity of $\dot E$ to be $2\times 10^{27}$\,erg s$^{-1}$. The observed radio luminosity of $\nu L_{\nu}\approx 10^{28}\ \rm d_{kpc}^2 \ \Omega_{4\pi}$\,erg/s and the unabsorbed 0.2--12\,keV X-ray luminosity of $L_{X}\approx 10^{29}\ \rm d_{kpc}^2 \ \Omega_{4\pi}$\,erg s$^{-1}$ (where we assume that it radiates into $4\pi\Omega_{4\pi}\,$steradian) implies that if \ulp\ is spin-down powered like pulsars \citep{ruderman76}, then the distance to the source has to be $<100$\,pc.


\subsection{What is the origin of the radio emission?}\label{sec:origin}
A steep spectrum (\S\ref{sec:mkt}) and narrowband spectral features rule out thermal emission. While incoherent emission can show low levels of polarisation (a few percent; \citealp{dulk1985}), the presence of highly polarised emission (with 100\% circular polarisation) implies the presence of strong ($\gtrsim$kG) magnetic fields and coherent emission. In addition, using a brightness temperature limit of $10^{12}\,$K we estimate the maximum distance to the source beyond which incoherent emission mechanisms can not explain the observed radio brightness. This depends on the size of the emitting radius ($r_{\rm em}$), which in the least constraining case can be as large as $775\,R_{\astrosun}$ from the light travel time. More informative but model dependent estimates can be made but they vary: $\approx$ 375\,$R_{\astrosun}$ from light cylinder radius; a few times (1--10) the source's radius for GHz cyclotron emission (e.g., 10$R_{\oplus}$ for a WD with $R_{\oplus}$); and $\approx0.7$\,$R_{\astrosun}$ from the Roche lobe if \ulp\ is a binary (1\,$M_{\astrosun}$+0.10\,$M_{\astrosun}$). Considering these, a limit on $r_{\rm em}$ of $R_{\astrosun}$ yields a maximum distance of $200$\,pc for incoherent emission for a range of possible hosts (see \S\ref{sec:progenitor}). If \ulp\ is more distant than this, then the radio emission has to be powered by a coherent mechanism.

Coherent emission in astrophysical sources is attributed to plasma radiation, ECME, or pulsar radiation \citep{melrose2017}. Both plasma and ECME emission are observed as coherent nonthermal radio emission from the Sun, flare stars, and dwarf stars (isolated or in binaries like MCVs;  \citealt{dulk1985,Hallinan2007,Hallinan2006,Kao2016,Barrett2020}). ECME is usually thought to arise from the outer layers of the magnetosphere (a few stellar radii) where the plasma density is low (and transparent), but the local magnetic field strength is sufficient (kG) to power GHz emission. ECME is intrinsically narrow-band and is emitted at the fundamental and first few (2--3) harmonics as 100\% circularly polarised emission \citep{Melrose1982,melrose2017}. Depolarisation at higher harmonics or due to propagation effects \citep{melrose2017}, which can give rise to lower circular polarisations. In addition, in the case of relativistic ECME, the relativistic beaming can produce highly linear polarised emission \citep{Qu2024}, resulting in elliptical polarisations. Spectral cut-offs are explained as the limit on the altitude below which the plasma frequency becomes comparable to the cyclotron frequency, quenching the maser instability. Broadband ECME is usually explained by emissions from different altitudes in the magnetosphere.

However, the observed band structure in \ulp\ means that we are observing either higher harmonics or fundamental emission from discrete altitudes/multiple populations of electrons. The observed periodicity of 17\,MHz corresponds to the fundamental frequency if the emission is non-relativistic. In the case of relativistic ECME, Doppler boosting of frequencies implies that the fundamental frequency can be much lower than this if the Lorentz factor of the electrons is $\gg1$. Alternately, if the emission arises from the fundamental frequency, it imposes an extremely stringent condition on the structure of the magnetosphere: that material is organized in the form of dense shells, or the emission arises from multiple populations of electrons with different Lorentz factors, both of which seem unlikely. However, in the case of higher harmonics, the strength of the maser drops precipitously at higher harmonics \citep{Melrose1982}, and hence explaining the sustained maser instability at higher harmonics can be difficult. We note that similar narrow band structures are seen in the dynamic spectra of Jupiter's decameter emission, which is attributed to the presence of a thin scintillating sheet of plasma intrinsic to the source \citep{Imai1992}, but this results in the pulses drifting in the dynamic spectrum, which are absent in the dynamic spectra of \ulp.   

Plasma emission also suffers similar limitations in explaining the harmonic structure of \ulp\ since it is intrinsically narrow-band and is limited to a few harmonics, much like  ECME. Similar difficulties arise with the pulsar mechanism, which is usually attributed to curvature radiation \citep{ruderman76}, in explaining the harmonic structure in \ulp.

Incoherent emission mechanisms, like gyro-synchrotron radiation, can also be modulated (rotationally if isolated, orbitally if in a binary), both in intensity and the level of polarisation, due to misaligned rotational and magnetic axes leading to varying longitudinal magnetic field strengths \citep{Franciosini1996,Hallinan2006,Leto2020}. However, in addition to distance limitation, this can also face similar difficulties in explaining the observed spectral structure. The observed narrow bands imply that the emission is produced (from higher harmonics) by a localized region, and the intensity decreases as we go to higher harmonics, becoming undetectable beyond 1500\,MHz. If this is the case, the system needs to be relatively close by (a few hundred pc) for the brightness temperature not to exceed $10^{12}$\,K.

\subsection{What is the nature of \ulp?}\label{sec:progenitor}
We start by investigating the nature of the source: whether \ulp\ is an isolated star or a binary.

\subsubsection{Isolated source --- Compact object}

We start by examining the possibility of \ulp\ being an isolated compact object --- a pulsar, a magnetar, or an isolated magnetic WD. The long period and the wide bursts make a standard pulsar mechanism unlikely to explain the observed emission. The period and period derivative of \ulp\ place it beyond the pulsar death valley \citep{chen1993}. In addition, if \ulp\ is at $>100$\,pc, the observed X-ray luminosity would be greater than the spin-down luminosity. This coupled with the UV/OIR detections (pulsars are usually very faint at optical wavelengths) and the optical variability, makes it difficult for \ulp\ to be an isolated pulsar. 

Magnetars, powered by the decay of their magnetic fields ($B\gtrsim10^{14}$\,G; \citealt{magnetarreview}), are very bright at X-rays and often show pulsed X-ray emission. The duty cycle of the pulses can vary from $\sim$20\%--100\%. Some of them are also radio-loud and can show pulse-to-pulse variability. Although rare, magnetars can be detectable at OIR \citep{Hulleman2000,Durant2005} and can show variability at these wavelengths \citep{Israel2002,Rea2004}. 
Magnetars are usually observed with periods between 2--10\,s, however, proposals involving an accretion phase to slow down the magnetar even further have been put forward to explain the periods observed in LPTs \citep{ronchi2022}. The radio spectrum in magnetars is usually flat or inverted ($\alpha \gtrsim 0$), although there can be periods where it is steep \citep{Bansal2023,Torne2015}. In addition, magnetars usually show very high levels of linear polarisation with little to no circular polarisation. The steep spectrum in \ulp, the high circular polarisation, the harmonic narrow-band structure, and the optical variability are inconsistent with \ulp\ being an isolated magnetar.

Although undetected so far, isolated WD pulsars are also proposed as the underlying sources for LPTs \citep{Zhang2005,Katz2022}. The increased moment of inertia of the WD ($\sim 10^{50}$\,g\,cm$^{-2}$) allows for longer periods and makes the spin-down luminosity adequate to power the radio pulses in a WD at $\sim$\,kpc. The broadband SED of \ulp\ can be consistent with the thermal emission from a $\sim$15\,000\,K WD at this distance. However, if the radio pulses are powered by pair production and their subsequent acceleration in the WD magnetosphere (similar to pulsars), then there should be  a ``death valley'' for isolated WD pulsars (similar to pulsars, see Figure~1 in \citealt{rea_long-period_2024}). Our weak constraint on $|\dot{P}|<2\times10^{-8}$ yields a weak constraint on the surface magnetic field $B<10^{12}$\,G (assuming that it is spindown powered), which is much higher than the extreme fields of $B\sim10^9$\,G invoked for WD pulsars. Even a lower field (B$\sim10^9$\,G) places it beyond the death valley for WDs \citep{rea_long-period_2024}. This, coupled with the observed optical variability and the harmonic narrow-band frequency structure, makes it unlikely that \ulp\ is an isolated magnetar, although we can not  rule it out.

\subsubsection{Isolated source --- Alternate scenarios}

If it is an isolated main sequence star, the observed period limits the mass of the star to $\lesssim 1\,M_{\astrosun}$, given centrifugal breakdown. Evolved stars (like Wolf-Rayet stars or sub-luminous dwarf stars) can be rotationally sustained; however, radio emission is rare \citep{sdbreview,Dougherty2000} and is usually powered by incoherent emission. Given the distance limit from the brightness temperature (a few hundred pcs), this implies that such objects have to be much brighter than observed and can be ruled out.  

A second possibility, if \ulp\  is an isolated star, is that \ulp\ is an ultra-cool dwarf (M7.5 or later, including L/T/Y; from here on unless explicitly stated, we refer to $>$M7.5/L/T/Y stars as ultra-cool dwarfs or UCDs). UCDs can explain the observed radio bursts and the period \citep{Berger2001,Berger2002,Hallinan2007,Kao2016,Vedantham2020,Rose2023}. In particular, M-dwarf stars, known for their flaring activity, can explain variability across wavelengths. During the flaring activity, M-dwarfs can appear bluer due to enhanced emission (both continuum and line) from their chromospheres \citep{uvceti_flares,Kowalski2013,Gunther2020}. Elliptically polarised radio bursts have been observed in sources like UV-Ceti \citep{Zic2019,Villadsen2019}. However, the SED, which peaks around NUV, is difficult to explain if \ulp\ is an isolated dwarf star.

\subsubsection{Binary}\label{sec:binary}
We then consider the possibility of a binary system. There can be two possibilities here: interacting binaries (like CVs, magnetic or non-magnetic) and non-interacting binaries (detached WD binaries). 

\textit{Non-interacting binaries:} Section \ref{sec:mm_spec} (and Figure~\ref{fig:sed_multi_bb}) shows that the SED can be described by a non-interacting binary system with a hot WD and a cold dwarf placed at a kpc distance. In this case, the radio emission would likely originate in the magnetosphere of the dwarf star. 
However, most of the radio detections from cool dwarfs (that are similar in flux density to \ulp) are from sources relatively close by (within 100\,pc; \citealt{Callingham2021,Kao2024}). This is reflected in the relative strength of radio to IR behavior of stars (see Figure~\ref{fig:j1448_xray_ir_pop}) which indicates that \ulp\ is unusually radio bright. The only comparable source to \ulp\  is BDR 1750+3809, which is a close-by radio-discovered UCD \citep{Vedantham2020}. For such a source to be detected at $\sim$\,kpc, the radio emission has to be extremely bright (brightness temperatures of $10^{17-18}$\,K), brighter than some of the observed pulsars/magnetars, making it unlikely. 

Alternately, if \ulp\ is a non-interacting binary like a WD pulsar \citep{marsh_radio-pulsing_2016,pelisoli_53-min-period_2023}, the interaction of the WD beam with the companion produces broadband emission at the beat frequency (spin and orbital periods). However, the lack of pulsations on time scales of $\sim$min may rule out the presence of a fast spinning WD. For a 1\,$M_{\astrosun}$, 6000\,km WD, centrifugal breakdown limits the spin period to $>$\,10\,s, which is greater than the integration time of our radio observations and hence we would not miss the detection of WD spin period if there was one.

\textit{Interacting binaries:} When it comes to interacting binaries, the most likely scenario is an MWD binary, like a polar or an intermediate polar. X-ray/UV emission in polars is dominated by the thermal emission in the accretion column \citep{Ferrario2020}, while the optical emission results from the cyclotron emission, and the IR emission is dominated by the companion. Circularly polarised radio emission is observed from MCVs \citep{Barrett2020} and is hypothesized to be the result of ECME. 
However, X-ray/UV/Optical emission from polars is usually modulated at the orbital (and hence the spin) period as the line of sight to the accretion column changes, but neither the X-ray light curve (from XMM) nor the optical light curve (from the Leesedi) show hint for strong variations. This implies that the inclination angle to the system is low, leading to weak or no modulations.

On the other hand, if the magnetic field of the WD is not too high (kG) to allow the formation of an accretion disk, the presence of an accretion disk will give rise to multi-wavelength emission, from UV (at the innermost edge of the disk) to IR (outer edges of the disk + possibly the M-dwarf). We get reasonable fits by modelling the observed SED with a disk spectrum (see \S\ref{sec:mm_spec}, \S\ref{sec:disk}), with the temperature at the inner radius (92\,000\,K) consistent with CVs \citep{accretion}. For the inner edge of the disk to penetrate close to the WD, the WD's surface magnetic field cannot be too high ($\sim$\,kG--MG). The observed X-ray emission is at least 3 orders of magnitude too bright to be powered by a 92\,000\,K disk, implying that it might be coming from the polar cap heating of the WD.

In both cases (polars or weakly magnetized WD binaries), the observed flare and its bluer appearance can also be explained either as an instability in the disk or a partial nova-like ejection episode from the WD. However, comparison of the \ulp's broadband SED with prototypical systems of such kinds (see Figure~\ref{fig:sed_comp} and \S\ref{sec:sed_comp}) shows that the SEDs are not similar to such known systems. The radio emission in \ulp\ is brighter by several orders of magnitude (a broader comparison with a sample of MCVs, see Figure~\ref{fig:j1448_xray_ir_pop}, leads to a similar conclusion\footnote{Given the optical variability and the radio pulse-to-pulse variability, if there are multiple states in \ulp, similar to ASKAP J1832$-$0911 \citep{Wang2025}, then care should be taken during the comparison. But given the lack of radio observations in the optically bright state in \ulp\ and a wider LPT population in general, more evidence will be required to address this.}) compared to known systems. In addition, the optical/X-ray portion of the SEDs suggests that either the X-ray emission is heavily suppressed or the optical emission is enhanced\footnote{The distance to \ulp\ is uncertain and hence we normalized the SEDs to have the same \textit{J}-band magnitude. One can also normalize them to the same X-ray flux, in which case the optical emission in \ulp\ will be much brighter compared to known systems.}, at least by 2 orders of magnitude. In either case, the radio emission in \ulp\ is too bright compared to known systems, suggesting that if \ulp\ is similar to an MWD binary, its discovery might represent a subpopulation of such sources with unusually bright radio emission (leading to radio discovery). 

The origin of radio emission still needs to be explained: if it is from an incoherent mechanism like gyro-synchrotron emission, radio emission is likely at multiple harmonics (although not observed so far). \cite{Chanmugam1982} proposed radio emission in MCV systems (like AM~Her) through such higher harmonics, although without evidence of clear narrow band features. However, this still has limitations to it --- as discussed in \S\ref{sec:origin}, we get a maximum distance of 200\,pc for incoherent emission, beyond which the brightness temperature exceeds $10^{12}$\,K. Using the upper limit on the amplitude in the case of a disk and the maximum distance, we can constrain the system's inclination to be almost perfectly edge-on. Such a configuration (unlikely but not impossible) might also help explain the non-discovery of such a nearby MCV system since the disk is nearly edge-on. Alternately, given the level of polarisation, radio emission is more likely from a coherent process (like ECME), although the observed narrow-band structure is difficult to explain in the paradigm of ECME.

Finally, we cannot  rule out other rare classes of objects like a transitional NS binary, similar to PSR J1023+0038 \citep{Bond2002,archibald2009}, although they are unlikely. PSR J1023+0038 was discovered as a variable radio source with a CV-like optical spectrum. The broadband SED of PSR J1023+0038 is shown in Figure~\ref{fig:sed_comp}, and it broadly resembles the observed spectrum of \ulp. Outbursts from PSR J1023+0038 were seen at X-ray and optical wavelengths during periods when the radio pulsations were absent \citep{bogdanov15,baglio25}, suggesting an accretion phase. The period of \ulp\ is shorter than the orbital periods of known transitional millisecond pulsars \citep[$\gtrsim 5\,$hr;][]{papitto22}, although it is consistent with the populations of accreting millisecond pulsars and the so-called `spider' pulsars \citep{spiders}. However, our search for radio pulsations has revealed no pulsar candidates, and the bursts from \ulp\ had very similar widths as a function of frequency.  In contrast,  pulsations were detected from PSR~J1023 during most phases of its orbit at higher frequencies, and the eclipse durations were very frequency dependent \citep{archibald2009}. While not definitive, this suggests that \ulp\ may not be similar to PSR J1023+0038 in the quiescent state.

\subsubsection{What is the source of energy in \ulp\ ?}\label{sec:sed_comp}
\begin{figure*}
    \centering
    \includegraphics[width=0.75\textwidth]{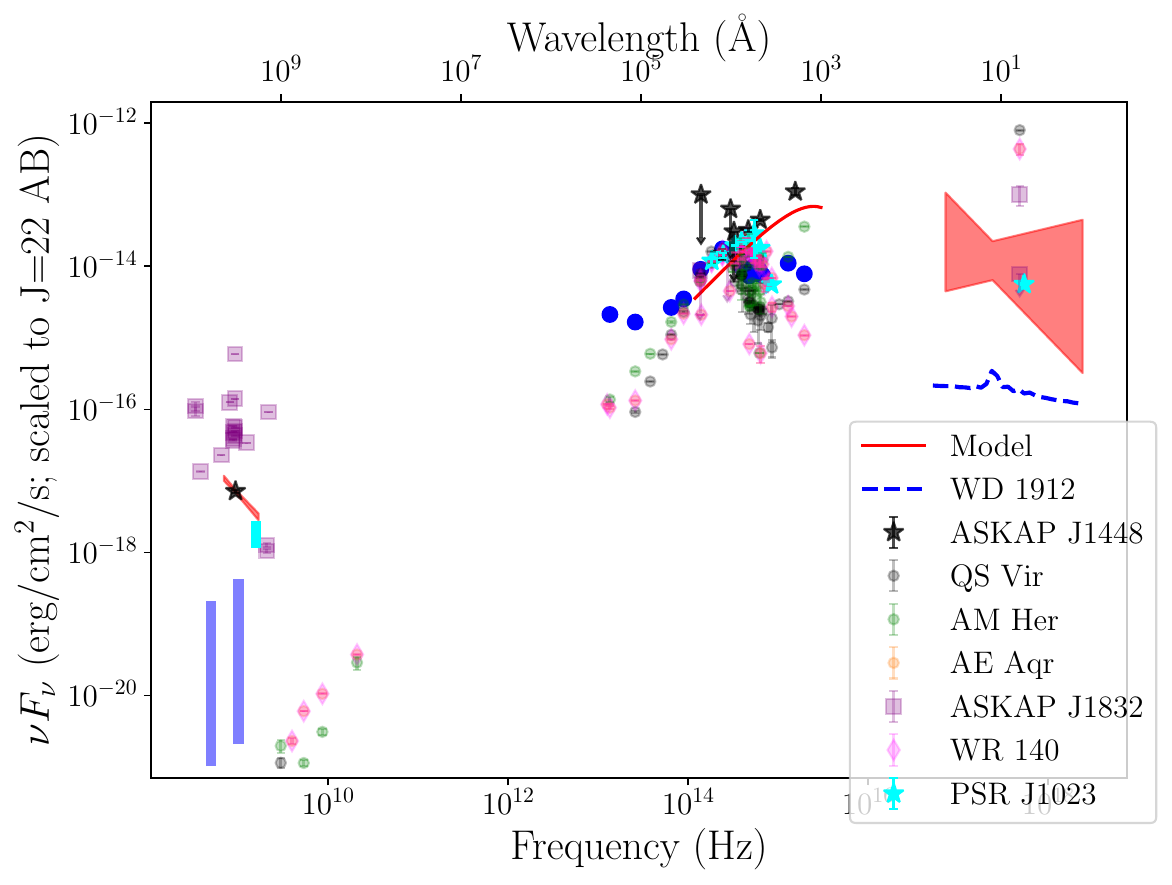}
    \caption{Multi-wavelength SED of \ulp\  showing the phase-averaged flux emitted at different wavelengths. The black stars show the observed data and the red curves show the model fits to the spectra (see \S\ref{sec:mm_spec}). For comparison, we show three typical CVs, a non-magnetic CV (QS Vir as black dots; \citealt{ridder2023}), a polar (AM Her as green dots; \citealt{Chanmugam1982}), and a propeller-type accreting CV (AE Aqr as orange dots). We also show a WD pulsar, WD 1912 (blue lines/dots; \citealt{pelisoli_53-min-period_2023}), the X-ray bright LPT, ASKAP J1832$-$0911 (purple squares; \citealt{Wang2025}), a transitional millisecond pulsar (PSR J1023+0038 in the `pulsar' state; \citealt{Archibald2010}), and a WR star (as magenta diamonds). All the sources are scaled to have the same $J$-band flux as \ulp, and all the data have been corrected for interstellar absorption.}
    \label{fig:sed_comp}
\end{figure*}

Figure~\ref{fig:sed_comp} shows the broadband multi-wavelength SED of \ulp\  from radio to X-rays. In the case of \ulp, this seems to hint that much of the energy is emitted at UV wavelengths. To compare the relative contributions to the SED of radio and other wavelengths, we show three typical CVs (a non-magnetic, a magnetic, and a propeller system) all scaled to have the same $J$-band flux), a WR star, the WD pulsar WD1912 \citep{pelisoli_53-min-period_2023}, the transitional MSP \citep{Archibald2010} and the X-ray bright LPT ASKAP~J1832$-$0911 \citep{Wang2025}.

When compared to CVs/MCVs, this suggests that the observed X-ray luminosity is much fainter than what is observed in CVs (for the same optical flux), while the observed radio emission is much brighter. This would mean that if \ulp\ was a similar system to CV/MCV, then much of the optical/X-ray flux has to be suppressed (probably due to geometric effects) in comparison to the radio flux, suggesting an edge on system. Comparing it to the WD pulsar WD 1912, both the X-ray and the radio emission are much brighter (for the same optical flux), which suggests that the emission mechanisms in LPTs and WD pulsars may be different; however, as addressed above, the presence of multiple states in \ulp\ makes it uncertain.
It is comparable to the X-ray bright LPT ASKAP J1832 in its X-ray faint state, although ASKAP J1832 was not detected at OIR wavelengths. 

The optical variability in \ulp\ suggests that there might be different states for \ulp\ as well. Although \ulp\ in its quiescent state is inconsistent with any of the source populations, its appearance in the optically bright state may be close to known analogs like CVs. If this is the case, the observed radio brightness implies that \ulp\ (and perhaps a fraction of LPTs in general) might represent a sub-population that is unusually radio bright in comparison to its multi-wavelength emission with respect to known classes of sources. However, given its radio discovery, in contrast to optical/X-ray discovery of CVs, the initial discoveries can unsurprisingly be extremely radio bright, similar to the radio discovery brown dwarf BDR 1750+3809 \citep{Vedantham2020}. Whether this is intrinsic to the source or a geometric effect is something that can be answered by discovering more LPTs that can be detected across wavelengths.

\section{Conclusions}\label{sec:conclusions}
We report the discovery of a new LPT, \ulp. Discovered as a 1.5\,hr periodic radio source, \ulp, shows a steep spectrum, elliptical polarisation, and periodic narrow-band emission that declines at frequencies above 1.5\,GHz. \ulp\  has also been detected across wavelengths, from X-rays to radio, with the spectrum peaking at NUV wavelengths. Multi-wavelength modeling of the SED combined with the radio properties indicates that \ulp\ may be an MWD binary with a magnetic field strength of $>1$\,kG. However, we could not conclusively rule out \ulp\ being an isolated WD pulsar or a tMSP-like system. The origin of the radio emission is also uncertain, with coherent emission (via ECME) or incoherent emission (via gyro-synchrotron emission) both plausible, depending on the distance to the source. The harmonic structure in the frequency spectrum likely corresponds to multiple harmonics (harmonic number of 40--60) of the fundamental frequency (17\,MHz). If it is incoherent gyro-synchrotron emission, using the limit of the brightness temperature, we deduce that the system has to be within 200\,pc. This implies a highly inclined, almost edge-on MCV system, but we see no evidence for orbital modulation of the optical light curve (although this is due to the sensitivity of our $g$-band observations). 

Putting \ulp\  in a broader context of astrophysical transients, if \ulp\  is indeed an MCV, it would be the first radio-discovered MCV and highlights its relative radio brightness as an indicator to discover more such systems. This hints at a (sub)-population of transients that, either due to their geometry or intrinsic radio brightness, can be better discovered with the current all-sky radio surveys. 
Combining \ulp\ with the growing number of long-period radio transients adds to the variety of multi-wavelength behavior and will help deepen our understanding of this emerging population (or, indeed, populations).

\section*{Acknowledgements}
The authors declare no conflict of interest. This work made use of data from the following facilities: \textit{ASKAP, ATCA, MeerKAT, ESO:VISTA, Magellan:Baade, DECam, SAAO:1m, Swift(XRT and UVOT), XMM (EPIC)}.

We thank Ingrid Pelisoli for providing the SED data for WD 1912 and Sarah Buchner for helping with the MeerKAT observations. 

We thank the anonymous referee for providing very useful feedback, which improved the manuscript. AA and DLK are partially supported by NSF grant AST-1816492. N.R. is supported by the European Research Council (ERC) via the Consolidator Grant “MAGNESIA” (No. 817661) and the Proof of Concept ``DeepSpacePulse" (No. 101189496), by the Catalan grant SGR2021-01269, the Spanish grant ID2023-153099NA-I00, and by the program Unidad de Excelencia Maria de Maeztu CEX2020-001058-M. N.H.-W. is the recipient of an Australian Research Council Future Fellowship (project number FT190100231). Parts of this research were conducted by the Australian Research Council Centre of Excellence for Gravitational Wave Discovery (OzGrav), project number CE170100004. This research has made use of the SVO Filter Profile Service "Carlos Rodrigo", funded by MCIN/AEI/10.13039/501100011033/ through grant PID2023-146210NB-I00.

This scientific work uses data obtained from Inyarrimanha Ilgari Bundara / the Murchison Radio-astronomy Observatory. 
We acknowledge the Wajarri Yamaji People as the Traditional Owners and native title holders of the Observatory site. 
The Australian SKA Pathfinder is part of the Australia Telescope National Facility (\url{https://ror.org/05qajvd42}) which is managed by CSIRO. Operation of ASKAP is funded by the Australian Government with support from the National Collaborative Research Infrastructure Strategy. ASKAP uses the resources of the Pawsey Supercomputing Centre. The establishment of ASKAP, the Murchison Radio-astronomy Observatory, and the Pawsey Supercomputing Centre are initiatives of the Australian Government, with support from the Government of Western Australia and the Science and Industry Endowment Fund. The Australia Telescope Compact Array is part of the Australia Telescope National Facility (grid.421683.a), which is funded by the Australian Government for operation as a National Facility managed by CSIRO. We acknowledge the Gomeroi people as the traditional owners of the Observatory site.

We thank SARAO for the approval of the MeerKAT DDT request DDT-20240719-AA-01. The MeerKAT telescope is operated by the South African Radio Astronomy Observatory, which is a facility of the National Research Foundation, an agency of the Department of Science and Innovation. Observations made use of the Pulsar Timing User Supplied Equipment (PTUSE) servers at MeerKAT, which were funded by the MeerTime Collaboration members ASTRON, AUT, CSIRO, ICRAR-Curtin, MPIfR, INAF, NRAO, Swinburne University of Technology, the University of Oxford, UBC, and the University of Manchester. This research is based on observations obtained with XMM-Newton, an ESA science mission with instruments and contributions directly funded by ESA Member States and NASA. We thank Norbert Schartel (the XMM PI) for accepting our DDT request for the XMM and the XMM team for scheduling and performing the observations. This article made use of data from the Swift telescope. We thank the Swift team for rapidly approving and scheduling these observations. This paper includes data gathered with the 6.5-meter Magellan Telescopes located at Las Campanas Observatory, Chile. This article used data products from observations made with ESO Telescopes at the La Silla or Paranal Observatories under ESO programme ID 179.B-2002. This paper uses observations made at the SAAO.

%





\section*{Data Availability}

All the data except for the optical data from the Lesedi telescope and the NIR data from the Magellan telescope are public. The data underlying this article is available on the respective telescope archives (ASKAP, MeerKAT, Swift, XMM, DECaPS, VVV) --- for example, calibrated ASKAP visibilities, images, and source catalogues are available from CASDA (\url{http://data.csiro.au}) and MeerKAT visibilities are available from the SARAO web archive (\url{https://archive.sarao.ac.za/}). The Lesedi/Magellan data can be shared by the authors upon reasonable request.



\bibliographystyle{mnras}
\bibliography{references} 




\appendix
\section{Data reduction}
\subsection{Radio data reduction}\label{sec:radio_data_analysis}
\subsubsection{ASKAP EMU}\label{sec:emudata}
Calibrated visibility data for the EMU observations are publicly available. Two EMU observations cover the field of \ulp. The first EMU observation was on June 15, 2023, and the second observation was on May 26, 2024. The EMU observations were carried out in a continuous 10\,hr window at a central frequency of 943\,MHz with a bandwidth of 288\,MHz. We used CASA \texttt{tclean} with \texttt{briggs} weighting, a \texttt{robust} parameter of 0.5 and 10\,000 iterations, to generate the sky model and subtracted it to get the model subtracted visibilities. We removed \ulp\ from the model so that it is retained in the model-subtracted visibilities. We performed a final round of deconvolution to generate images in all four Stokes parameters. To generate the time and frequency-resolved intensities, also known as a \textit{the dynamic spectrum}, we used \texttt{phaseshift} to rotate the phase center of the model-subtracted visibilities to the \ulp's position. We then averaged the emission over all baselines using \texttt{dstools} \citep{dstoools} to generate the dynamic spectra.

Figure~\ref{fig:emu_ds} shows the dynamic spectra from both EMU observations. The corresponding light curves, obtained by frequency averaging, are presented in Figure~\ref{fig:emu_lc}. To look for periodic bursts (in time), we performed autocorrelation on the light curves which revealed a period of $\sim1.5$\,hrs. We obtained the frequency spectrum by averaging across the bursts in time and performed autocorrelation on these spectra to look for any harmonic/narrow-band emission. Figure~\ref{fig:burst_period_emu} shows the corresponding correlations and it can be seen from the bright bursts that there is a hint for harmonic emission with a fundamental frequency of $\sim$20\.MHz.

\begin{figure*}
    \centering
    \includegraphics[width=2\columnwidth]{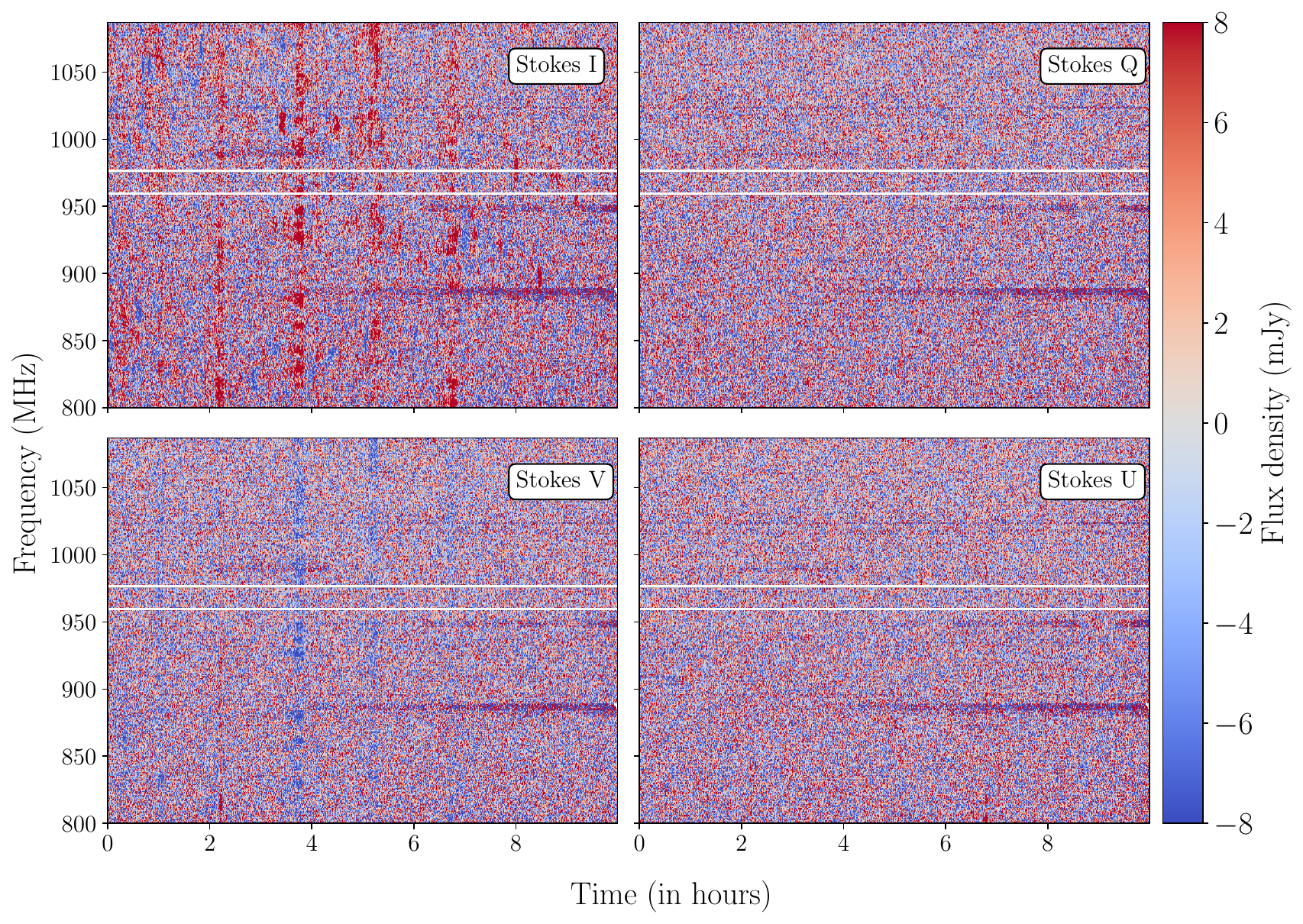}
    \caption{Dynamic spectra of all four Stokes parameters with a 10\,s time resolution and 1\,MHz frequency resolution from the second 10\,hr EMU observation on May 26, 2024, and 799--1090\,MHz bandpass. We can see multiple radio bursts in all four polarisations. The color scale on the right represents the flux density. The horizontal gaps (white lines) in the spectra are the flagged frequency channels. The two persistent narrow band features (around 900 and 950\,MHz) are likely due to un-flagged radio frequency interference (RFI).}
    \label{fig:emu_ds_app}
\end{figure*}

\begin{figure}
    \centering
    \includegraphics[width=\columnwidth]{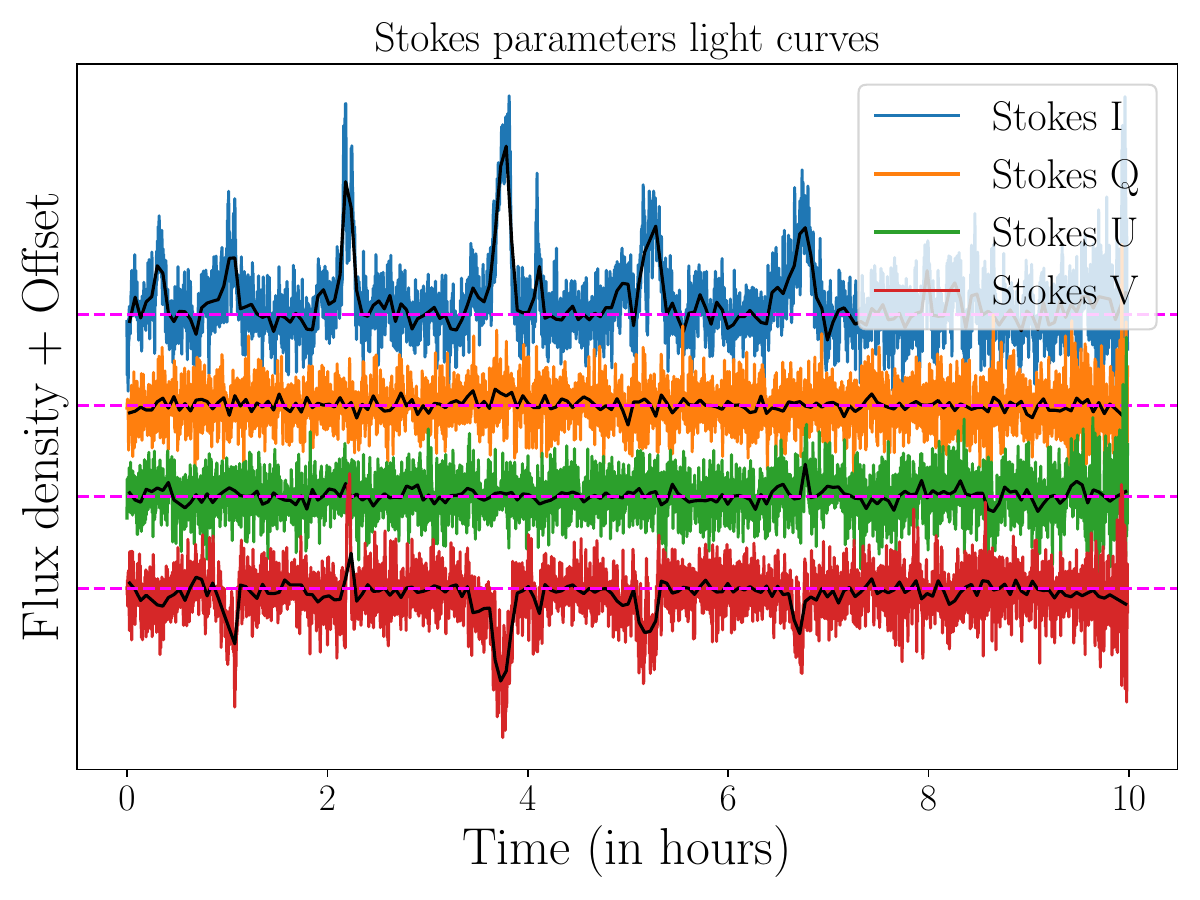}
    \caption{Light curves of the Stokes parameters from the second EMU observation obtained by averaging the dynamic spectra along the frequency axis. The native resolution of the data is 10\,s and the overlaid black lines show the light curves rebinned at 200\,s. The dashed magenta line shows zero intensity levels.}
    \label{fig:emu_lc_app}
\end{figure}

\begin{figure*}
    \centering
    \includegraphics[width=2\columnwidth]{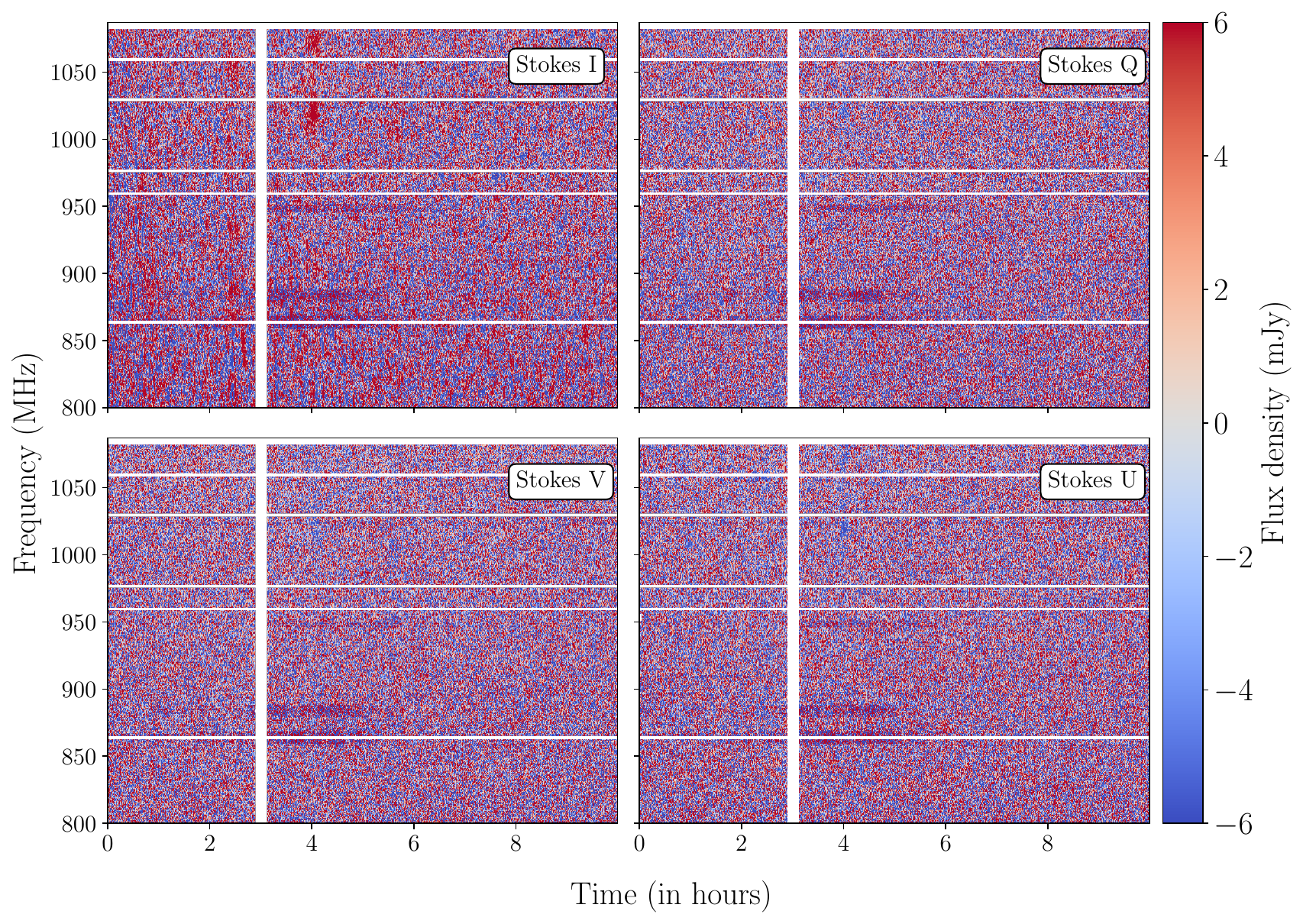}
    \caption{Dynamic spectra of all four Stokes parameters with a 10\,s time resolution and 1\,MHz frequency resolution from the third 10\,hr EMU observation on December 25, 2024, and 799--1090\,MHz bandpass. We can see one bright and a couple of faint bursts in all four polarisations. The color scale on the right represents the flux density. The horizontal gaps (white lines) in the spectra are the flagged frequency channels. The two persistent narrow band features (around 900 and 950\,MHz) are likely due to un-flagged radio frequency interference (RFI).}
    \label{fig:emu_ds_3_app}
\end{figure*}

\begin{figure}
    \centering
    \includegraphics[width=\columnwidth]{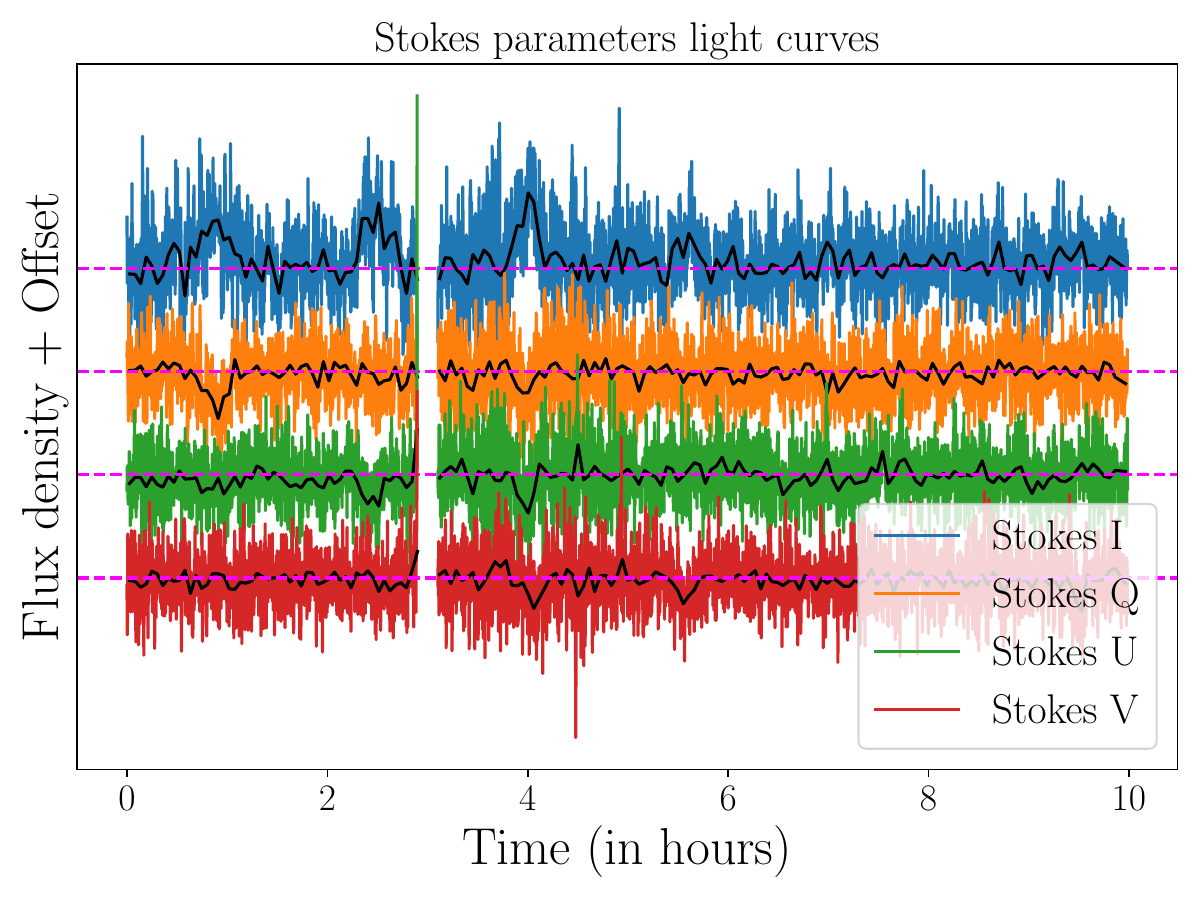}
    \caption{Light curves of the Stokes parameters from the third EMU observation obtained by averaging the dynamic spectra along the frequency axis. The native resolution of the data is 10\,s and the overlaid black lines show the light curves rebinned at 200\,s. The dashed magenta line shows zero intensity levels.}
    \label{fig:emu_lc_3_app}
\end{figure}

\begin{figure}
    \centering
    \includegraphics[width=\columnwidth]{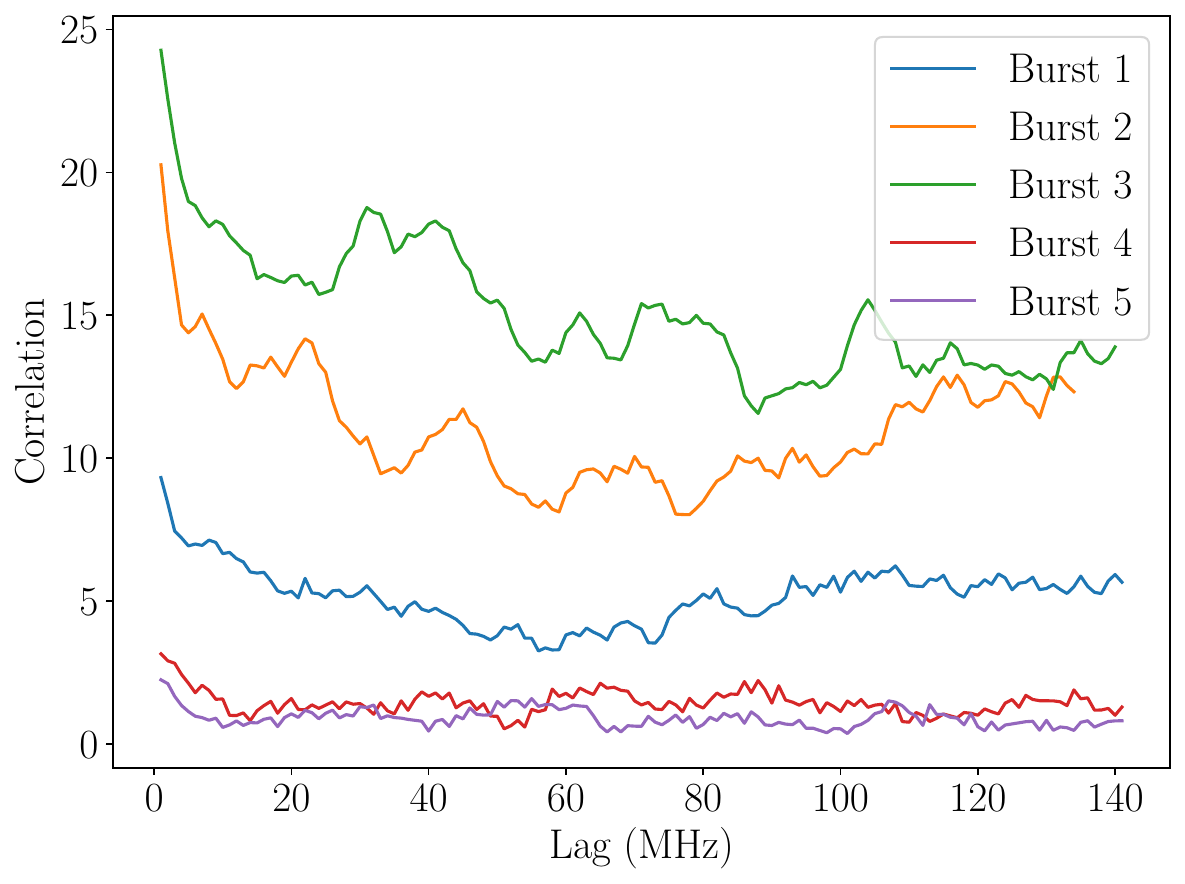}
    \includegraphics[width=\columnwidth]{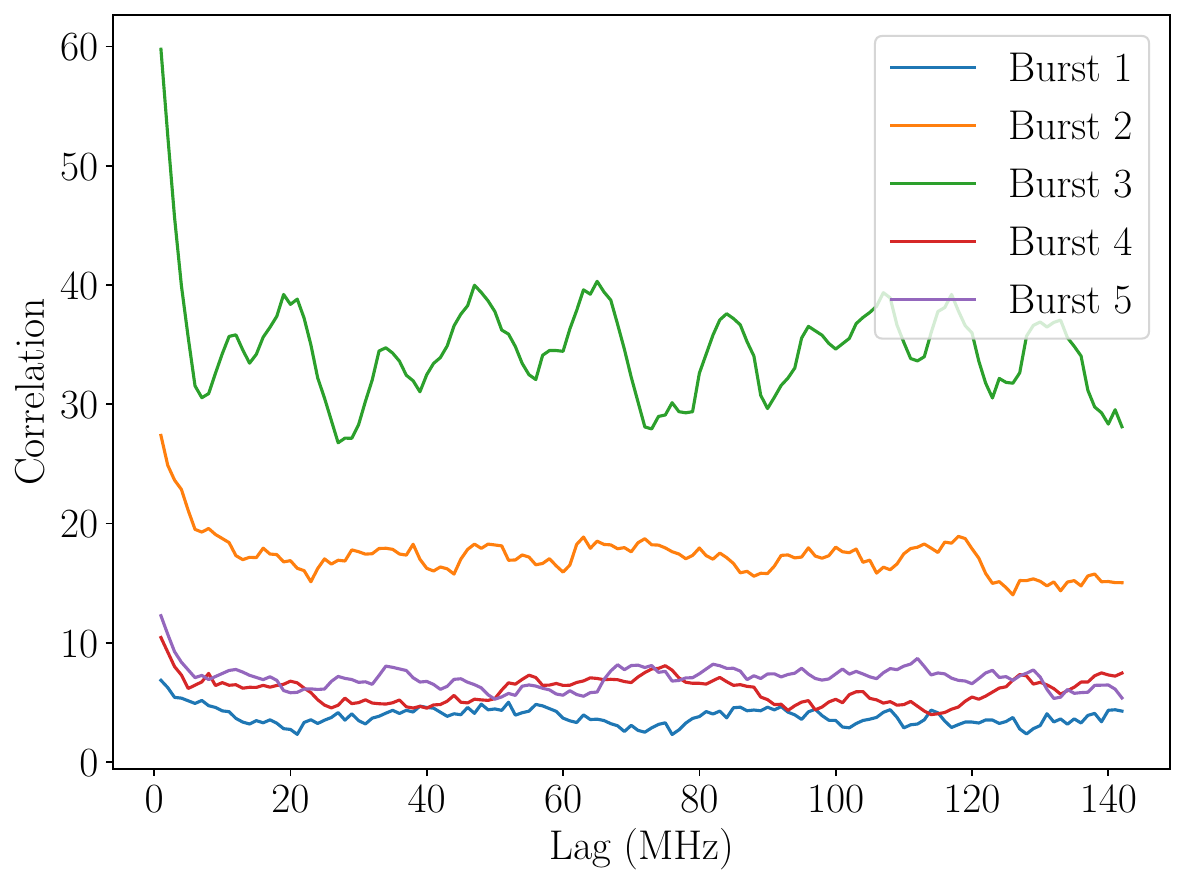}
    \caption{Plots showing the correlation coefficient of the 1D frequency spectrum of all the bursts in the EMU data (similar to Figure~\ref{fig:burst_period_mkt}). The left panel shows the correlation coefficient for the data from the first observation and the right panel shows the same for the second observation. A periodic feature at the $\sim$17\,MHz (fundamental) and its harmonics can be seen in at least one of the bursts from the EMU second observation. A similar (but less pronounced) feature can be seen in the data from the first observation.}
    \label{fig:burst_period_emu}
\end{figure}

\subsubsection{Australia Telescope Compact Array}\label{sec:atcadata}
\begin{figure*}
    \centering
    \includegraphics[width=2\columnwidth]{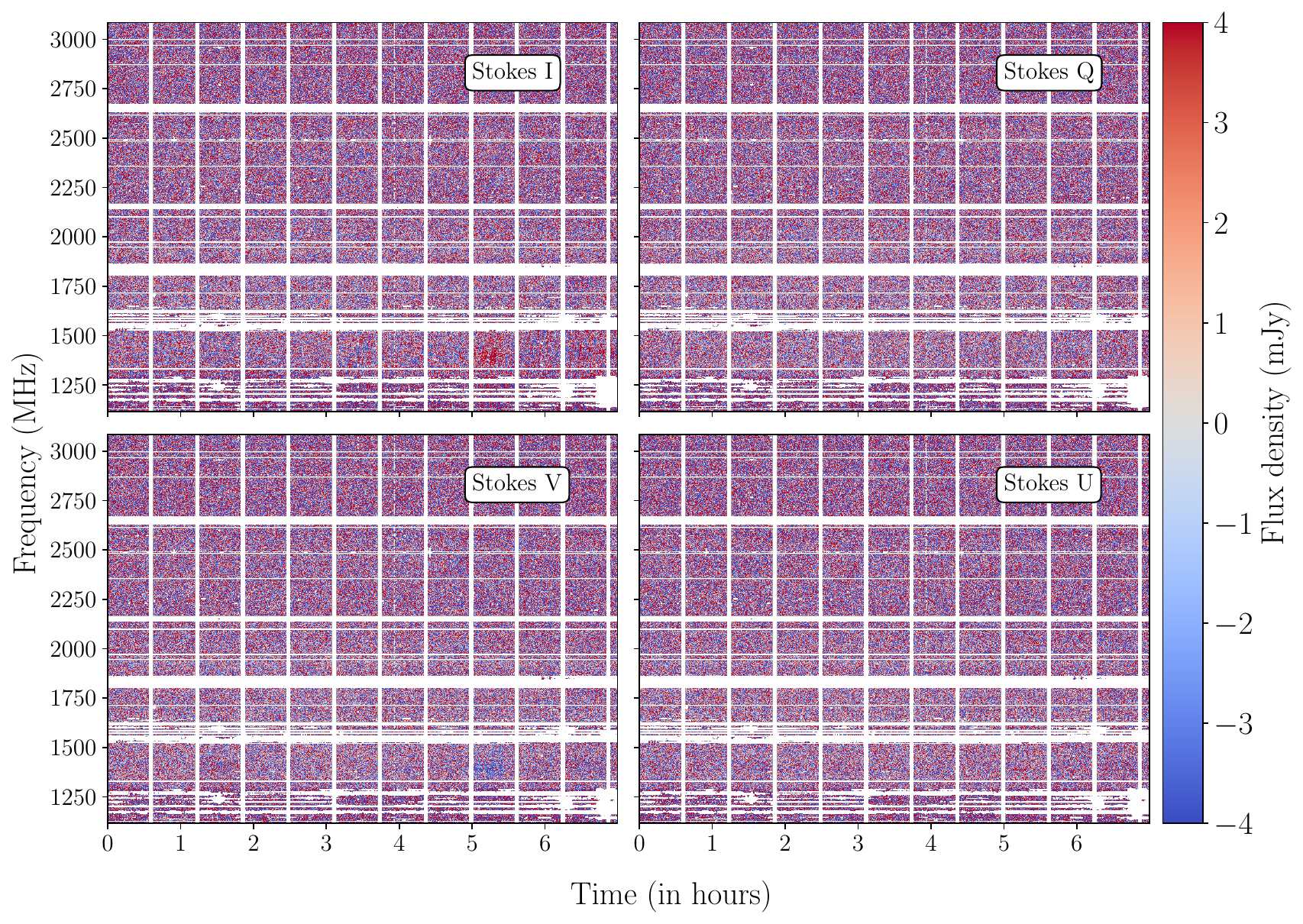}
    \caption{Dynamic spectra of all four Stokes parameters for the ATCA observation.  The color scale on the right represents the flux density for the Stokes I data. The horizontal white gaps are the flagged frequency channels, and the vertical gaps correspond to phase calibrator scans.}
    \label{fig:atca_ds}
\end{figure*}

\begin{figure}
    \centering
    \includegraphics[width=\columnwidth]{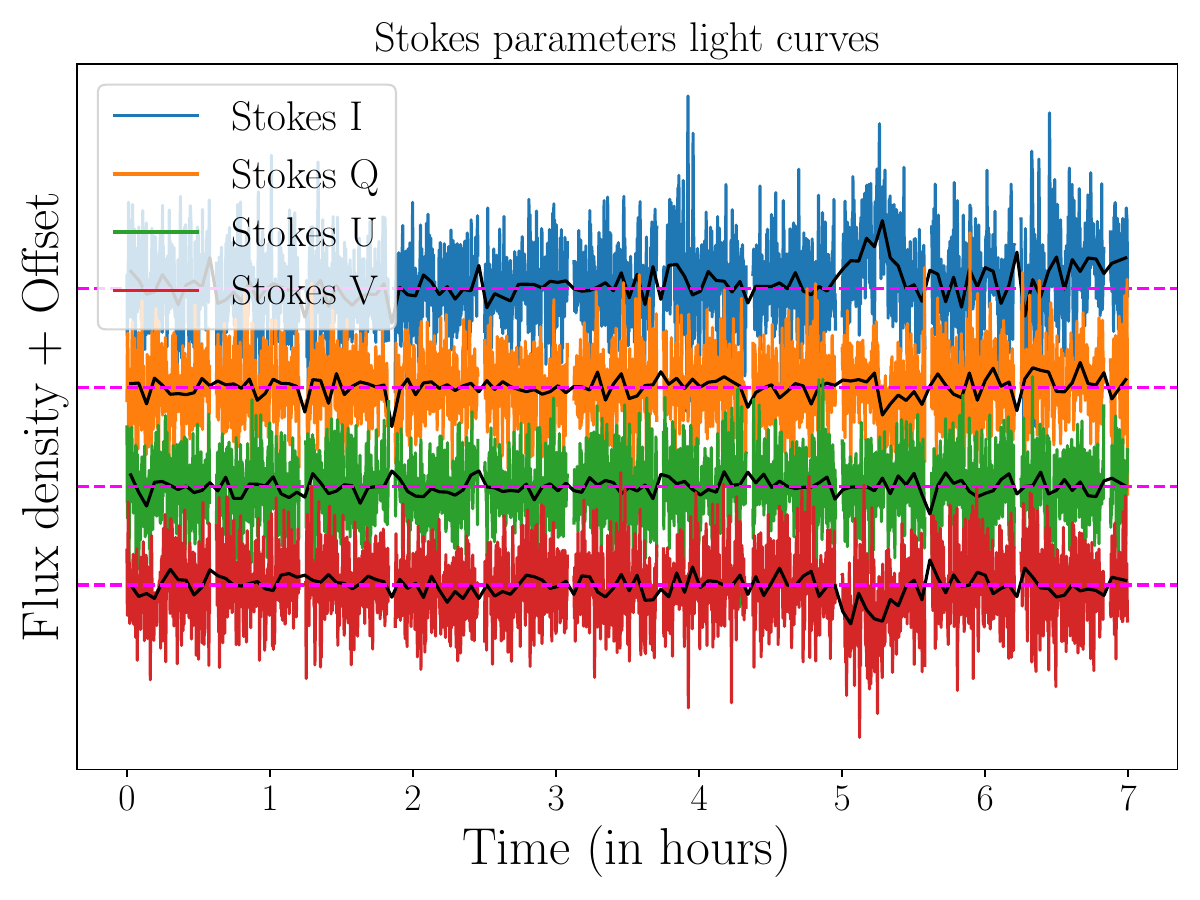}
    \caption{Light curves of the Stokes parameters obtained by averaging the dynamic spectra along the frequency axis from the ATCA observation.  The native resolution of the data is 10\,s, and the overlaid black lines show the light curves rebinned at 200\,s. The dashed magenta line shows the zero intensity level.}
    \label{fig:atca_lc}
\end{figure}

Observations were carried out in the L/S band (1.1-3.1\,GHz) for 6\,hrs, with a time resolution of 10\,s, on June 27, 2024, roughly 2 months after the second EMU observation. J1934$-$638 was used as the bandpass calibrator and J1329$-$665 was used as a phase calibrator. We used \textit{CASA} to flag frequency channels that were strongly affected by radio frequency interference (RFI). Calibrated visibilities were generated after applying the instrumental bandpass and phase corrections. We used CASA's \texttt{tclean} with \texttt{briggs} weighting, a \texttt{robust} parameter of 0.5 and 10\,000 iterations, to generate the sky model and subtracted it to get the model subtracted visibilities. We removed \ulp\ from the model so that it is retained in the model-subtracted visibilities. We performed a final round of deconvolution to generate images in all four Stokes parameters. To generate the time and frequency-resolved intensities, also known as a \textit{the dynamic spectrum}, we used \texttt{phaseshift} to rotate the phase center of the model-subtracted visibilities to the \ulp's position. We then averaged the emission over all baselines using \texttt{dstools} to generate the dynamic spectra. We did not detect any point source (in any of the Stokes parameters) in the full-time integrated images, but upon examining the dynamic spectra (Figure~\ref{fig:atca_ds}) we found a single burst, with lightcurve in Figure~\ref{fig:atca_lc}.

\subsubsection{MeerKAT}\label{sec:mktdata}
\begin{figure*}
    \centering
    \includegraphics[width=2\columnwidth]{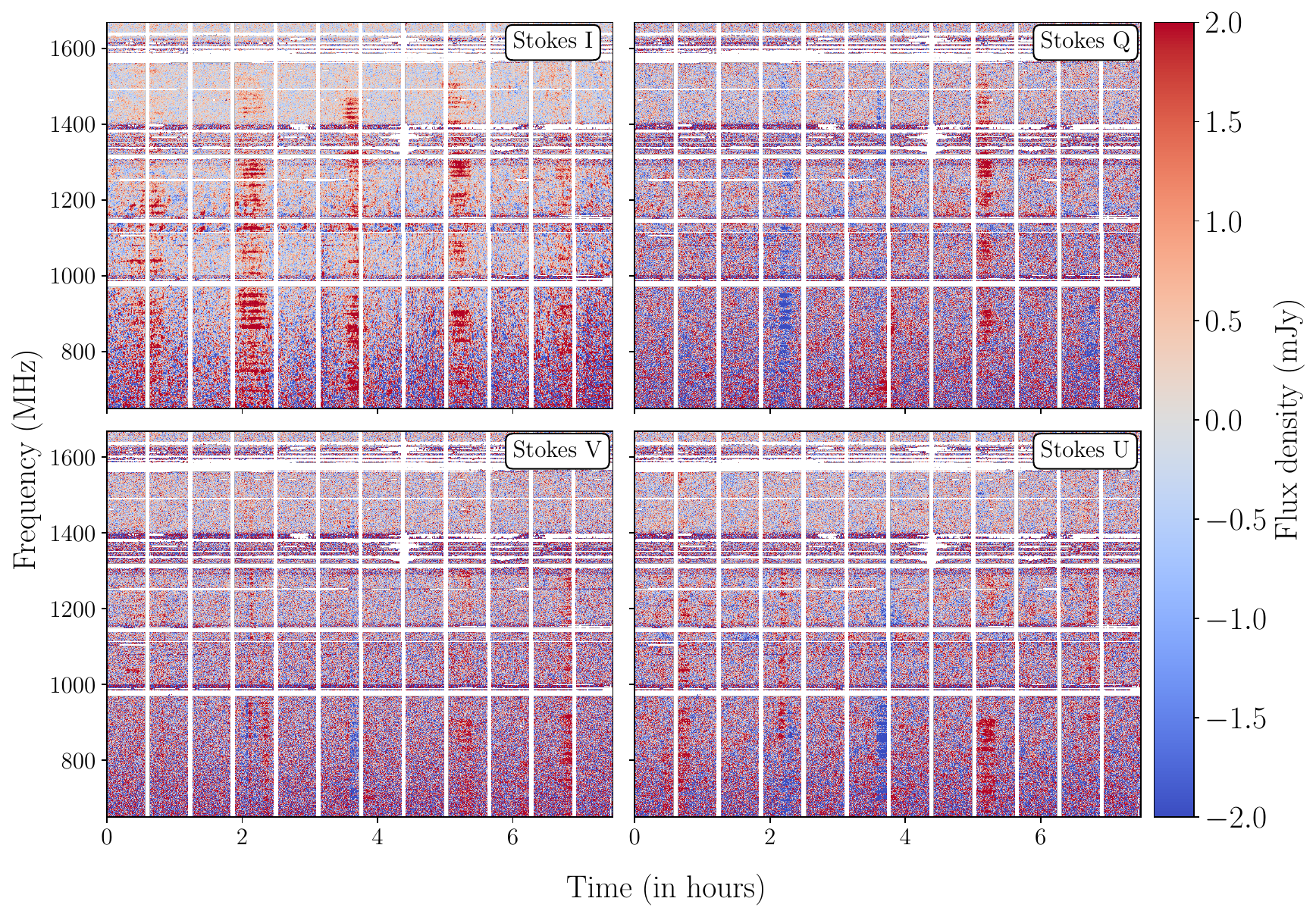}
    \includegraphics[width=2\columnwidth]{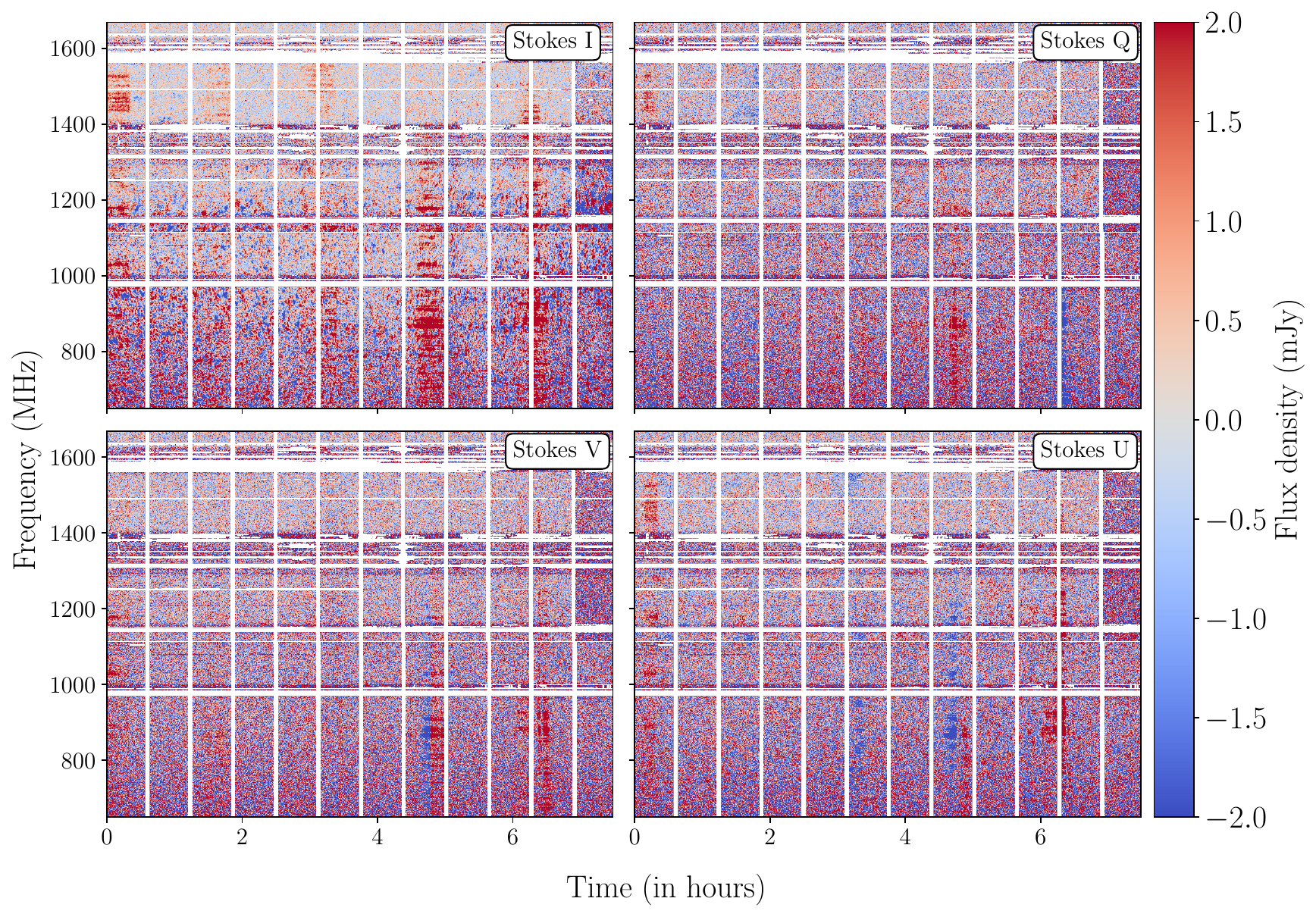}
    \caption{Dynamic spectra of all four Stokes parameters for the second 8\,hr MeerKAT observation over the entire band (UHF+L-band) from 650-1700\,MHz. The color scale on the right represents the flux density.}
    \label{fig:mkt_ds}
\end{figure*}

\begin{figure}
    \centering
    \includegraphics[width=\columnwidth]{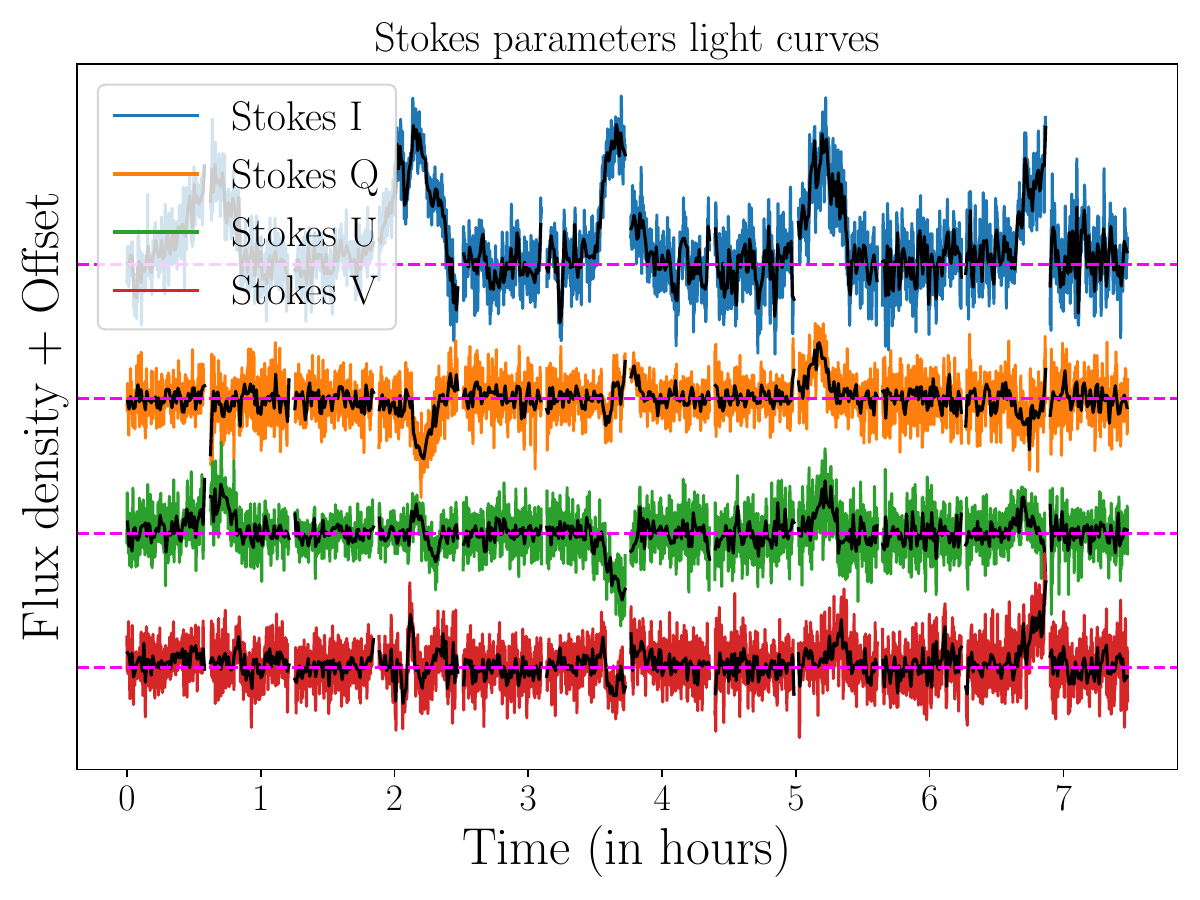}
    \includegraphics[width=\columnwidth]{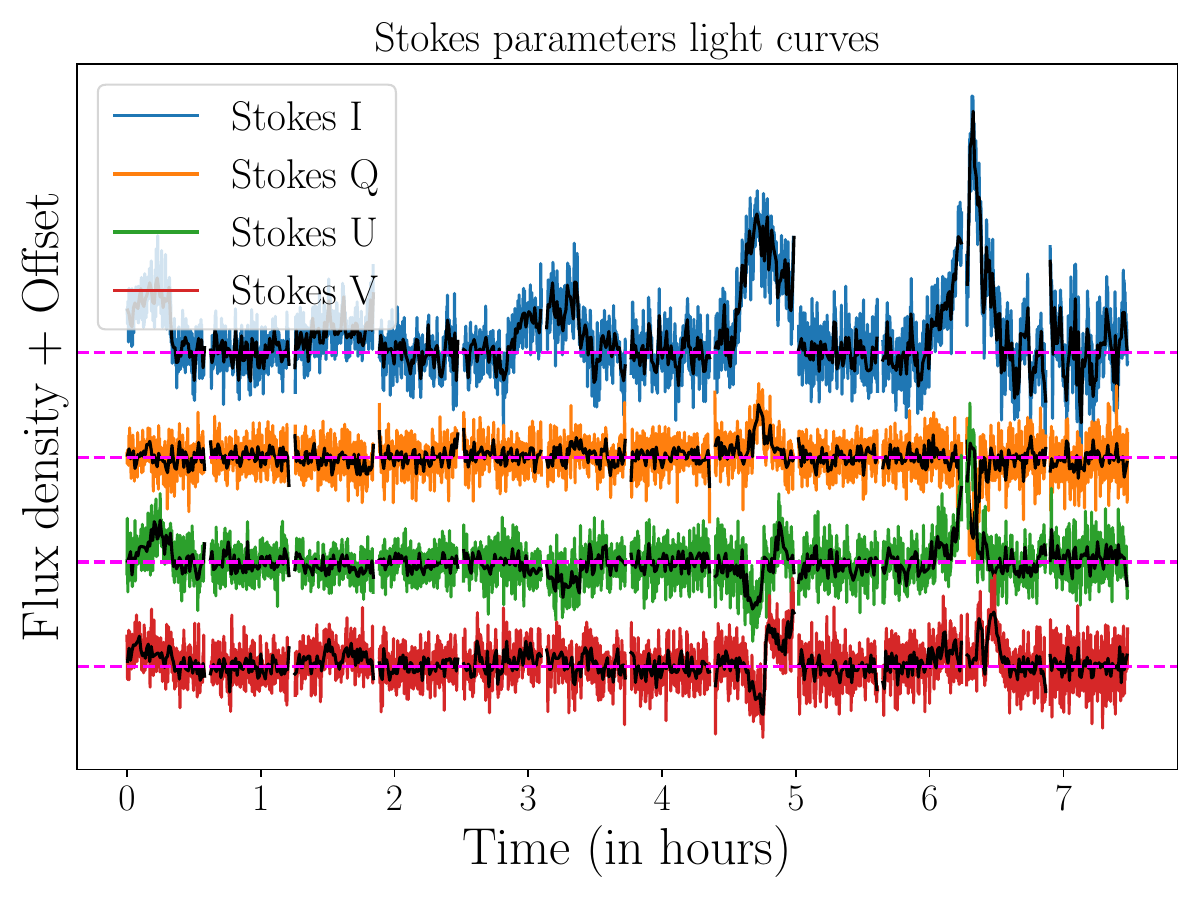}
    \caption{Light curves of the Stokes parameters obtained by averaging the dynamic spectra along the frequency axis. Shown on the left are the light curves from the first MeerKAT observation and on the right are the ones from the second MeerKAT observation. The native resolution of the data is 10\,s and the overlaid black lines show the light curves rebinned at 50\,s. The dashed magenta line shows zero intensity level.}
    \label{fig:mkt_lc}
\end{figure}

We observed \ulp\ with the MeerKAT radio telescope twice on July 30 and 31, 2024 for 8\,hr on each day. The array was used in split-array mode with half the antennae observing at the UHF band (544--1088\,MHz) and the other half at the L-band (856--1712\,MHz) at 8\,s resolution. J1939$-$6342 was used as a bandpass calibrator at the start of the observations and J1619$-$8418 was used as a phase calibrator during the observations. We used CASA to remove RFI, perform bandpass and gain calibrations. We used CASA's \texttt{tclean} with \texttt{briggs} weighting, a \texttt{robust} parameter of 0.5 and 10\,000 iterations, to generate the sky model and subtracted it to get the model subtracted visibilities. Model visibilities were generated independently at UHF and L-bands and dynamic spectra were generated for all four Stokes parameters. We removed \ulp\ from the model so that it is retained in the model-subtracted visibilities. We performed a final round of deconvolution to generate images in all four Stokes parameters. Dynamic spectra were generated using \texttt{dstools}. Figure~\ref{fig:mkt_ds} shows the MeerKAT spectra in all four Stokes parameters from both the observations. The corresponding light curves are shown in Figure~\ref{fig:mkt_lc}.

\begin{figure}
    \centering
    \includegraphics[width=\columnwidth]{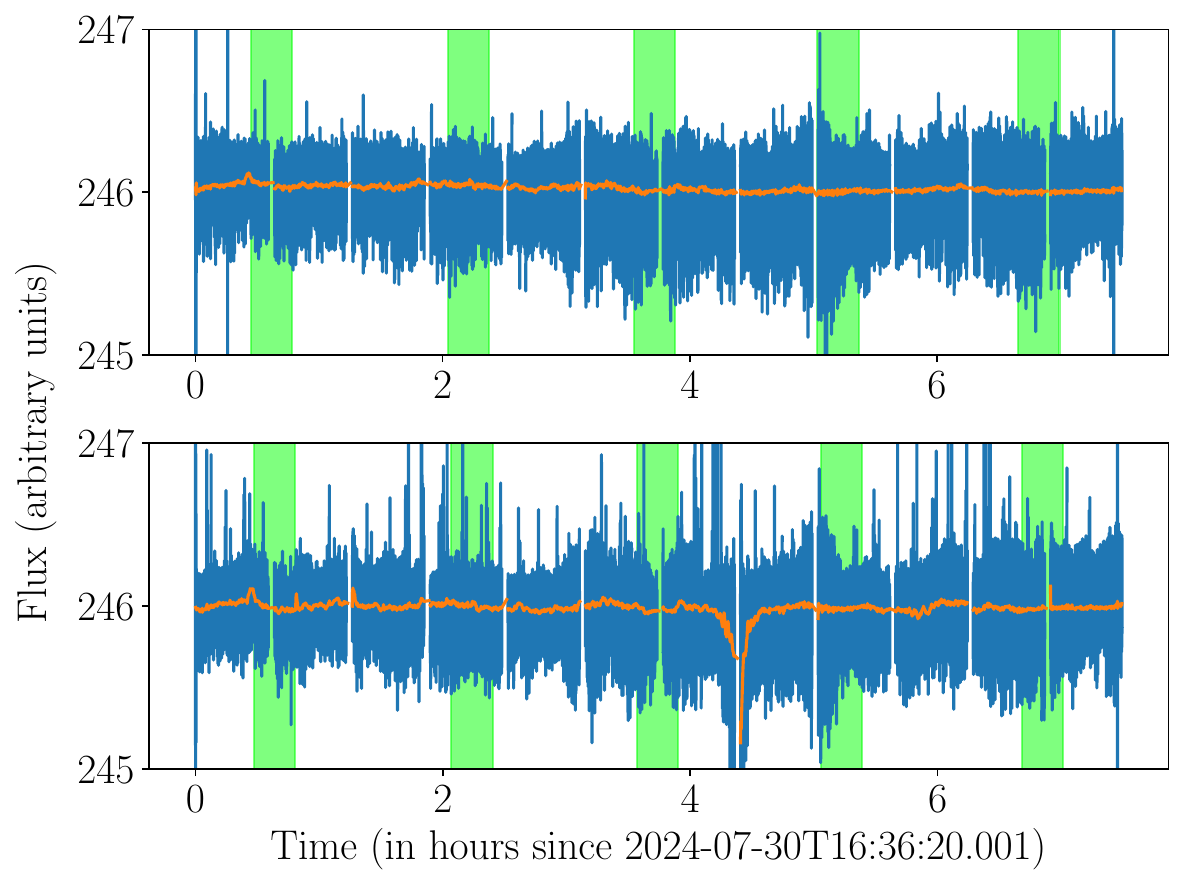}
    \caption{Light curve of \ulp\  using the uncorrelated beamformed data from the MeerKAT observations. Shown in blue is the light curve at 12.5\,ms resolution, and in the overlaid orange curve is the light curve resampled at 10\,s resolution. The underneath lime vertical bands show the pulses as seen in the correlated data. We do not see a clear sign of detection in the beamformed data (in both MeerKAT observations).}
    \label{fig:ptuse_lc}
\end{figure}

Our MeerKAT observations recorded simultaneous search mode data at both UHF-band and L-band at 100\,$\mu$s sampling. We conducted a standard pulsar search using \texttt{PRESTO} \citep{presto} to look for pulsations up to a maximal DM of 550\,PC cm$^-3$, and acceleration of \texttt{zmax}=200. If the observed 1.5\,hr period is the orbital period, then using the entire 8\,hr observation to conduct pulsar searches will result in a null detection as the orbital motion will smear the pulses. Hence, we extracted the data when radio bursts were observed in the correlated data (roughly spanning 30\% of the 1.5\,hr period) and searched for pulsation in these ten data blocks. No candidates were found in the search. We then looked at the raw data to look for any bright single pulses that could help with DM estimation, but did not detect any. This can likely be due to one of the two reasons (or both): the lack of short timescale emission ($<$ 100\,ms) can degrade the sensitivity in PTUSE data and the lack of a phase-up calibration at the start of the observation can lead to improper phase calibration of the antennae which can also decrease the sensitivity. Figure~\ref{fig:ptuse_lc} shows this light curve at a resolution of 12.5\,ms and 10\,s, and we see that these radio bursts are too weak to be detected in the beamformed data. Hence, a robust estimation of DM (and hence distance) was not possible.

\subsection{Swift X-ray data reduction}\label{sec:swiftdata}
We observed \ulp\ with the \textit{Swift} telescope between July 9, 2024, and July 11, 2024, for a total of 8\,ks. Data were collected in photon counting mode during nine different scans, and we used the standard reduction techniques (using \textit{HEASoft}'s \texttt{xrtpipeline}) to generate good time intervals and flag bad pixels. We barycentered the recorded events (using \texttt{barycorr}) and generated a combined image and exposure map (using \texttt{ximage} and \texttt{xselect}). We used \texttt{detect} from \texttt{ximage} to detect point sources in the image and found a weak detection at $\alpha=14^{\rm h}48^{\rm m}34\fs4\pm1\fs3$ and $\delta=-68\degr56\arcmin45\arcsec\pm5\arcsec$, 3\arcsec\ away from the Meerkat position.

\subsection{XMM data reduction}\label{sec:xmmdata}
We observed \ulp\ using \textit{XMM-Newton} for 30\,ks on July 31, 2024. We used the standard reduction techniques\footnote{See \url{https://www.cosmos.esa.int/web/xmm-newton/sas-threads} for more details.} to remove time intervals dominated by background particle flaring and flag bad pixels which resulted in a net exposure time of 9.7\,ks for \texttt{pn} detectors. We barycentered the recorded events (using \texttt{barycen}) and generated full-band (0.2--12\,keV) images. We used \texttt{edetect-chain} to detect point sources in the image and found a detection at $\alpha=14^{\rm h}48^{\rm m}34\fs 3\pm 0\fs3$ and $\delta=-68\degr 56\arcmin 43\arcsec\pm 1\arcsec$, 2\arcsec\, away from the Meerkat position (see Figure~\ref{fig:xmm}). We used \texttt{evselect} to select all the photons in a 20\arcsec radius for estimating the spectrum. The source is closer to the edge of the chip, so we chose the background region to be a circular region at the same distance (as the source) from the readout node on the same chip (instead of a typical annular region) \footnote{\url{https://www.cosmos.esa.int/web/xmm-newton/sas-thread-pn-spectrum}.}. To increase the quality of the fit, we grouped the data (using \texttt{specgroup}) to have at least 5 counts per energy bin. Figure~\ref{fig:xmm} shows the resulting EPIC-PN image with a point source detected at the position of \ulp.

\subsection{Swift ultraviolet data reduction}\label{sec:swiftuvotdata}
\begin{figure}
    \centering
    \includegraphics[width=\columnwidth]{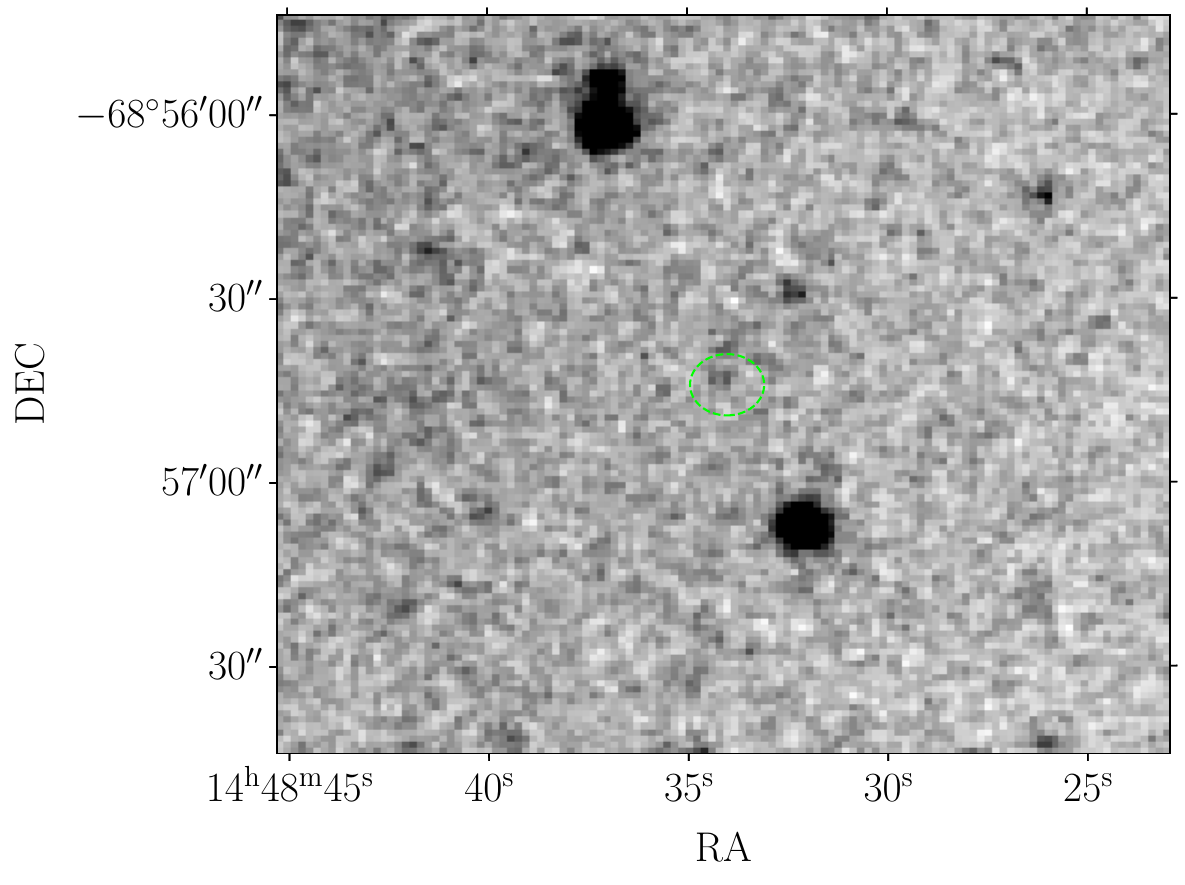}
    \caption{2.5\arcmin\, cut out of UVOT UVW2-band image showing a weak point source at the Meerkat position. The lime circle is 5\arcsec\, in radius.}
    \label{fig:uvot}
\end{figure}

Simultaneous UVOT data in the UVW2 filter were obtained with the Swift XRT observations. Standard data reduction techniques were used to reduce the data to generate images and exposure maps. We ran \texttt{uvotsource} on the individual images to estimate the aperture flux (in a 5\arcsec\, aperture), however, this resulted in non-detections in all the individual images. We stacked the data (using \texttt{uvotimsum}) and exposure maps to look for a faint source in the combined image which resulted in the detection of a weak (4.2-$\sigma$) source (see Figure~\ref{fig:uvot}) with magnitude 22.6$\pm$0.3\,mag (AB). Figure~\ref{fig:uvot} shows the stacked image showing a weak point source detected at \ulp's position.

\subsection{DECam data reduction}\label{sec:decamdata}
For the archival DECaPS data, we retrieved the instrument-calibrated images and catalogs for DECaPS\footnote{For more information, see \url{http://decaps.skymaps.info/}.} in all five bands (\textit{g, r, i, z, Y}) to estimate the source magnitude over this period. A point source was detected in multiple bands over multiple epochs. We used the individual catalogs and the standard zeropoints (from the DECaPS reductions) to estimate the apparent magnitude of the source\footnote{More details can be found at \url{http://decaps.skymaps.info/release/datamodel/files/DR2_REDUX/zps.html}.}. For the images that resulted in non-detections, there can be a faint source below the detection threshold (5-$\sigma$), and also, the limiting magnitude may not be the true estimate of local noise. In this case, we used \texttt{photutils} \citep{photutils} to look for faint sources\footnote{Using an aperture radius of $\frac{2}{3} \times$(seeing).} and estimate the local noise. Aperture corrections are estimated using \texttt{sextractor} \citep{sextractor}. Figure~\ref{fig:decaps} shows the light curve indicating variability at optical wavelengths.

\subsection{Lesedi data reduction}
\label{sec:lesedidata}
We obtained multiple 480\,sec exposures in the \textit{g}-band with the 1-m Lesedi telescope. Images were bias-corrected, flat-fielded, and astrometrically calibrated with \texttt{astrometry.net} \citep{astrometry} using the GAIA DR2 index files. We visually inspected the images and did not find a point source at the position of \ulp\ in the individual images. We used \texttt{swarp} \citep{swarp} to stack the individual images, which resulted in the detection of a faint point source (7200\,sec of exposure time). We used \texttt{sextractor} \citep{sextractor} and the corresponding DECaPS field to perform photometric calibration and estimate the zero-point. We performed aperture photometry at the position of \ulp\ and obtained a \textit{g}-band magnitude of 22.0$\pm$0.2.

\subsection{Magellan data reduction}
\label{sec:magellandata}
Data were reduced using \citet{Kelson2014}, which corrects for dark currents and non-linearity; the sky background is subtracted using a bivariate wavelet model. Individual 14.56 s frames were stacked, accounting for camera distortion and variations in sky background. The observing conditions (seeing) varied from 0.7-2\arcsec\ between the observations, with the worst seeing on Jul 22, which resulted in the non-detection of the source on July 22. We performed aperture photometry using \texttt{sextractor}. We used corresponding VHS catalogs to perform photometric calibration. The source's magnitude was 22.0$\pm$0.2, and 22.1$\pm$0.2 on July 18 and July 23, 2024, respectively.

\section{Radio burst properties and profiles}
\label{sec:bursts}
\begin{figure*}
    \includegraphics[width=\textwidth]{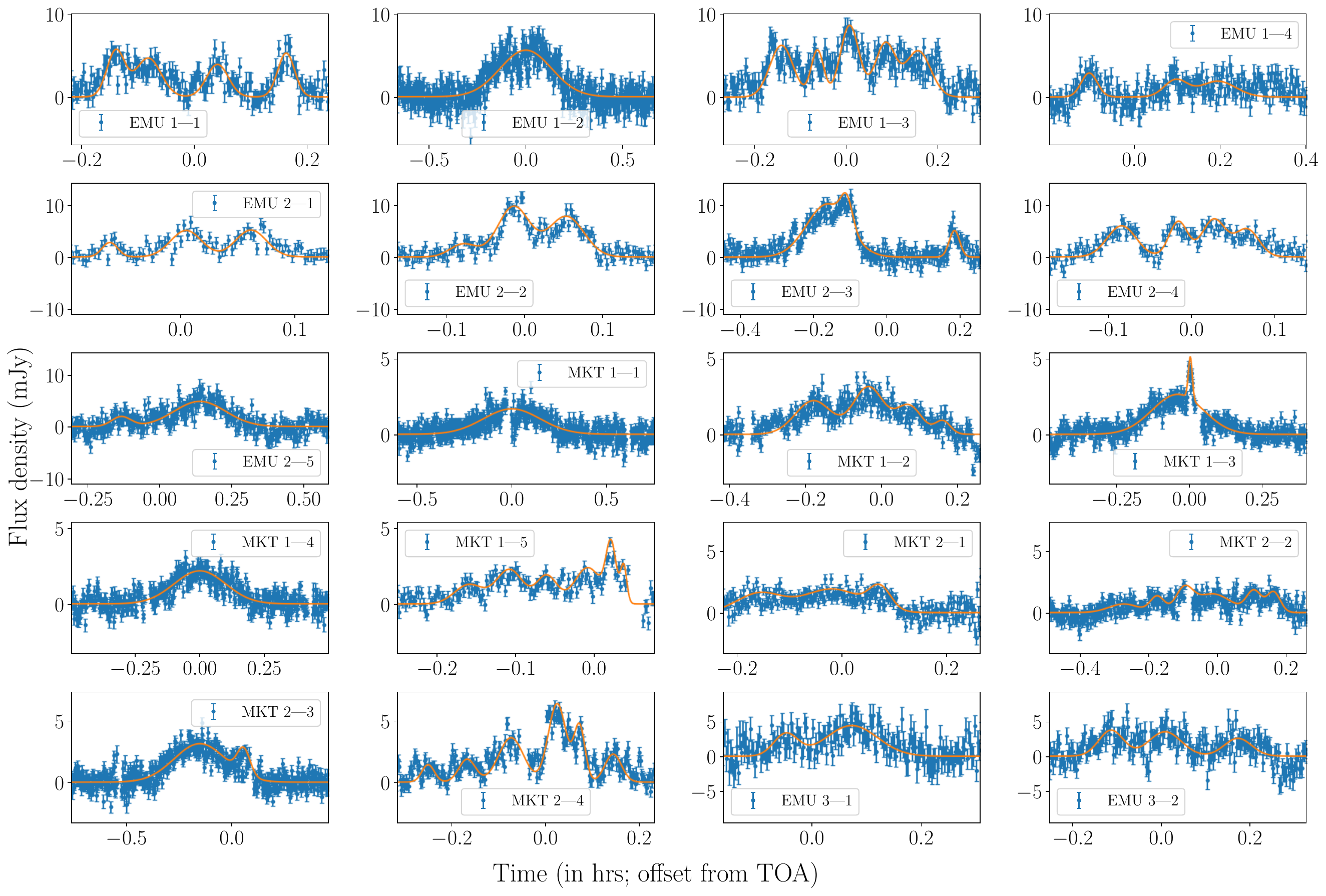}
    \caption{Figure showing the individual radio pulses of \ulp\  and their model fits. The blue data points show the pulses. Pulses are ordered chronologically with ``MKT 1--4'' referring to the fourth pulse in the first observation at MeerKAT. The overlaid orange curves are the model fits to the pulses. The corresponding TOAs are given in Table~\ref{tab:pol}.}
    \label{fig:profiles}
\end{figure*}

Here we provide the properties (times, flux densities, polarisation information) for all bursts in Table~\ref{tab:pol}, and the lightcurves of the individual bursts in Figure~\ref{fig:profiles}. Our model profile fits to the individual bursts are also shown in Figure~\ref{fig:profiles}.

\begin{table*}
\centering
\caption{Polarisation properties of bright (5-$\sigma$) bursts from EMU and MKT observations.}
\label{tab:pol}
\begin{tabular}{ccclrrrrrr}
\hline
Telescope & Date & Band & TOA & $I$ & $Q$ & $U$ & $V$ & $L$$^a$ & $C$ $^b$ \\
 & & & (MJD) & (mJy) & (mJy) & (mJy) & (mJy) & (\%) & (\%) \\
\hline
ASKAP & 2023-06-15 & EMU & 60110.47030(61) & 6.27(76) & 2.73(74) & 1.95(76) & $-$3.62(75) & 54(14) & $-$58(14)\phantom{c} \\
ASKAP & 2023-06-15 & EMU & 60110.53359(20) & 6.88(80) & $-$2.72(77) & 4.91(75) & $-$3.74(75) & 82(15) & $-$54(13)\phantom{c} \\
ASKAP & 2023-06-15 & EMU & 60110.59488(33) & 7.95(80) & $-$4.06(72) & 4.19(76) & $-$3.31(75) & 73(12)$^c$ & $-$42(10)\phantom{c} \\
ASKAP & 2024-05-26 & EMU & 60456.48671(52) & 5.49(81) & $<0.70$ & $<0.76$ & $-$5.80(77) & $<19$ & $-$100(21)\phantom{c} \\
ASKAP & 2024-05-26 & EMU & 60456.53589(32) & 11.52(80) & $<0.76$ & $<0.72$ & 5.38(71) & $<9$ & 47(\phantom{0}7)$^c$ \\
ASKAP & 2024-05-26 & EMU & 60456.6032(13) & 11.08(79) & $<0.76$ & $<0.74$ & $-$7.77(71) & $<9$ & $-$70(\phantom{0}8)$^c$ \\
ASKAP &2024-05-26 & EMU & 60456.66297(19) & 6.76(76) & $<0.77$ & $<0.74$ & $-$4.34(75) & $<15$ & $-$64(13)$^c$ \\
ASKAP & 2024-05-26 & EMU & 60456.7198(12) & 6.44(92) & $<0.80$ & $<0.86$ & $-$4.13(77) & $<16$ & $-$64(15)$^c$ \\
ASKAP & 2025-12-25 & EMU & 60669.85399(41) & 5.9(1.1) & $<1.40$ & $-$3.65(89) & $<1.2$ & 62(19) &  $<20$\phantom{c} \\
ASKAP & 2025-12-25 & EMU & 60669.79307(79) & 5.35(92) & 2.58(84) & $-$3.06(85) & $-$2.30(82) & 75(20)  & $-$43(17)\phantom{c} \\
\hline
ATCA & 2024-06-27 & L/S & ... & 3.1(1.1) & $<1$\phantom{.00} & $<1$\phantom{.00} & 2.6(0.9) & $<45$ & 84 (41)\phantom{c} \\
\hline
MeerKAT & 2024-07-30 & UHF & 60521.71856(24) & 4.47(28) & 1.09(31) & 2.30(29) & 1.60(28) & 57(7)\phantom{c} & 36(\phantom{0}7)\phantom{c} \\
MeerKAT & 2024-07-30 & UHF & 60521.78505(78) & 5.30(21) & $-$2.69(21) & $-$1.18(18) & 2.10(19) & 55(4) & 40(\phantom{0}4)$^c$ \\
MeerKAT & 2024-07-30 & UHF & 60521.847541(50) & 5.60(19) & 1.48(18) & $-$3.18(20) & 1.44(19) & 63(4) & 26(\phantom{0}4)\phantom{c} \\
MeerKAT & 2024-07-30 & UHF & 60521.90951(17) & 5.05(22) & 1.98(26) & 2.09(24) & 1.60(26) & 57(5) & 32(\phantom{0}5)\phantom{c} \\
MeerKAT & 2024-07-30 & UHF & 60521.97725(25) & 4.42(29) & $-$1.35(26) & 1.35(31) & 2.29(29) & 43(7) & 52(\phantom{0}7)$^c$ \\
MeerKAT & 2024-07-31 & UHF & 60522.82280(66) & 3.61(30) & 1.18(25) & $-$1.71(26) & 1.43(25) & 57(9) & 40(\phantom{0}8)\phantom{c} \\
MeerKAT & 2024-07-31 & UHF & 60522.89375(46) & 8.10(40) & 3.52(39) & $-$3.37(37) & $-$3.54(37) & 60(6) & $-$44(\phantom{0}5)\phantom{c} \\
MeerKAT & 2024-07-31 & UHF & 60522.95185(74) & 12.48(52) & $-$5.08(49) & 6.75(47) & 3.70(52) & 68(5) & 30(\phantom{0}4)$^c$ \\
\hline
\end{tabular}
\text{$^a$ Linear polarisation. Upper limits correspond to 1-$\sigma$.}\\
\text{$^b$ Circular polarisation. Upper limits correspond to 1-$\sigma$.}\\
\text{$^c$ Polarisation reaches 100\% at least once during the burst.}
\end{table*}

\section{UV to NIR SED modeling}\label{sec:mm_spec_app}

\subsection{Determination of quiescent state SED}\label{sec:quiescent_sed}
The DECaPS light curve shows an occurrence of an outburst (\S\ref{sec:decam}), and so estimating quiescent SED requires the knowledge of quiescent state source fluxes. However, the source seems to have returned to its quiescent state towards the end of the data span. Evidence for this comes from the relatively stable $g$ and $r$-band magnitudes on two different epochs separated by $\sim$200\,days. In addition, our $g$-band observations by the 1-m Lesedi telescope (see \S\ref{sec:mookodi}) also yielded a flux measurement similar to the quiescent flux observed by DECaPS. Hence we consider that the final observations of DECaPS correspond to the source flux in its quiescent state.

At NIR wavelengths, VHS observations predating the DECaPS outburst resulted in non-detections at both \textit{J} and $K_s$ bands. However, our follow-up deeper observations by the FourStar, long after the DECaPS outburst, at the $J$-band resulted in detection, consistent with VHS observations, and hence we adopt a combination of these results for the SED at NIR wavelengths (\S\ref{sec:vhs}). Similarly, at UV wavelengths, we assume that the source's quiescent flux is the same as observed by the UVOT (\S\ref{sec:uvot}). Table~\ref{tab:sed} shows the broadband flux density measurements from UV to NIR. We fit the broadband SED using different models (more details on synthetic photometry are provided in \S\ref{sec:syn_phot}).

In order to fully model the SED, we need to correct for foreground interstellar extinction.  The 3D dust extinction map from \citet{Zucker2025} predicts the reddening to saturate at $E(B-V)\approx 0.2\,$mag beyond 3\,kpc (or extinction $A_V\approx 0.6\,$mag) in the direction of \ulp. Although we do not restrict $A_V$ to this value during fitting, we evaluate the likelihood of the fit by comparing the $A_V$ values from the fits to this.

\subsection{SED modeling methodology}\label{sec:syn_phot}
We consider several models for the SED, but to compare them to the data we use a common methodology. For a given model spectrum $F_{\lambda}(\lambda)$, and a filter transmission curve\footnote{Transmission curves for various instruments were obtained from \url{http://svo2.cab.inta-csic.es/theory/fps/}.} $S_{\lambda,k}(\lambda)$ \citep{svo}, the model predicted flux is given by \\
\[
F_{\lambda,k} = \frac{\int_{\lambda} S_{\lambda,k}(\lambda) F_{\lambda}(\lambda) d\lambda}{\int_\lambda S_{\lambda,k}(\lambda) d\lambda}
\]
where `$k$' represents different wavebands (UV to NIR). We simultaneously use both detections and upper limits in the fit by defining the likelihood function as 
\[
L(\rm \{x_i\}) = \prod_k p_k(\{x_i\})
\]
where $x_i$ are the model parameters, and $p_k$ is the probability of the model yielding the observed data point `$k$' (UV to NIR observations). In the case of detections, it reduces to a Gaussian probability 
\[
p_k(\{x_i\}) \propto \exp\left[-\left(\frac{F_{\lambda,k}^{\rm obs} - F_{\lambda,k}^{\rm model}\{x_i\}}{F_{\lambda,{\rm err},k}^{\rm obs}}\right)^2 \right]
\]
and in the case of non-detections, it corresponds to the probability of the model yielding a value less than the observed flux, i.e.,
\[
p_k(\{x_i\}) = \frac{1}{2}\left[ 1 + \mathrm{erf} \left(\frac{5F_{\lambda,{\rm noise},k}^{obs} - F_{\lambda,k}^{\rm model}\{x_i\}}{F_{\lambda,{\rm err},k}^{\rm obs}}\right) \right]
\]
where `erf' is the error function.

\subsection{Stellar Atmospheres}\label{sec:bb}
We first tried to model the source spectrum assuming it is a star. The source SED, in this case, can be parameterized by three model parameters, the effective temperature, surface gravity, and the overall amplitude ($T_{\rm eff}$, $\log g$, $A_\star$) in addition to metallicity. We obtained standard stellar model atmospheres\footnote{From \url{https://www.stsci.edu/hst/instrumentation/reference-data-for-calibration-and-tools/astronomical-catalogs/castelli-and-kurucz-atlas}.} \citep{ck04models}. Given the sparse broadband data, the resulting fit will be insensitive to $\log g$, and metallicity and hence we performed separate fits ($T_{\rm eff}$, $A_\star$, $A_V$) for discrete values of $\log g$ and metallicity. The results do not change, as expected, as we vary $\log g$ and metallicity. 

\begin{figure*}
    \includegraphics[width=0.95\columnwidth]{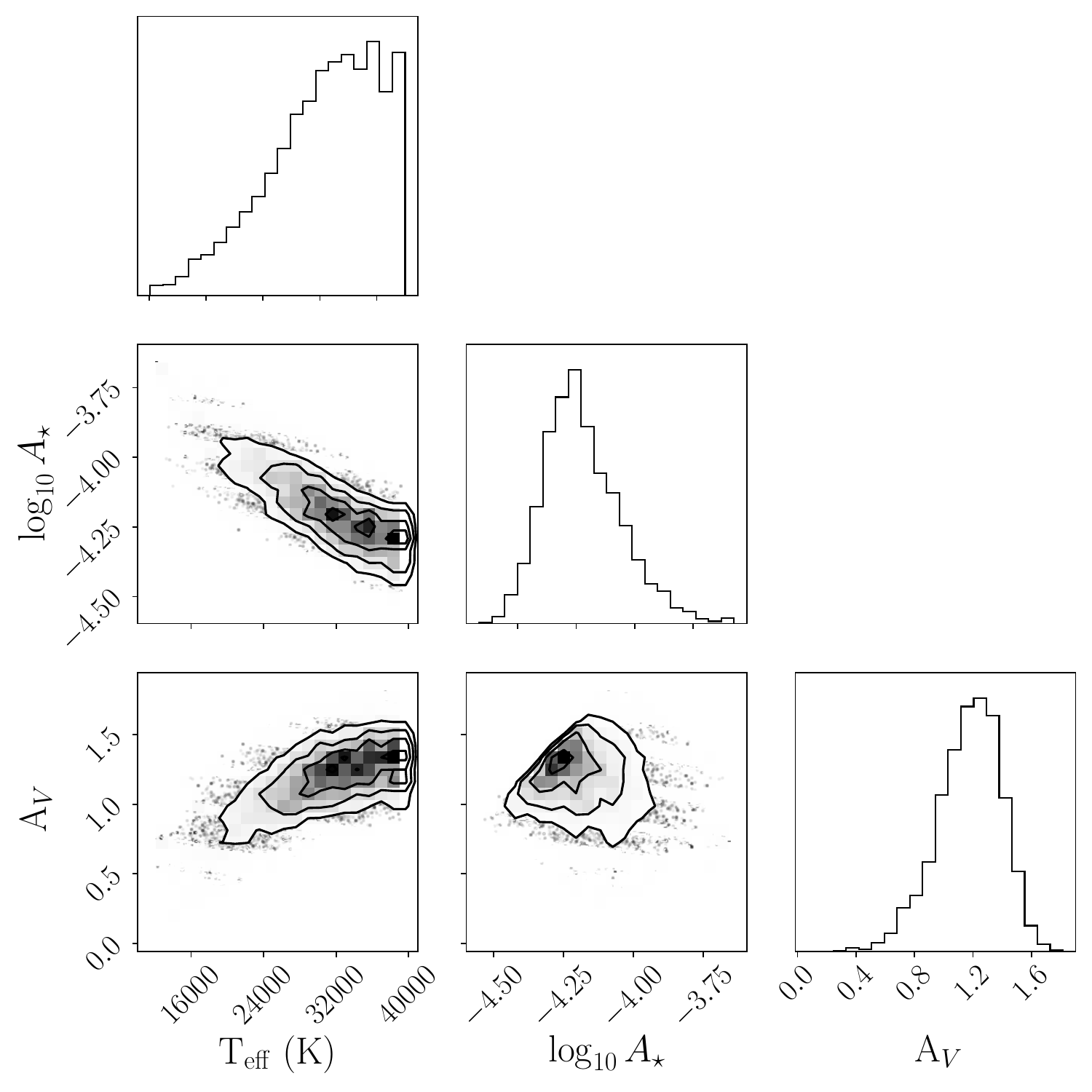}
    \includegraphics[width=0.95\columnwidth]{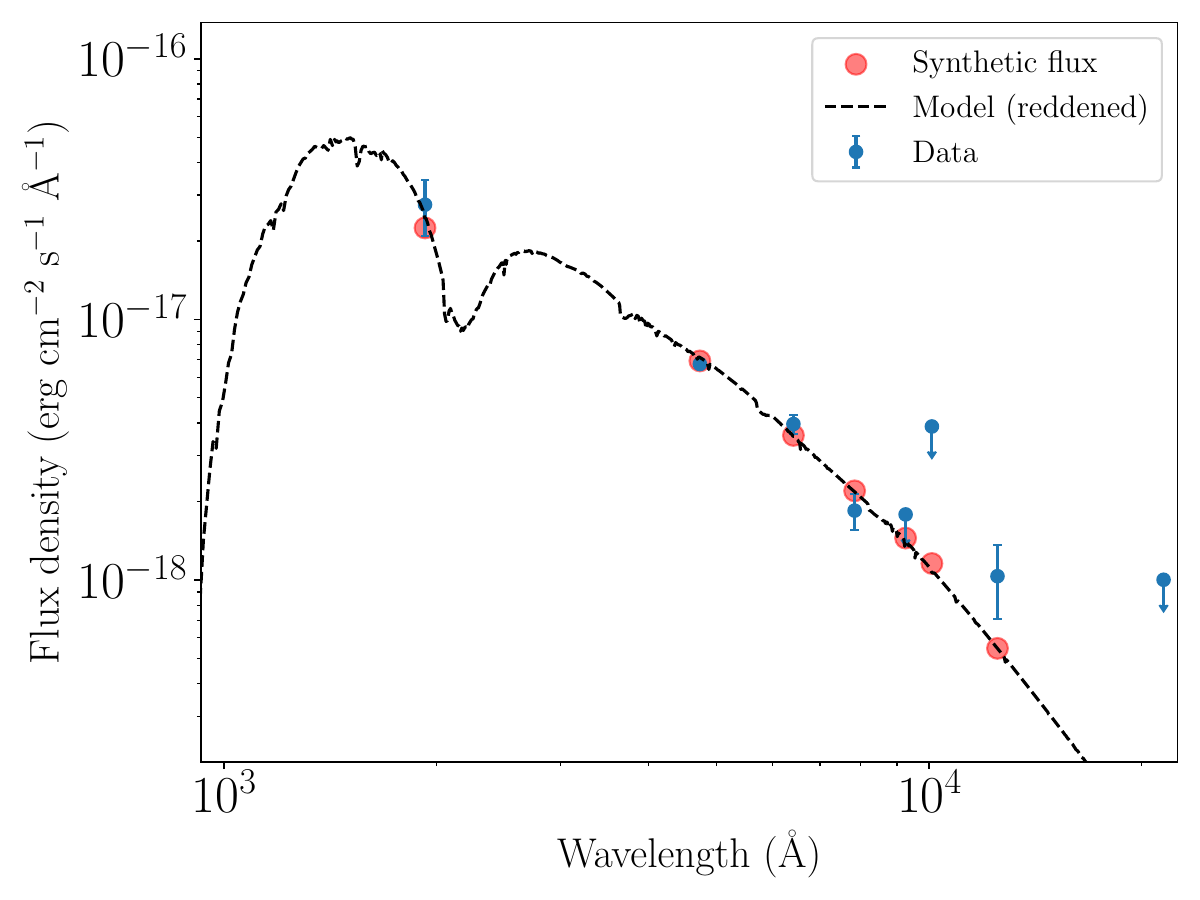}
    \caption{An example of a stellar atmospheric model fit to the SED of \ulp\  for $\log g=4$ and solar metallicity. Shown in the left panel are the posteriors for the model parameters  ($T_{\rm eff}, A_\star, A_V$). $A_\star$ is provided in units of $(R_{\astrosun}/{\rm kpc})^2$. The (reddened) model corresponding to the best-fit parameters is shown in the right panel.}
    \label{fig:sed_star}
\end{figure*}

Figure~\ref{fig:sed_star} shows an example model fit. Two problems arise: the first is the extremely high values of $T_{\rm eff}$, which push against the upper edge of the model atmospheres (40000\,K), which would imply a hot star. The second issue was that these model atmospheres could not simultaneously model the UV and NIR data. The observed data were flatter than the steeper models. Hence, any model that adequately fits the NIR data, overpredicts the extinction to be $A_V\approx$1.5, likely trying to flatten the model to simultaneously account for the UV data. The resulting amplitude (which is $(R_{\star}/d)^2$, with $R_\star$ the radius of the star in solar radii and $d$ the distance in kpc) implies that the star has to be at a distance of $>$50\,kpc (for a half-solar radius star; \citealt{Moffat1996,Crowther2007}).

\subsection{Interacting binary}\label{sec:disk}
Next, we consider the case of an interacting binary (through active mass transfer) to explain the SED by modeling it as a multi-temperature blackbody (akin to an accretion disk). In general, three parameters characterize such systems (that are not covariant), temperature at the inner radius $T_{\rm in}$, the ratio of outer to inner radius ($r_{\rm out}$/$r_{\rm in}$), and an overall amplitude (combination of inner radius, distance and the inclination angle). For the temperature and wavelengths that we are interested in, we can assume large values of $r_{\rm out}$/$r_{\rm in}$, which has an effect on the spectrum only at far IR wavelengths. In this case, we follow \citep{diskbb} and define the spectrum as 
\[
F_{\nu}(\nu) = A_{\rm disk} \int_1^{\infty} B_{\nu}(T(r), \nu) dr 
\]
where $A_{\rm disk}=2\pi\left(r_{\rm in}/d\right)^2\,\cos(i)$, $i$ is the inclination angle of the system, and $B_{\nu}(T(r), \nu)$ is a blackbody spectrum with temperature $T(r)$ with the temperature evolution given by 
\[
T(r) = T_{\rm in}  r^{-3/4}  (1 - r^{-1/2})^{1/4}
\]

\begin{figure*}
    \includegraphics[width=0.95\columnwidth]{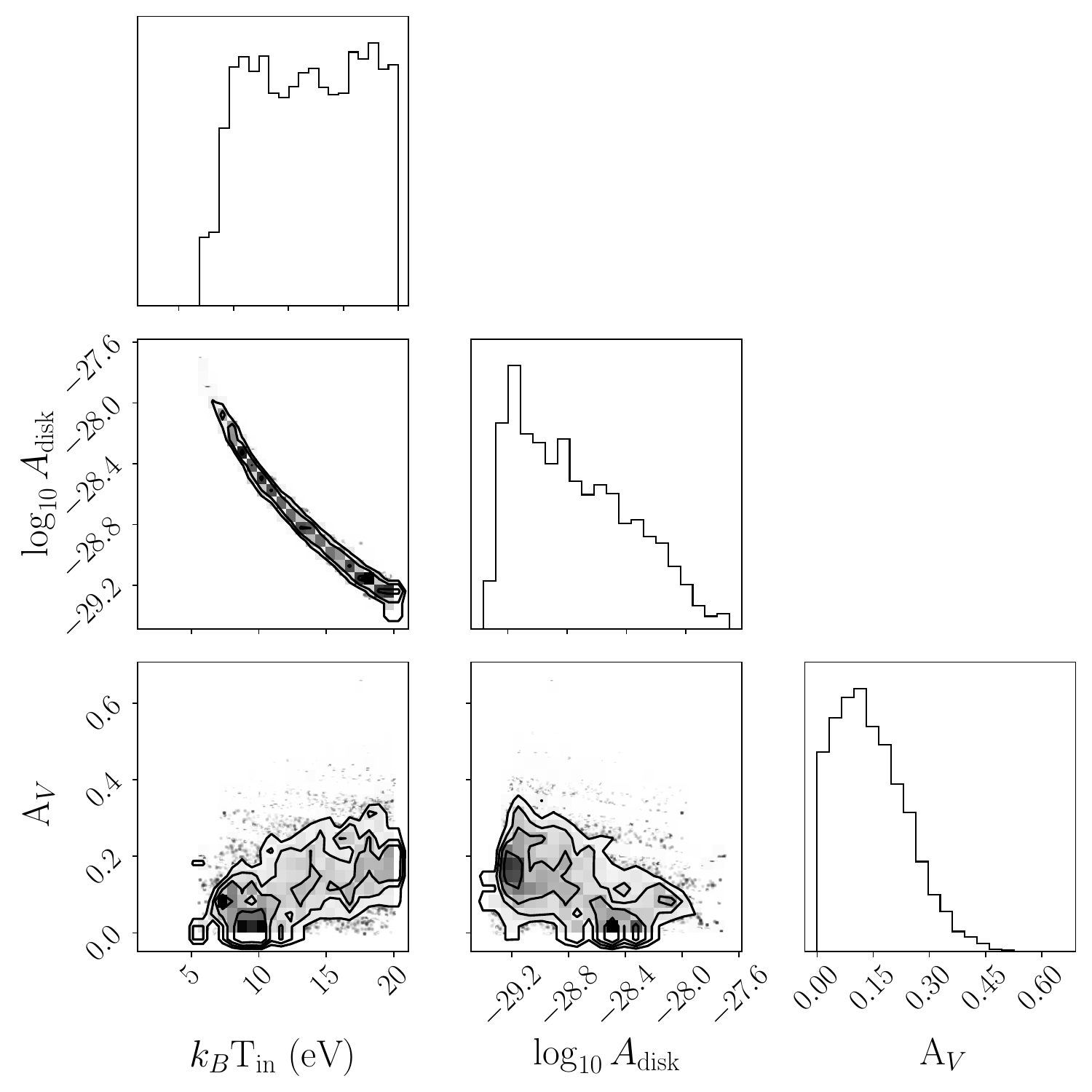}
    \includegraphics[width=0.95\columnwidth]{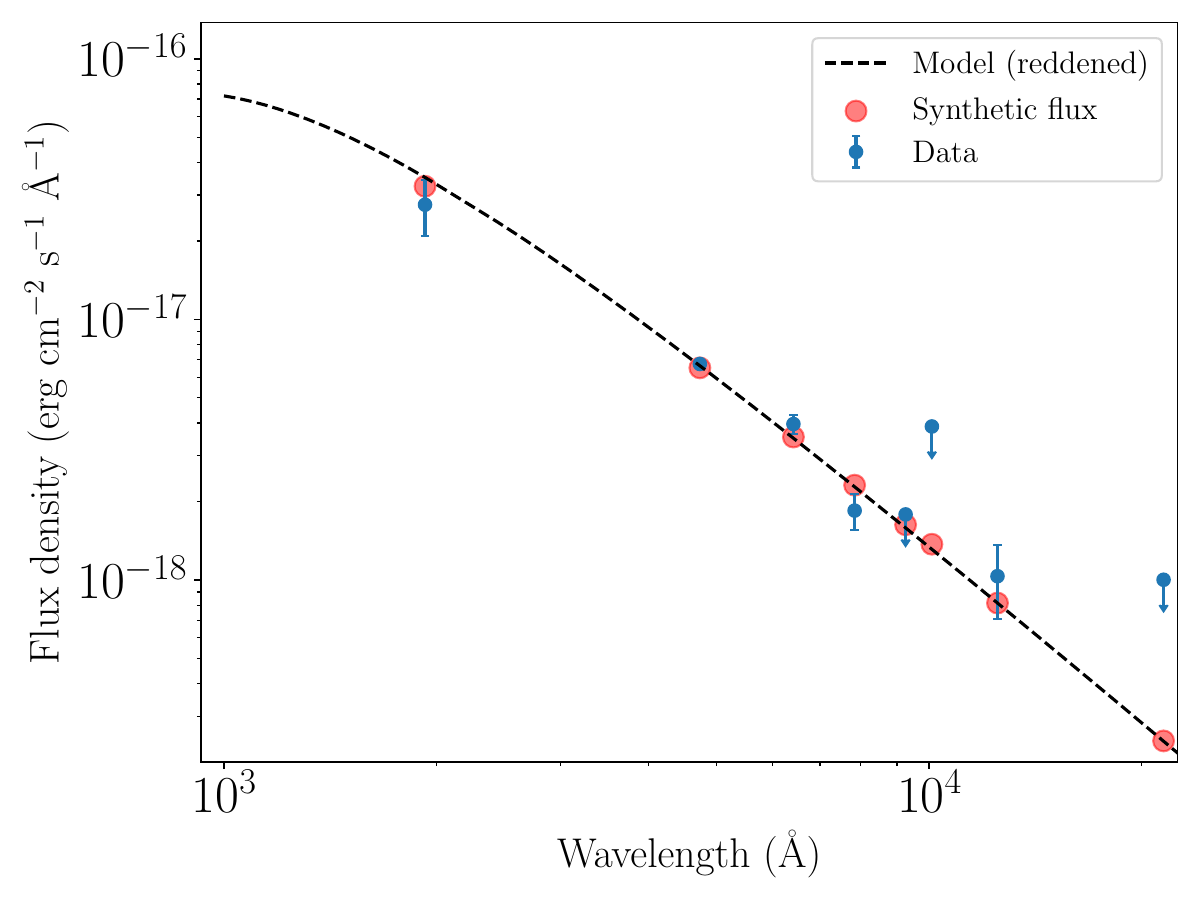}
    \caption{\textit{Left:} Posteriors for the model parameters from a separate Bayesian fit. The spread of $k_B T_{\rm in}$ over a large range and the covariance between $k_B T_{\rm in}$ and $A_{\rm disk}$ results in a tail for the amplitude posterior. \textit{Right}: Multi-temperature accretion disk fit for the data. The blue dots show the observed data, the black dashed line shows the model (reddened) spectrum and the red dots show the synthetic flux derived from the model. }
    \label{fig:sed_disk}
\end{figure*}

A simple least-squares fit estimated the model parameters to be $T_{\rm in}$=92\,000$\pm$3\,000\,K, $\log_{10} A_{\rm disk}=-28.2\pm0.2$, and $A_V=0.01\pm0.11$. The corresponding fit is shown in the right panel of Figure~\ref{fig:sed_disk}. However, looking at the fit, this also suggested that we are in a regime where $T_{\rm in}$ and $A_{\rm disk}$ can be covariant. Disk blackbody spectra have three distinctive components: an exponential tail at very high energies ($ h\nu\gg k_B T_{\rm in}$), a $\nu^{1/3}$ dependence at intermediate energies, and $\nu^2$ tail at very low energies. Our wavelength sampling indicates that we are in the intermediate regime and hence changing the temperature $T_{\rm in}$ scales the spectrum, similar to $A_{\rm disk}$. This effect can be seen in the 2D posterior ($T_{\rm in}, A_{\rm disk}$) in the left panel in Figure~\ref{fig:sed_disk}. Even with this effect, it is clear that $T_{\rm in}$ values below 92\,000\,K are not favored (despite the prior allowing for it). Hence we consider this the lower limit on the inner radius of the disk $T_{\rm in}\geq92\,000$\,K. The distance to the source can then be estimated for a given inner radius and an inclination angle ($i$). For inclination angles that are favorable for the radio emission (closer to edge-on than face-on), and an inner radius $r_{\rm in}$ of $\sim R_{\oplus}$\footnote{We consider the inner edge of the disk to be at twice the radius of a $1M_{\astrosun}$ WD ($0.5R_{\oplus}$).}, we get a maximum distance of $\sim 6$\,kpc (for $i=80^{\circ}$).

\subsection{Detached binary}\label{sec:wd_md}
Given \ulp's radio resemblance to other LPTs \citep{de_ruiter_white_2024,hurley-walker_29-hour_2024}, we considered the scenario in which \ulp\ is a detached binary (with no active mass transfer). In this case, the emission arises from the individual components --- something like  a hot WD and a cold brown dwarf (BD) star. We parameterize the total spectrum using 5 parameters, $2 \times$ (effective temperature, amplitude) and $A_V$. We used cooling models from \cite{Koester2010} for the WD component\footnote{\url{http://svo2.cab.inta-csic.es/theory/newov2/syph.php?model=koester2}} and BT-DUSTY models\footnote{\url{http://svo2.cab.inta-csic.es/theory/newov2/syph.php?model=bt-dusty}} \citep{Allard2012} for the brown dwarf component. Given our sparse data set, we are insensitive to $\log g$ and metallicity and hence we freeze $\log g=7.0$ for the WD, and $\log g=5.0$ with solar metallicity for the dwarf.

\begin{figure}
    \centering
    \includegraphics[width=\columnwidth]{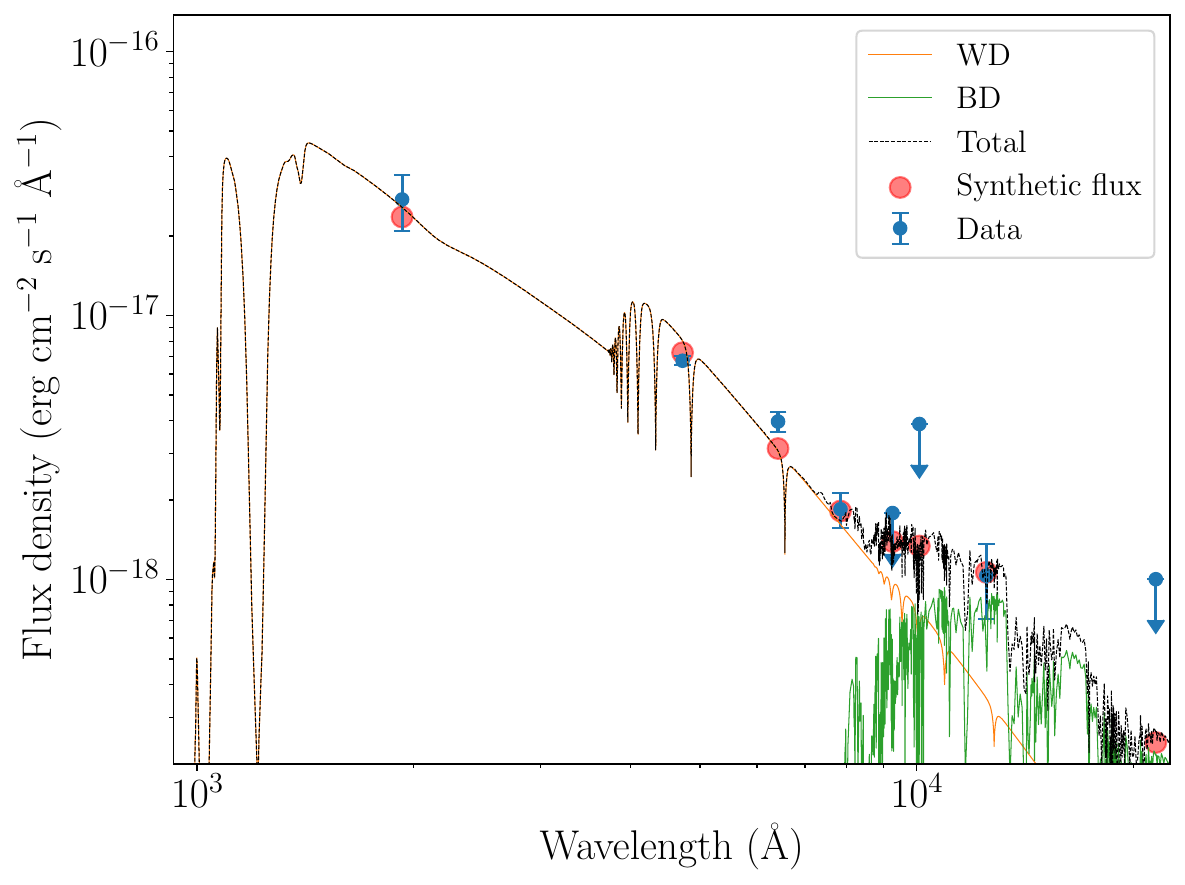}
    \caption{Two-component fit to the SED of \ulp. The blue points show the observed data, the orange line shows the first (WD) component, the green line shows the second (brown dwarf) component and the overlaid black dashed line shows the combined spectrum.}
    \label{fig:sed_multi_bb}
\end{figure}

Figure~\ref{fig:sed_multi_bb} shows the resulting fit to the observed SED --- with the best fit parameters $T_{\rm eff, 1}=13286\pm570$\,K, $T_{\rm eff, 1}=2300\pm108$\,K, $\left(R_1/d\right)^2=(1.2\pm0.2)$\,$R_{\oplus}^2$\,$\rm kpc^{-2}$, $\left(R_2/d\right)^2=(0.01\pm0.07)$\,$R_{\astrosun}^2$\,$\rm kpc^{-2}$, $A_V=0.10\pm0.13$. The poor fit of the brown dwarf component is expected since we are using 1 detection ($J$-band) and 1 non-detection ($K_s$-band) to constrain the model, and hence parameter covariances can not be broken. If we assume the WD to be the size of Earth ($R_{\oplus}$), then this places the system at 0.9\,kpc and yields a radius estimate of the brown dwarf to be 0.09 $R_{\astrosun}$ (consistent with a cool dwarf). Given the period, the semi-major axis of the system (for a $1\, M_{\astrosun}$ WD and a $0.1\, M_{\astrosun}$ dwarf) can be 1\,$R_{\astrosun}$, 10 times the combined radii. Hence such a system is allowed by the data to be the progenitor of \ulp. It is worth reiterating that the goal here is to probe whether such a system is plausible given the data, and not to robustly constrain the SED (our data set is too sparse to do that).

\section{Are the narrow-band features a result of scintillation?}

As shown in Figure~\ref{fig:burst_period_mkt}, the MeerKAT frequency spectra show a periodic narrow-band structure. This could either be intrinsic to the source or caused by the interstellar propagation, due to scintillation. It is clear, given the frequency variations that we would have to be in the strong diffractive scintillation regime if this were due to scintillation. For estimating various parameters, we adopt the simplified thin screen model, a thin screen placed midway between the source and the Earth, i.e., at $d/2$ \citep{handbook}. Scatter broadening due to inhomogeneities increases the angular size of the source to $\Delta\theta_d\propto d^2/\nu^4$. We use this to compute the scintillation bandwidth, by demanding that the observed bandwidth of $\Delta\nu=$17\,MHz corresponds to two consecutive constructive interference peaks, i.e. $\Delta\nu_{\rm DISS}d\phi/d\nu=2\pi$, which yields $\Delta\nu\propto\nu^4/d^2$. The first hint that the observed frequency variation is not due to scintillation comes from the consistent bandwidth (see Figure~\ref{fig:burst_period_mkt}) at both UHF and L-bands. In the case of diffractive scintillation, the bandwidth at L-band should be at least 4 times larger than that observed at UHF.

We can also estimate the scintillation timescale $\Delta t_{\rm DISS}$, using the scintillation bandwidth and the relative transverse velocity of the sources $v_s$ as 
\[
\frac{\Delta t_{\rm DISS}}{1s} = \nu_{\rm GHz}^{-1} \ \Delta\nu_{\rm MHz}^{1/2} \ d_{\rm kpc}^{1/2}\  \left(\frac{v_s}{10^4\, \mathrm{km}\ \mathrm{s}^{-1}}\right)^{-1}
\]
We exploit the consistency between the Meerkat position and optical position (using DECaPS), separated by 6\,yrs, to within 2.5\arcsec\, to estimate the velocity of the source to be less than 2000\,km s$^{-1}$ (for a nominal distance of 1\,kpc; see \S\ref{sec:mm_spec}). Using this, we estimate $\Delta t_{\rm DISS}$ to be 20\,s, greater than the integration time. However, looking at the light curve (see Figure~\ref{fig:mkt_lc}), we do not see any oscillatory variations on this timescale. Rather, the light curve is dominated by the systematic rise and decay of the bursts. Both these inconsistencies --- frequency-independent scintillation bandwidth and the absence of temporal variation, makes it unlikely that the observed narrow-band structure is caused by diffractive scintillation. There is a bright pulsar, B1451$-$68, in the field of view at both UHF and L bands. B1451$-$68 is at a DM of 8.6, hence, it is likely that it will be affected by diffractive scintillation (predicted bandwidth is 93\,MHz at 1\,GHz, and timescale is 1600\,s for a velocity of $100\,{\rm km\,s}^{-1}$). We created a dynamic spectrum at the position of PSR~B1451$-$68, treating it just as a continuum source (i.e., ignoring the pulse period).  When inspected the dynamic spectrum, we identified the presence of stochastic ``scintles" or local brightness maxima in the time and frequency domains, with parameters similar to those predicted. For this pulsar, the scintillation behavior is as expected, the scintillation bandwidth is larger at L-band than at UHF and the pattern does not repeat between maxima. Comparing the dynamic spectra of PSR B1451$-$68 and \ulp\ we reach the same conclusion that the highly structured narrow-band emission seen in \ulp\ is unlikely to be caused by the scintillation seen in PSR~B1451$-$68. 

Temporal flux density variations can also be caused by refractive interstellar scintillation (RISS), however, RISS affects flux densities on longer timescales ($\sim$months;  \citealt{Bhat1999}) and the flux variations are usually modest (few tens of \%). Hence, it is unlikely that the observed structure is caused by RISS.

\bsp	
\label{lastpage}
\end{document}